\documentclass[twocolumn, twocolappendix]{aastex63}
\usepackage{amsmath,amsfonts,amsthm}
\usepackage{lineno}
\usepackage{ulem}
\shorttitle{Milky Way Nuclear Cluster Star Formation History}
\shortauthors{Chen et al.}
\graphicspath{{./}{figures/}}

\begin{document}

\title{THE STAR FORMATION HISTORY OF THE MILKY WAY'S NUCLEAR STAR CLUSTER}

\correspondingauthor{Zhuo Chen}
\email{zhuochen@astro.ucla.edu}

\author[0000-0002-3038-3896]{Zhuo Chen}
\affiliation{Department of Physics and Astronomy, University of California, Los Angeles, CA 90095-1547, USA}

\author[0000-0001-9554-6062]{Tuan Do}
\affiliation{Department of Physics and Astronomy, University of California, Los Angeles, CA 90095-1547, USA}

\author[0000-0003-3230-5055]{Andrea Ghez}
\affiliation{Department of Physics and Astronomy, University of California, Los Angeles, CA 90095-1547, USA}

\author[0000-0003-2874-1196]{Matthew Hosek Jr.}
\altaffiliation{Brinson Prize Fellow}
\affiliation{Department of Physics and Astronomy, University of California, Los Angeles, CA 90095-1547, USA}

\author[0000-0002-0160-7221]{Anja Feldmeier-Krause}
\affiliation{Max Planck Institute for Astronomy, K\"{o}nigstuhl 17, D-69117 Heidelberg, Germany}

\author[0000-0003-3765-8001]{Devin Chu}
\affiliation{Department of Physics and Astronomy, University of California, Los Angeles, CA 90095-1547, USA}

\author[0000-0001-7017-8582]{Rory Bentley}
\affiliation{Department of Physics and Astronomy, University of California, Los Angeles, CA 90095-1547, USA}

\author[0000-0001-9611-0009]{Jessica R. Lu}
\affiliation{Department of Astronomy, University of California, Berkeley, CA 94720-3411, USA}

\author[0000-0002-6753-2066]{Mark R. Morris}
\affiliation{Department of Physics and Astronomy, University of California, Los Angeles, CA 90095-1547, USA}






\begin{abstract}

We report the first star formation history study of the Milky Way’s nuclear star cluster (NSC) that includes observational constraints from a large sample of stellar metallicity measurements. These metallicity measurements were obtained from recent surveys from Gemini and VLT of 770 late-type stars within the central 1.5 pc. These metallicity measurements, along with photometry and spectroscopically derived temperatures, are forward modeled with a Bayesian inference approach.  Including metallicity measurements improves the overall fit quality, as the low-temperature red giants that were previously difficult to constrain are now accounted for, and the best fit favors a two-component model. The dominant component contains 93\%$\pm$3\% of the mass, is metal-rich ($\overline{[M/H]}\sim$0.45), and has an age of 5$^{+3}_{-2}$ Gyr, which is $\sim$3 Gyr younger than earlier studies with fixed (solar) metallicity; this younger age challenges co-evolutionary models in which the NSC and supermassive black holes formed simultaneously at early times. The minor population component has low metallicity ($\overline{[M/H]}\sim$ -1.1) and contains $\sim$7\% of the stellar mass. The age of the minor component is uncertain (0.1 - 5 Gyr old). Using the estimated parameters, we infer the following NSC stellar remnant population (with $\sim$18\% uncertainty): 1.5$\times$10$^5$ neutron stars, 2.5$\times$10$^5$ stellar mass black holes (BHs) and 2.2$\times$10$^4$ BH-BH binaries. These predictions result in 2-4 times fewer neutron stars compared to earlier predictions that assume solar metallicity, introducing a possible new path to understand the so-called ``missing pulsar problem”.  Finally, we present updated predictions for the BH-BH merger rates (0.01-3 Gpc$^{-3}$yr$^{-1}$).

\end{abstract}



\section{Introduction} \label{sec:intro}

The innermost region of most galaxies is occupied by a spectacularly dense and massive assembly of stars, which forms a nuclear star cluster (NSC). The star formation in this region is believed to be affected by the central supermassive black hole (SMBH), but the physical mechanisms behind it are not entirely known. The center of the Milky Way galaxy is host to the closest example of a SMBH ($4.2 \times 10^6 M_{\odot}$, e.g., \citealt{Ghez et al. 2008, Gillessen et al. 2009, Do et al. 2019}), embedded in a NSC ($\sim 2.5 \times 10^7 M_{\odot}$, e.g., \citealt{Launhardt et al. 2002, Schodel et al. 2014}). Given its proximity, the Milky Way NSC provides a unique opportunity to resolve the stellar population and to study phenomena and physical processes which may be happening in other galactic nuclei.

The star formation history of the NSC is crucial to our understanding of how the formation of stars connects to the formation of the central SMBH, the surrounding nuclear stellar disc (NSD) and the Galactic bulge. The NSC and the NSD are found to be composed of different stellar populations and star formation histories (e.g., \citealt{Schodel et al. 2020, Nogueras-Lara et al. 2021}), their relation and formation, however, are not fully understood. Previous studies have suggested that the star formation history of the NSC is complex. The stellar population of the NSC is composed of cool, evolved giants; and hot, young main-sequence/post-main-sequence stars. \citet{Blum et al. 2003} combined spectroscopic and photometric observations of the 79 most luminous asymptotic giant branch (AGB) and supergiant stars in the central 5 pc. They constructed the Hertzsprung-Russell (H-R) diagram from CO and H$_2$O molecular absorption features in H- and K-band spectra, and claimed that $\sim$75\% of stars formed more than 5 Gyr ago. \citet{Maness et al. 2007} reported the first study using adaptive optics (AO) observations of 329 giants in the central 1 pc, including helium-burning red clump stars, red giants, and AGB stars. These stars with longer-lived evolutionary phases are better understood by theoretical evolutionary models. They derived stellar effective temperature ($T_{eff}$) using the prominent $^{12}$CO 2.2935 $\mu$m $\nu$ = 2 - 0 rovibrational bandheads, and favored a continuous star formation over the last 12 Gyr with a top-heavy initial mass function (IMF). \citet{Pfuhl et al. 2011} presented AO observations of 450 giants (central 1 pc), and claimed a maximum star formation rate $\sim$10 Gyr ago to a deep minimum 1-2 Gyr ago, followed by a significant increase during the last few hundred Myrs. They favored a ``canonical" Chabrier/Kroupa IMF \citep{Kroupa 2001} which was found in the local universe and is consistent across different star formation regions, and reported that $\sim$ 80\% of the stellar mass formed more than 5 Gyr ago.

The limitation in our current understanding of the NSC star formation history is that previous spectroscopic studies assumed that all stars have solar metallicity. However, possible degeneracies between stellar age and metallicity in the star formation history may potentially cause biases in the age estimates. Earlier works have noted such degeneracies (e.g., \citealt{Blum et al. 2003, Maness et al. 2007, Pfuhl et al. 2011}), but were not able to account for them due to limited metallicity measurements. Recent spectroscopic surveys have revealed a significant spread in metallicity of late-type stars from the NSC, which motivates us to revisit the star formation history and its implications for the formation and evolution of the NSC. \citet{Do et al. 2015} reported an AO-fed sample (R $\sim$5,400) of 83 red giants with scaled solar metallicity measurements (henceforth described by $[M/H]$), ranging from sub-solar ($[M/H]$ $<$ -1.0) to metal-rich stars ($[M/H]$ $>$ +0.5). \citet{Feldmeier-Krause et al. 2017,Feldmeier-Krause et al. 2020} confirmed the broad distribution ($[M/H]$ $<$ -1.0 to $[M/H]$ $>$ +0.3) on a larger sample (R $\sim$4,000), covering roughly half of the enclosed area of the NSC ($R_{eff}\sim$ 4.2 pc, \citealt{Schodel et al. 2014,Gallego-Cano et al. 2020}). \citet{Ryde Schultheis 2015} and \citet{Rich et al. 2017} also reported a broad distribution with $[Fe/H]$ measurements (R $\sim$24,000), spanning -0.5 $<$ $[Fe/H]$ $<$ +0.5 for at least 15 M-giants belonging to the NSC.

Knowing the star formation history of the NSC is important because it allows us to make more accurate predictions of the number of compact objects, including stellar-mass black holes (SBHs), neutron stars (NSs) and white dwarfs (WDs) at the Galactic center, and their rates of mergers for interpreting gravitational wave detections like those from LIGO. Such predictions have been explored assuming different mass profiles (e.g., \citealt{Baumgardt et al. 2004, Alexander et al. 2007}). \citet{Morris 1993} reported a total mass of remnants of 0.4 - 5 $\times$ $10^6$ $M_{\odot}$, assuming a high low-mass cut-off to the IMF (1 $M_{\odot}$). \citet{Maness et al. 2007} expected a significant mass of dark remnants from a top-heavy IMF. \citet{Lockmann et al. 2010} favored a canonical IMF and predicted $\sim$2.5 $\times$ 10$^4$ SBHs and NSs for every 1.5 $\times$ 10$^6$ $M_{\odot}$ of total cluster mass. \citet{Hailey et al. 2018} reported observations of a dozen quiescent X-ray binaries which contain a SBH, and estimated conservatively $\sim$600 - 1000 quiescent BH low-mass X-ray binaries (qBH-LMXBs) in the inner 1 pc (or $\sim$300 - 500 if some observed sources are rotation-powered millisecond pulsars, rMSPs). \citet{Generozov et al. 2018} predicted 1 - 4 $\times$ 10$^4$ BHs within the central parsec today, and $\sim$60 - 200 accreting BH-XRBs currently in the central parsec that formed from tidal capture of stars by BHs. \citet{Mori et al. 2021} further confirmed these X-ray sources and reported a lower predicted number of BH-LMXBs with $\sim$500 - 630 (or $\sim$240 - 300) in the central parsec. These predictions of BH X-ray binaries provide a lower limit to the total number of BHs in the central parsec. The current limitation is that predictions of compact objects and their merger rates have assumed a canonical IMF and a solar metallicity, which may have large impacts on the resulting compact remnant properties.

In this work, we construct the star formation history of the NSC with the first metallicity constraints. We make updated predictions of the number of compact objects and the resulting gravitational wave merger rates at the Galactic center. The datasets used in this work are described in section \ref{sec:datasets}. Section \ref{sec:methods} presents the methods we use to model the cluster and fit the star formation history. Section \ref{sec:results_all} reports the results of the star formation history and the impacts of the metallicity constraints on the cluster age. Section \ref{sec:discussion} further discusses the implications and impacts of the resulting star formation history on the number of compact objects and their merger rates. We conclude with a summary in section \ref{sec:conclusion}. 

\section{Datasets} \label{sec:datasets}
The data for late-type stars used in this work to construct the star formation history of the NSC are from a combination of AO and seeing-limited observations.

\subsection{AO dataset}\label{sec:dataset1}

\begin{figure}[ht!]
\includegraphics[width=85mm]{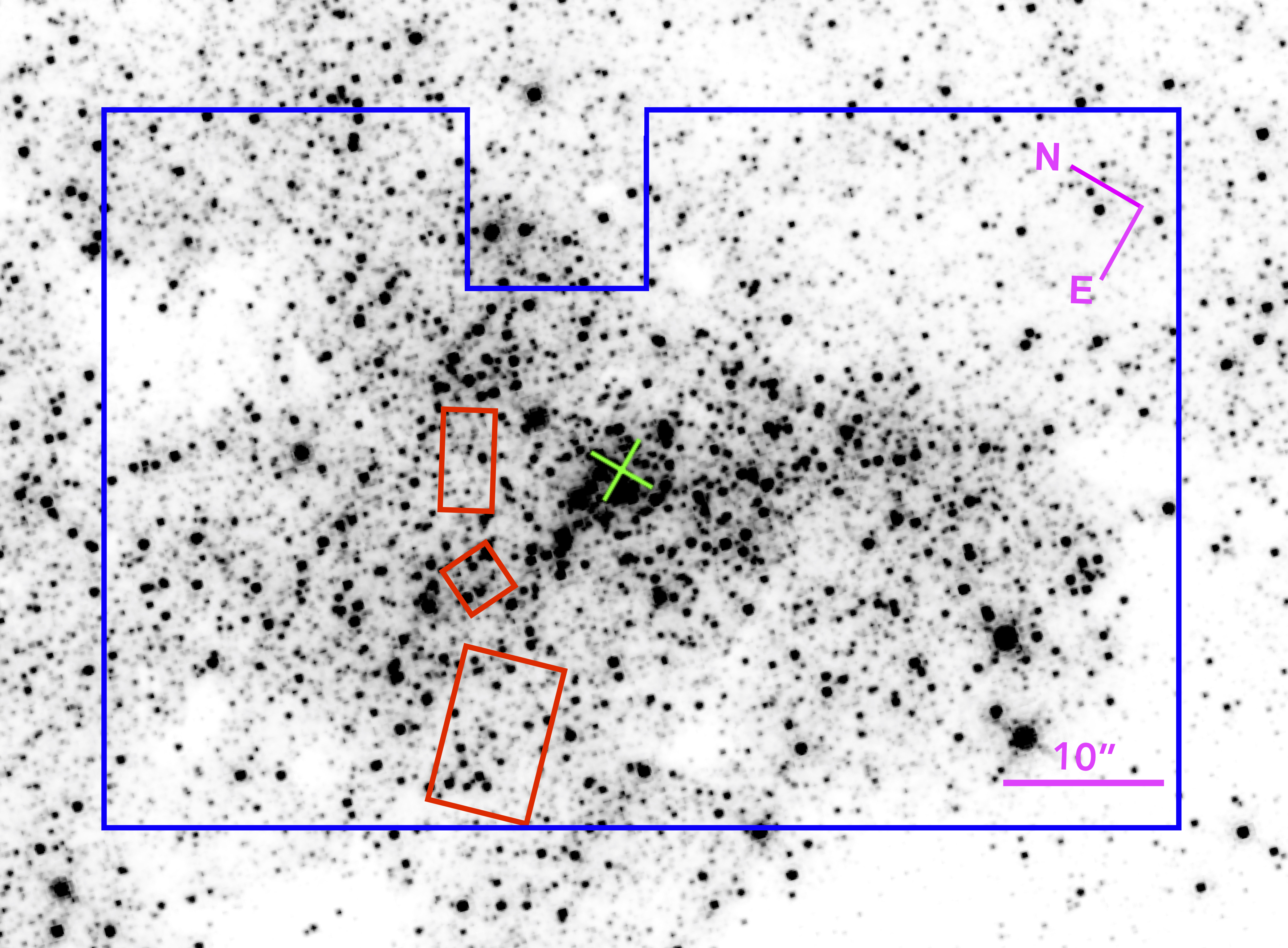}
\centering
\caption{Datasets for late-type stars used in this work. Red regions are the Gemini NIFS AO observations \citep{Do et al. 2015}, while the blue region shows the VLT KMOS seeing-limited observations \citep{Feldmeier-Krause et al. 2017}. The orientation of the Galactic plane runs horizontally through the figure. The green cross in the center shows the position of Sgr A*. The background image is from the HST WFC3-IR observations of the NSC (GO-12182, PI Do). \label{fig:datasetproj1}}
\end{figure}

A spectroscopic survey of a sample of 83 late-type stars (F-type or later) within a radius of 1 pc from the central SMBH yielded metallicity measurements, $[M/H]$, for all of those stars (\citealt{Do et al. 2015}, also see details in \citealt{Stostad et al. 2015}). The original spectra were obtained with the medium-spectral-resolution Near-Infrared Integral Field Spectrograph (NIFS) on the Gemini North telescope with the natural-guide-star and laser-guide-star AO system ALTAIR. NIFS provides the observed spectra in the K broadband filter (1.99-2.40 $\mu$m) with a spectra resolution of R $\sim$ 5,000 and a spatial resolution of 115-165 mas. The observations between 2012 May and 2014 May span a projected radius of 8-22 arcsec (0.3-0.9 pc) from Sgr A*, covering a total surface area of 81 $arcsec^2$, approximately 0.15 $pc^2$ at a distance of 8 kpc. See Figure \ref{fig:datasetproj1} for the location of the fields. 

Late-type stars were classified as stars that exhibit strong CO bandheads and Na I doublet absorption lines at 2.2062 and 2.2090 $\mu$m, and more precisely classified using the equivalent width (EW) measurements of the lines of these features. For each stellar spectrum, physical parameters were fitted simultaneously with the MARCS synthetic spectral grid \citep{Gustafsson et al. 2008} using the STARKIT code \citep{Kerzendorf Do 2015}: $T_{eff}$, log g, $[M/H]$, and radial velocity ($v_z$). We also report the temperature measurements of CO-$T_{eff}$ as derived from the calibrated $T_{eff}-EW_{CO}$ relation (see Appendix \ref{sec:app-teff} for details). The stars in the sample have a large metallicity range (-1.27 $<$ $[M/H]$ $<$ 0.96), with a mean uncertainty of 0.32 dex. All 83 stars are confirmed to be NSC members after color analysis to exclude foreground or background sources \citep{Stostad et al. 2015}, and considering different contamination sources and potential biases \citep{Do et al. 2015}. See section \ref{sec:sys_contam} for further discussion.

The photometry and extinction correction of the dataset were obtained and reported by \citet{Stostad et al. 2015}. Here is the summary. The K$_s$-band and H-band photometry was obtained by matching the spectroscopic detections to the photometric catalog from \citet{Schodel et al. 2010}. The matching process was performed by searching stars with location and estimated K magnitudes (see details in \citealt{Stostad et al. 2015}). The photometry was corrected for dust extinction, $A_{Ks}$, using the extinction map and extinction law of \citet{Schodel et al. 2010}. We correct for observational incompleteness of the field using the overall completeness curve derived in \citet{Stostad et al. 2015}. The overall completeness (both photometric and spectroscopic) is the average likelihood of detecting and classifying stars as a function of stellar brightness. The average total photometric and spectroscopic completeness across the whole field is $\sim$74\% at K$_s$ = 15.5 mag. \citet{Do et al. 2015} restricted the analysis to stars with signal-to-noise ratios (S/Ns) greater than 35. We obtained the completeness by multiplying a ratio at each magnitude bin, calculated as the fraction of stars studied by \citet{Do et al. 2015} divided by the number of stars in the whole sample. The resulting completeness curve for this dataset shows 50\% completeness at K$_s$ $\sim$15.5 mag.

\begin{figure*}[ht!]
\includegraphics[width=170mm]{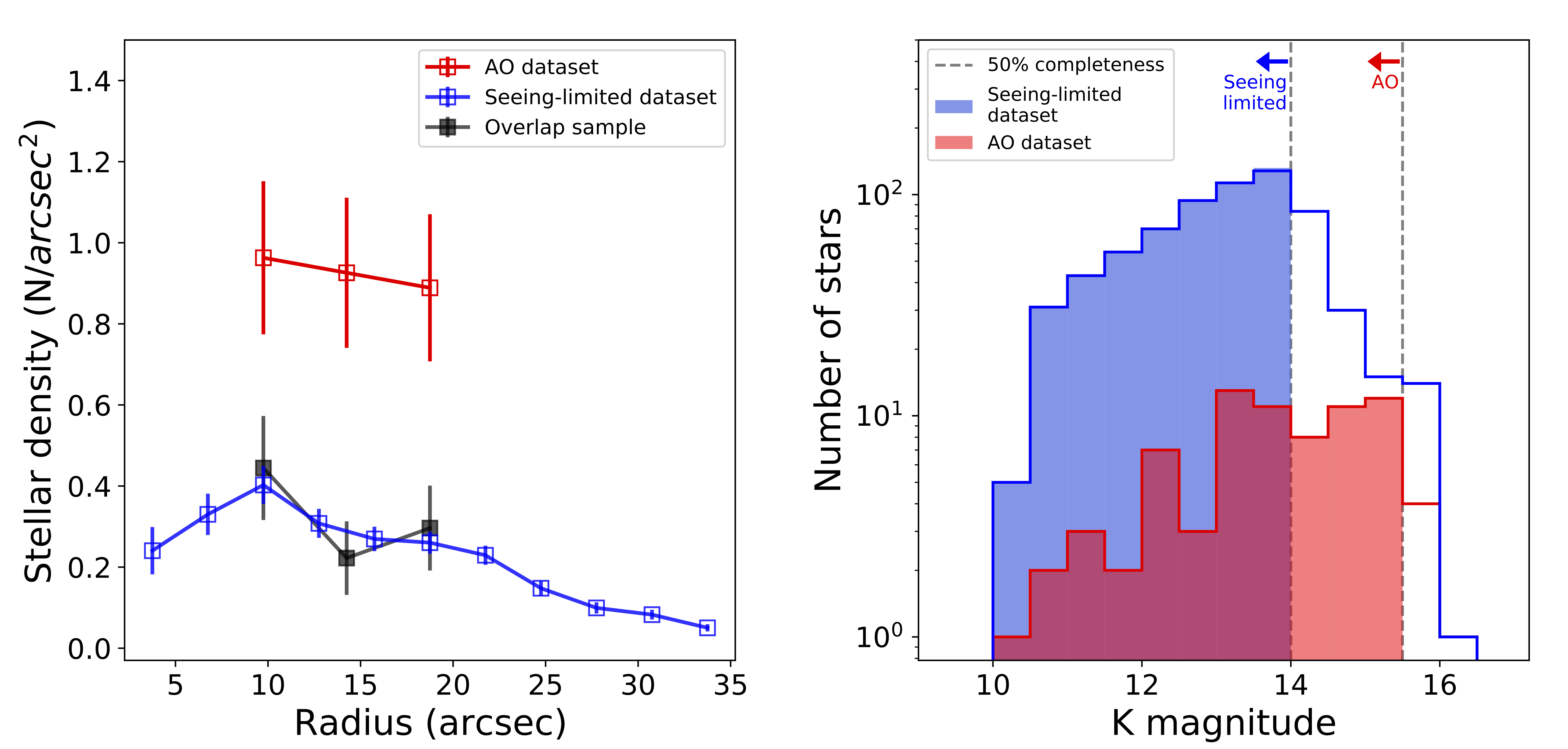}
\centering
\caption{\textbf{Left}: Stellar density for two datasets and the overlap sample as a function of distance from Sgr A*. The uncertainties are calculated as $\sqrt{N_{stars}}$/AREA in each radial bin. The seeing-limited dataset \citep{Feldmeier-Krause et al. 2017} enables a larger sample with a much wider coverage, but is limited to a shallower depth. The AO dataset \citep{Do et al. 2015} is deeper in spectroscopic sensitivity, with a much higher stellar density in the observed region. \textbf{Right}: Background non-shaded histograms (solid-line edges) show the observed luminosity function of all stars from the AO dataset (red) and the seeing-limited dataset (blue) respectively. Vertical dashed lines mark the detection limits at 50\% completeness for each dataset. Shaded regions represent the stars used in this work that are above the 50\% completeness. \label{fig:datadepth}}
\end{figure*}

\begin{deluxetable*}{llccc}
\tablenum{1}
\tablecaption{Datasets \label{tab:datasets_sum}}
\tablewidth{0pt}
\tablehead{
\colhead{} & \colhead{Properties} & \colhead{Dataset1} & \colhead{Dataset2} & \colhead{Overlap}}
\startdata
Spatial-Related  & Atmospheric Correction & Yes (AO) & No (Seeing-limited) & -- \\
& Angular Resolution, Average ($arcsec$) & 0.14 $\pm$ 0.03 & 1.0 $\pm$ 0.3 & -- \\
& Limiting K mag (50\% completeness) & 15.5 & 14.0 & 14.0\\
& Sky Coverage ($arcsec^2$) & 81 & 2700 & 81\\
& Number of Stars & 83 & 687 & 27\\
\hline
Spectral-Related & Spectral Resolution & 5,400 & 4,000 $\pm$ 700 & -- \\
& Spectral Range ($\mu$m) & 1.99 - 2.40 & 1.93 - 2.46 & --  \\
& Spectral Grid & MARCS & PHOENIX & -- \\
\hline
Reference & & \citet{Do et al. 2015} & \citet{Feldmeier-Krause et al. 2017} & -- 
\enddata
\end{deluxetable*}

\subsection{Seeing-limited dataset}\label{sec:dataset2}

Metallicity determinations were made by \citet{Feldmeier-Krause et al. 2017} using a spectroscopic survey of 687 late-type stars within a radius of 1.5 pc from the central SMBH.  The original spectra were obtained with seeing-limited observations using the medium-resolution integral-field spectrograph KMOS at the Very Large Telescope (VLT) in the K band filter ($\sim$1.934-2.460 $\mu$m). The spectral resolution, R, varies between 3310 and 4660 for 23 different active IFUs on the KMOS detectors with a standard deviation of 300 over all IFUs. The observations on 2013 September 23 covered an area of 2700 $arcsec^2$, approximately 4 $pc^2$ at a distance of 8 kpc. See Figure \ref{fig:datasetproj1} for the location of the fields.

Late-type stars were classified as stars that exhibit prominent CO bandheads and the Na I doublet absorption lines at 2.2062 and 2.2090 $\mu m$, and further confirmed by the measurements of $EW_{CO}$ and $EW_{Na}$. Each spectrum was fitted with synthetic PHOENIX grid \citep{Husser et al. 2013} using the STARKIT code \citep{Kerzendorf Do 2015}. And the stellar effective temperature CO-$T_{eff}$ was further measured using the calibrated $T_{eff}-EW_{CO}$ relation (see Appendix \ref{sec:app-teff}). The sample stars have a large metallicity range (-1.25 $<$ $[M/H]$ $<$ 1.00), with a mean uncertainty of 0.27 dex. All 687 stars are confirmed to be NSC members after color analysis to exclude foreground or background sources, and considering different contamination sources (\citealt{Feldmeier-Krause et al. 2015, Feldmeier-Krause et al. 2017}). See section \ref{sec:sys_contam} for further discussion. 

The photometry and extinction correction of the dataset were obtained and reported by \citet{Feldmeier-Krause et al. 2017}. Here is the summary. The K$_s$-band and H-band photometry was obtained by matching spectroscopic detections to the photometric catalogs from \citet{Schodel et al. 2010}, \citet{Nogueras-Lara et al. 2018a} and \citet{Nishiyama et al. 2006}. The photometry was corrected for dust extinction using the extinction map and extinction law of \citet{Schodel et al. 2010}. Stellar extinction values ($A_{Ks}$) were extracted from the extinction map, which covers 70\% of the sample. For stars outside the field of view of the \citet{Schodel et al. 2010} extinction map, the photometry was corrected using the \citet{Nogueras-Lara et al. 2018a} extinction map. See section 2.3 in \citet{Feldmeier-Krause et al. 2017} for more details. We then correct for observational incompleteness using the completeness from \citet{Feldmeier-Krause et al. 2015} and determine separately for stars at different projected radial distances from Sgr A*. The resulting completeness curve for the dataset shows a 50\% completeness at K$_s$ $\sim$14.0 mag. 

\subsection{Comparison between the two datasets}

The seeing-limited dataset (50\% complete at K$_s$ = 14.0 mag) presents a larger spectroscopic sample ($N_{stars}=687$) with a much wider coverage ($\sim$2700 $arcsec^2$) than the AO sample. The large sample is essential to obtain robust measurements of the stellar population across the whole field of view. The AO dataset (50\% complete at K$_s$ = 15.5 mag) presents a deeper spectroscopic sample with fewer stars ($N_{stars}=83$) and a smaller coverage ($\sim$81 $arcsec^2$). The AO spectroscopy is most useful in the innermost dense region and achieves a depth 1.5 magnitudes fainter than the seeing-limited spectroscopy. In the overlapped region between the two datasets, 27 stars were detected in both surveys. See Table \ref{tab:datasets_sum} for the summary. The left panel of Figure \ref{fig:datadepth} presents the stellar density for two datasets and the overlap sample as a function of distance from Sgr A*. The right panel of Figure \ref{fig:datadepth} presents the luminosity functions of observed stars from the AO and the seeing-limited dataset respectively. In this work, we only use the stars brighter than the 50\% completeness in each dataset.

\section{Methodology} \label{sec:methods}
In this section, we describe how we model the properties of the NSC by generating synthetic clusters and applying a Bayesian framework (section \ref{sec:modeling} and \ref{sec:bayesian}). We introduce the prior on the parameters (section \ref{sec:prior}) and the sampling technique (section \ref{sec:sampling}). We present five star-formation history models in section \ref{sec:sfh} and the model selection criteria in section \ref{sec:modelselection}. The fitter tests on simulated clusters are summarized in section \ref{sec:sim_test}.   

\subsection{Generating a synthetic cluster}\label{sec:modeling}

\begin{figure}[ht!]
\includegraphics[width=85mm]{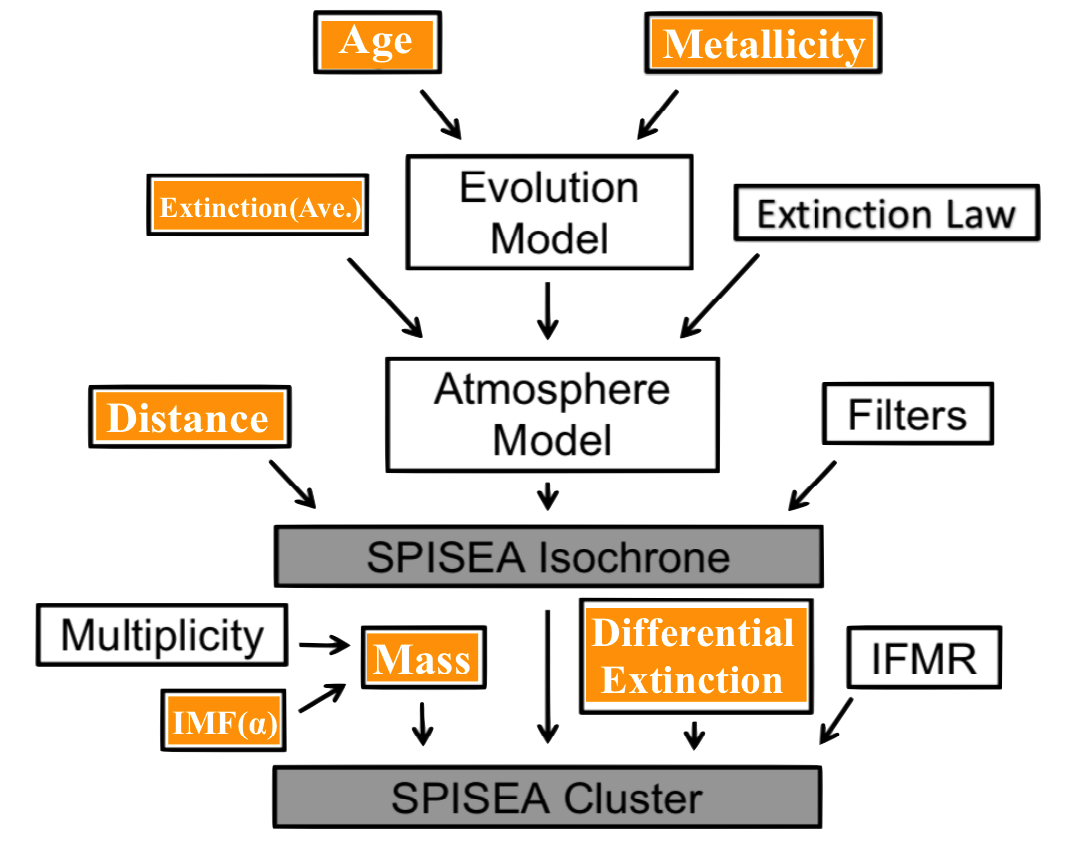}
\centering
\caption{Diagram of the SPISEA code \citep{Hosek et al. 2020}. The white boxes represent the fixed inputs specified in the modeling while the orange boxes represent the model variables (see Table \ref{tab:model_para}). The grey boxes represent the primary SPISEA outputs of the Isochrone and Cluster objects.\label{fig:pop_flow}}
\end{figure}

We use a forward-modeling approach to derive the cluster properties by comparing the observational input data to a synthetic cluster within a Bayesian framework. We start with the example of generating a single-age cluster. We use SPISEA, an open-source Python package \citep{Hosek et al. 2020} for simulating simple stellar populations (SSPs), to generate a cluster. The advantage of SPISEA is the ability to control 13 input parameters when generating a cluster. Intrinsic properties ($T_{eff}$, $log(g)$, etc) and synthetic photometry are assigned for stars spanning the range from pre-main sequence to post-main sequence types. Figure \ref{fig:pop_flow} presents the top-level diagram of the SPISEA code workflow. The variables used in the cluster modeling are: the cluster age ($log(t)$), cluster metallicity ($\overline{[M/H]}$), total cluster mass ($M_{cl}$), IMF slope ($\alpha$), distance to the cluster ($d$), average extinction ($\overline{A_{Ks}}$), residual differential extinction after the extinction map correction ($\Delta A_{Ks}$), and the minimum and maximum stellar mass ($m_{min}$, $m_{max}$) of the IMF. Here we only consider a one-segment IMF with a slope of $\alpha$ between the stellar mass of $m_{min}$ and $m_{max}$. See Table \ref{tab:model_para} for the summary. We also specify the fixed inputs used in the cluster modeling: stellar evolution model, atmosphere model, extinction law, photometric filters, multiplicity and initial-final mass relation (IFMR).

We use SPISEA to build a theoretical isochrone, which defines the stellar properties at a given age as a function of initial stellar mass, for a given set of model variables and specified inputs. We use the most recent MESA Isochrones and Stellar Tracks (MIST) of the v1.2 stellar evolution model with rotation (\citealt{Choi et al. 2016, Dotter 2016}), to determine the stellar physical properties. MIST is the only end-to-end self-consistent stellar evolution model to produce stars from pre-main-sequence to the post-main-sequence, which avoids merging multiple models; it also agrees broadly with the observations for less massive stars. We use a merged atmosphere model: an ATLAS9 grid \citep{Castelli Kurucz 2004} for $T_{eff}$ $>$ 5500 K and a PHOENIX grid (version 16; \citealt{Husser et al. 2013}) for $T_{eff}$ $<$ 5000 K; and the average in the $T_{eff}$ transition region. The intrinsic spectral energy distributions (SEDs) are generated from the atmosphere model, and applied with the total extinction ($\overline{A_{Ks}}$) and the extinction law from \citet{Schodel et al. 2010}. The synthetic photometry is then calculated by convolving the extinguished SEDs with the K$_s$ and H filter transmission functions. We assume no binary systems as default in the modeling since very few late-type giants can be binaries. The fraction of binary stars declines over time as a result of merging and evaporating (e.g., \citealt{Stephan et al. 2016, Rose et al. 2020}). The late-type giants would harbor little to no binary systems at their age ($<$ 2\% for stars older than 3 Gyr, \citealt{Stephan et al. 2016}). The additional tests with other multiplicity properties (e.g., \citealt{Lu et al. 2013, Moe et al. 2017}) show that the choice of multiplicity in the modeling of this work makes negligible difference on the results.

We use SPISEA to generate a star cluster, given an isochrone, $M_{cl}$, $\Delta A_{Ks}$, IMF, and multiplicity within the defined IMF stellar mass range. SPISEA also simulates the differential extinction of stars by perturbing the photometry by a random step from a Gaussian distribution ($\mu$ = 0, $\sigma = \Delta A_{Ks}$) at each filter.

\begin{deluxetable}{lcc}
\tablenum{2}
\tablecaption{Model Cluster Variables \label{tab:model_para}}
\tablewidth{0pt}
\tablehead{
\colhead{} & \colhead{Parameter} & \colhead{Description}}
\startdata
Input data & $K_s$ & Stellar $K_s$ magnitude\\
 & $color$ & Stellar ($H$ - $K_s$) color\\
 & $T_{eff}$ & Stellar effective temperature \\
 & $A_{Ks}$ & Stellar extinction\\
 & $[M/H]$ & Stellar metallicity \\
 & $N_{obs}$ & Number of observed stars\\
\hline
Model variable &  $log t$ & Cluster age\\
 & $\overline{[M/H]}$ & Cluster Metallicity \\
 & $M_{cl}$ & Total initial cluster mass$^{a}$ \\
 & $\alpha$ & IMF slope \\
 & $d$ & Distance to the cluster\\
 & $\overline{A_{Ks}}$ & Average extinction \\
 & $\Delta A_{Ks}$ & Differential extinction \\
 & $m_{min}$ & IMF minimum stellar mass \\
 & $m_{max}$ & IMF maximum stellar mass \\
\enddata
\tablecomments{\\
$^{a}$ $M_{cl}$ models the cluster mass with stellar mass between $m_{min}$ and $m_{max}$ (0.8 and 120 $M_{\odot}$, respectively), over which the IMF is sampled in the cluster modeling. }
\end{deluxetable}

\subsection{Bayesian Analysis}\label{sec:bayesian}

We develop a Bayesian inference approach to derive the star formation history and cluster properties. The input observational data includes: stellar K$_s$ magnitude, (H - K$_s$) $color$, effective temperature ($T_{eff}$), extinction value ($A_{Ks}$), metallicity measurement ($[M/H]$) of individual stars, and the total number of observed stars ($N_{obs}$). See Table \ref{tab:model_para} for summary. The detailed methodology has been described in \citet{Lu et al. 2013} and \citet{Hosek et al. 2019}. The task is to consider the parameter degeneracy and observational uncertainties, and fit the cluster parameters simultaneously. We expand and improve the methodology to a 3-dimensional fitting that for the first time includes stellar measurements of K$_s$ magnitude, $color$ and $T_{eff}$ in the modelings, and considers metallicity constraints.

In order to correct for differential extinction, we differentially deredden the observed stellar K$_s$ magnitude and (H - K$_s$) $color$ to the average extinction value $\overline{A_{Ks}}$ of the input dataset using the extinction map. We define the K$_{s,dered}$ and $color_{dered}$ as the differentially dereddened magnitude and color of the observed stars.

We use Bayes theorem to derive the best-fit cluster model,

\begin{equation}
  \begin{aligned}
    &P(\Theta|\boldsymbol{k}_{obs}, N_{obs}, [M/H]) = \\
    &\frac{\mathcal{L}(\boldsymbol{k}_{obs}, N_{obs}, [M/H]|\Theta)\cdot P(\Theta)}{P(\boldsymbol{k}_{obs}, N_{obs}, [M/H])}
  \end{aligned}
\end{equation}

where $\{\boldsymbol{k}_{obs}, N_{obs}, [M/H]\}$ is the input data (see Table \ref{tab:model_para}), and $\boldsymbol{k}_{obs}$ is the set of $\{K_{s,dered},~ color_{dered},~ T_{eff}\}$ measurements for the $N_{obs}$ stars observed. $\Theta$ is the cluster model defined by the set of model variables $\Theta$ = $\{t, ~\overline{[M/H]}, ~M_{cl}, ~\alpha, ~d, ~\overline{A_{Ks}}, ~\Delta A_{Ks}\}$. $\mathcal{L}(\boldsymbol{k}_{obs}, N_{obs}, [M/H]|\Theta)$ is the likelihood function of observing the data given the model $\Theta$, $P(\Theta)$ captures the prior knowledge on the model variables, and $P(\boldsymbol{k}_{obs}, N_{obs}, [M/H])$ is the sample evidence as a normalizing factor. This calculation results in the posterior probability distributions $P(\Theta|\boldsymbol{k}_{obs}, N_{obs}, [M/H])$ for the given model variables $\Theta$. 

The likelihood function is composed of three independent components, 
\begin{equation}\label{equ:likelihood}
  \begin{aligned}
    &\mathcal{L}(\boldsymbol{k}_{obs}, N_{obs}, [M/H]|\Theta) = \\
    &p(\boldsymbol{k}_{obs}|\Theta) \cdot p(N_{obs}|\Theta) \cdot p([M/H]|\Theta).
  \end{aligned}
\end{equation}

\begin{itemize}
 \item {$p(\boldsymbol{k}_{obs}|\Theta)$}: the probability of observing the distribution of stars in the $\boldsymbol{k}_{obs}$ = \{$K_{s,dered}$, $color_{dered}$, $T_{eff}$\} space. 
 \item {$p(N_{obs}|\Theta)$}: the probability of detecting the number of observed cluster stars $N_{obs}$ given the observational completeness.
 \item {$p([M/H]|\Theta)$}: the probability of measuring the observed $[M/H]$ values for the observed stars. 
\end{itemize}

For the first term $p(\boldsymbol{k}_{obs}|\Theta)$, we calculate the probability of observing the sample of stars by multiplying the individual observed stars’ probabilities,

\begin{equation}\label{equ:likeli_cmd}
  p(\boldsymbol{k}_{obs}|\Theta) = \prod_{i=1}^{N_{obs}} p(\boldsymbol{k}_{obs,i}|\Theta)
\end{equation}
The probability of observing the $i^{th}$ star $p(\boldsymbol{k}_{obs,i}|\Theta)$, given the observed $\{K_{s,dered}, color_{dered}, T_{eff}\}$, is obtained by the probability distribution derived from synthetically ``observing" a simulated cluster. We first calculate the intrinsic probability distribution $p(\boldsymbol{k}_{int}|\Theta)_{int}$ for stars in the synthetic cluster. The synthetic cluster is generated given the model $\Theta$ with model parameters described in section \ref{sec:modeling} and Table \ref{tab:model_para},
\begin{equation}\label{equ:cmd_int}
  p(\boldsymbol{k}_{int}|\Theta)_{int} = Simulated~Cluster (\Theta) \circledast G(\boldsymbol{\mu}, \boldsymbol{\sigma})
\end{equation}

where $\boldsymbol{k}_{int}$ =  $\{K_{s},~ color,~ T_{eff}\}$ is the distribution of synthetic stellar properties in the model cluster; $G(\boldsymbol{\mu}, \boldsymbol{\sigma})$ is a Gaussian distribution with the mean as the generated values $\boldsymbol{k}_{int}$, and standard deviation from observational errors $\boldsymbol{\sigma}$ = \{$\sigma_{Ks}$, $\sigma_{color}$, $\sigma_{T_{eff}}$\}. We bin the $Simulated~Cluster$ stars in 3 dimensions: $K_s$, $color (H - K_s)$ and $T_{eff}$. In each dimension, we represent each star as a Gaussian distribution with the mean equal to the generated value, and standard deviation equal to the expected measurement uncertainties based on observations. To reduce the stochastic sampling effects and obtain a more accurate estimate of the probability distribution, all model clusters are generated with a total mass of $5 \times 10^7 M_{\odot}$.

The intrinsic probability distribution $p(\boldsymbol{k}_{int}|\Theta)_{int}$ is multiplied by the completeness cube $C(\boldsymbol{k}_{int})$ to match the data, and then normalized, to give the probability distribution of observing a star in the model cluster, 
\begin{equation}\label{equ:cmd_int_obs}
  p(\boldsymbol{k}_{int}|\Theta)_{obs} =  \frac{p(\boldsymbol{k}_{int}|\Theta)_{int}\cdot C(\boldsymbol{k}_{int})}{\iiint_{V}p(\boldsymbol{k}_{int}|\Theta)_{int}\cdot C(\boldsymbol{k}_{int})d\boldsymbol{k}_{int}}
\end{equation}
where the completeness cube $C(\boldsymbol{k}_{int})$ is constructed from the observational completeness curve (as a function of $K_s$, see sections \ref{sec:dataset1} and \ref{sec:dataset2}) and is applied to the 3-dimensional binned simulated cluster $\{K_s, color, T_{eff}\}$, assuming consistency along the axis of color and $T_{eff}$. 

The probability of observing $\boldsymbol{k}_{obs,i}$ for a given star in the input observed data is then calculated by
\begin{equation}
  p(\boldsymbol{k}_{obs,i}|\Theta) = \boldsymbol{k}_{obs,i} \cdot p(\boldsymbol{k}_{int}|\Theta)_{obs}
\end{equation}

The resulting first term of the likelihood is calculated from feeding all stars’ probabilities $p(\boldsymbol{k}_{obs,i}|\Theta)$ into equation (\ref{equ:likeli_cmd}).

For the second term, $p(N_{obs}|\Theta)$, we calculate the probability of obtaining the number of stars we could observe given the cluster model. We apply the observational completeness cube to the synthetic cluster to get the total number of stars ($N_{sim}$) that we would expect to observe from the model. We then linearly scale the number of stars to the mass of the cluster model to obtain the expected number of observed stars, $N_{e}$:
\begin{equation}\label{equ:num_exp}
  N_{e} = N_{sim} \times \frac{M_{cl}}{5 \times 10^{7}}
\end{equation}
The likelihood of observing the number of cluster stars $N_{obs}$ is then taken as a Poisson distribution:

\begin{equation}\label{equ:num_cl}
  p(N_{obs}|\Theta)  = \frac{N_e^{N_{obs}} \times e^{-N_e}}{N_{obs}!}
\end{equation}

For the last term, $p([M/H]|\Theta)$, we model the cluster metallicity distribution as a Gaussian from stellar metallicity measurements $[M/H]$. For each star $i$, the likelihood of measuring $[M/H]_i$ is
\begin{equation}\label{equ:likeli_feh_i}
  \begin{aligned}
    &p([M/H]_{i}|\Theta)  =  \\
    &\frac{1}{\sqrt{2\pi}\sigma_{total,i}} \times exp(-\frac{([M/H]_{i} - \overline{[M/H]})^2}{2\sigma_{total,i}^2})
  \end{aligned}
\end{equation}
where $[M/H]_i$ and $\sigma_{[M/H], i}$ are the measured stellar metallicity and uncertainty. $\overline{[M/H]}$ is the cluster model metallicity. $\sigma_{\overline{[M/H]}}$ is the intrinsic metallicity dispersion of the NSC, and is conservatively estimated from the standard deviation of the observed sample (0.32). 

$\sigma_{total,i} = \sqrt{\sigma^{2}_{[M/H], i} + \sigma_{\overline{[M/H]}}^2}$. The overall likelihood of metallicity measurements is calculated by multiplying the likelihoods of individual stars together:
\begin{equation}\label{equ:likeli_feh}
  p([M/H]|\Theta) = \prod_{i=1}^{N_{obs}} p([M/H]_i|\Theta)
\end{equation}

\begin{deluxetable*}{clll}
\centering
\tablenum{3}
\tablecaption{Star Formation History Models\label{tab:SFH_model}}
\tablewidth{0pt}
\tablehead{
\colhead{Model}& \colhead{Name} & \colhead{Description} & \colhead{Fitting Parameters} }
\startdata
1&Single burst & One burst of star formation & log(t), $\overline{[M/H]}$, $M_{cl}$, $\alpha$, d, $\overline{A_{Ks}}$, $\Delta A_{Ks}$\\
2&Two bursts & Two bursts of star formation & log(t$_{1}$), log(t$_{2}$), $\overline{[M/H]}_{1}$, $\overline{[M/H]}_{2}$, Fraction$_{M, 1}$, $M_{cl}$,\\
& & & $\alpha$, d, $\overline{A_{Ks}}$, $\Delta A_{Ks}$\\
3&Three bursts & Three bursts of star formation & log(t$_{1}$), log(t$_{2}$), log(t$_{3}$), $\overline{[M/H]}_{1}$, $\overline{[M/H]}_{2}$, $\overline{[M/H]}_{3}$, \\
& & & Fraction$_{M, 1}$, Fraction$_{M, 2}$, $M_{cl}$, $\alpha$, d, $\overline{A_{Ks}}$, $\Delta A_{Ks}$\\
4& Linear SFR & Continuous star formation with a linearly & m, $\overline{[M/H]}$, $M_{cl}$, $\alpha$, d, $\overline{A_{Ks}}$, $\Delta A_{Ks}$\\
& & increasing/decreasing SFR$^{a}$\\
5& Exponential SFR & Continuous star formation with an & $\lambda$, $\overline{[M/H]}$, $M_{cl}$, $\alpha$, d, $\overline{A_{Ks}}$, $\Delta A_{Ks}$\\
& & exponentially increasing/decreasing SFR$^{b}$\\
 \\
\enddata
\tablecomments{\\
$^{a}$ SFR(t) $\propto$ $mt$, where $t$ is the elapsed lookback time starting at 30 Myr and extending as far as 13 Gyr. \\ $^{b}$ SFR(t) $\propto$ $e^{-\lambda t}$, where $t$ is the elapsed lookback time starting at 30 Myr and extending as far as 13 Gyr. 
}
\end{deluxetable*}

\subsection{Prior knowledge on the model variables}\label{sec:prior}

We use uniform priors on the model variables which we aim to measure independently in this work: cluster age, cluster metallicity, total cluster mass, and differential extinction. The lower and upper limits of the cluster age are set from the typical age range of late-type stars (30 Myr to 13 Gyr). The upper limit of cluster metallicity ($\overline{[M/H]}$ = +0.5) is set from the theoretical stellar evolutionary models \citep{Choi et al. 2016}. The upper limit of the cluster differential extinction ($\Delta A_{Ks}$ = 0.5) is set with a conservative 5-$\sigma$ limit, which is 5 times the total uncertainty (systematic and statistical) of the extinction map \citep{Schodel et al. 2010}.

We include informative priors for some model parameters including the distance to the cluster and the cluster average extinction ($\overline{A_{Ks}}$). In this work, a Gaussian distributed prior is applied to the distance ($\mu$ = 8030 pc, $\sigma$ = 200 pc) as obtained from the accurate measurements of the Galactic center distance in the literature (\citealt{Gravity 2019, Do et al. 2019}). The average extinction ($\overline{A_{Ks}}$) adopts a Gaussian-distributed prior with the mean, $\mu$, being the average of stellar extinction values $A_{Ks}$ of the dataset, and the standard deviation, $\sigma$, being the total uncertainty (systematic and statistical) of the extinction map.

Simulated synthetic clusters are used to identify possible degeneracies between parameters, and probe the impact of the prior on the fitting results. Several parameters show correlations. The moderate correlation between the cluster age and IMF slope also results in a correlation between the total cluster mass and the cluster age, or the IMF slope. The most massive stars have disappeared at older ages, and thus the total cluster mass would increase to match the observed number of stars brighter than the detection limit. We note that all stars in our sample are late-type giants. There are no massive stars in the sample. The observed late-type stars occupy such a small range of stellar mass that the IMF slope is not constrained in the independent fit. Therefore, we assume that the IMF slope is either a Kroupa IMF ($\alpha = -2.3 \pm 0.36$ for stars with $m > 0.5 M_{\odot}$, \citealt{Kroupa 2001}) or a top-heavy IMF ($\alpha = -1.7 \pm 0.20$, \citealt{Lu et al. 2013}), and use the corresponding Gaussian distribution as the IMF prior in the fits.

\subsection{Sampling Posterior Probability Distributions with MultiNest}\label{sec:sampling}

We use a nested sampling technique \citep{Skilling 2004} called MultiNest (\citealt{Feroz Hobson 2008, Feroz et al. 2009}), which is a publicly available multi-modal nested sampling algorithm, to obtain detailed probability distribution for cluster parameters given limited observations. This method accounts for the biases from stochastic sampling of stellar masses, and is less computationally expensive ($\sim$ 5 - 10 times shorter than using the Markov chain Monte Carlo method) with more accuracy in our cases \citep{Lu et al. 2013}. For each round of iteration, MultiNest fixes a number of live points to sample the parameter space and calculate the established Bayesian evidence at each point position. The same number of points converge into smaller and smaller patches around the center of the most probable regions until the change of evidence is no longer higher than the selected tolerance value. Here we adopt 600 live points, an evidence tolerance of 0.5, and a sampling efficiency of 0.8 to perform this simulation with a well-sampled parameter space and high efficiency. This MultiNest algorithm is executed by using the python wrapper module PyMultinest \citep{Buchner et al. 2014}.

\subsection{Deriving the Star Formation History}\label{sec:sfh}

We fit several star formation history models:

\begin{itemize}
 \item One burst of star formation, similar to the single age population in the bulge \citep{Genzel et al. 2003}. The age of the burst can vary smoothly between 30 Myr and 13 Gyr. See model 1 in Table \ref{tab:SFH_model}.
 \item Multiple bursts of star formation. The age of each burst can vary smoothly and independently between 30 Myr and 13 Gyr. Within each burst, we assume a single metallicity. We fit up to three bursts in order to distinguish between theoretical models under the current observational uncertainties. See model 2, 3 in Table \ref{tab:SFH_model}. Please refer to the continuous star formation below for scenarios with more than three bursts.
 \item Continuous star formation between 30 Myr and 13 Gyr ago (e.g., \citealt{Figer et al. 2004}). The star formation rate (SFR) is either linearly or exponentially increasing/decreasing. See model 4, 5 in Table \ref{tab:SFH_model}. 
\end{itemize}

\subsection{Model selection and information criteria}\label{sec:modelselection}

We perform model selection among different star formation history models based on the Bayesian information criterion (BIC). BIC is independent of the prior and penalizes the complexity of the model (number of parameters). For each model, BIC is defined as:
\begin{equation}\label{equ:BIC}
  BIC = -2ln(\widehat{\mathcal{L}}) + kln(N)
\end{equation}
where $\widehat{\mathcal{L}}$ is the achieved maximum value of the likelihood function for each model, $k$ is the total number of free parameters used in each model, and $N$ is the number of observed data points used in the modeling. BIC is minimized in the model selection, e.g. the model with the lowest BIC is preferred. Furthermore, we also use the Bayesian evidence (also called ``Bayes factor") and the Akaike information criterion (AIC, e.g., \citealt{Gelman et al. 2013}) to further confirm our selection of star formation history models.

\begin{deluxetable}{cccccc}
\tablenum{4}
\tablecaption{Model selection between star formation history models \label{tab:bayes_do}}
\tablehead{
\colhead{Dataset} & \colhead{Fit} & &\colhead{Model} & & \colhead{$\Delta$BIC$^{a}$}}
\startdata
AO & 1 & &single burst & & 0\\
& 2 & &two bursts &  & \textbf{-10.9}\\
& 3 & &three bursts &  & 0.3\\
& 4 & &linear SFR &  & 9.8\\
& 5 & &exponential SFR &  & 6.4\\
\hline
Seeing-&  1 & &single burst &  & 0 \\
limited & 2 & &two bursts &  & \textbf{-3.3} \\
& 3 & &three bursts &  & 12.0 \\
& 4 & &linear SFR &  & 22.7 \\
& 5 & &exponential SFR &  & 16.9\\
\enddata
\tablecomments{\\
$^{a}$ We compare the BIC within each dataset. BIC of models is minimized in the model selection, e.g. the mode with the lowest BIC is preferred.}
\end{deluxetable}

\subsection{Testings on Simulated Clusters} \label{sec:sim_test}
We test our Bayesian methodology by generating a synthetically ``observed" cluster, and inputting the simulated sample back to the fitter to derive the probability distribution function for each parameter using the Bayesian inference techniques as described above. See Appendix \ref{sec:fitter_testings} for details on the fitter tests. Figure \ref{fig:test_single} in the Appendix shows the output posterior probability distribution for simulated single-age cluster's properties. Each input parameter falls well within the 68\% (1$\sigma$ equivalent) confidence interval of the posterior probability density function. We further examine the fitter on synthetic clusters with different ages, IMFs, multiplicity, metallicity properties, and star formation history models. Our Bayesian inference methodology is always able to recover the input properties with no significant systematic biases in the tests on synthetic clusters.

\begin{figure*}[ht!]
\includegraphics[width=180mm]{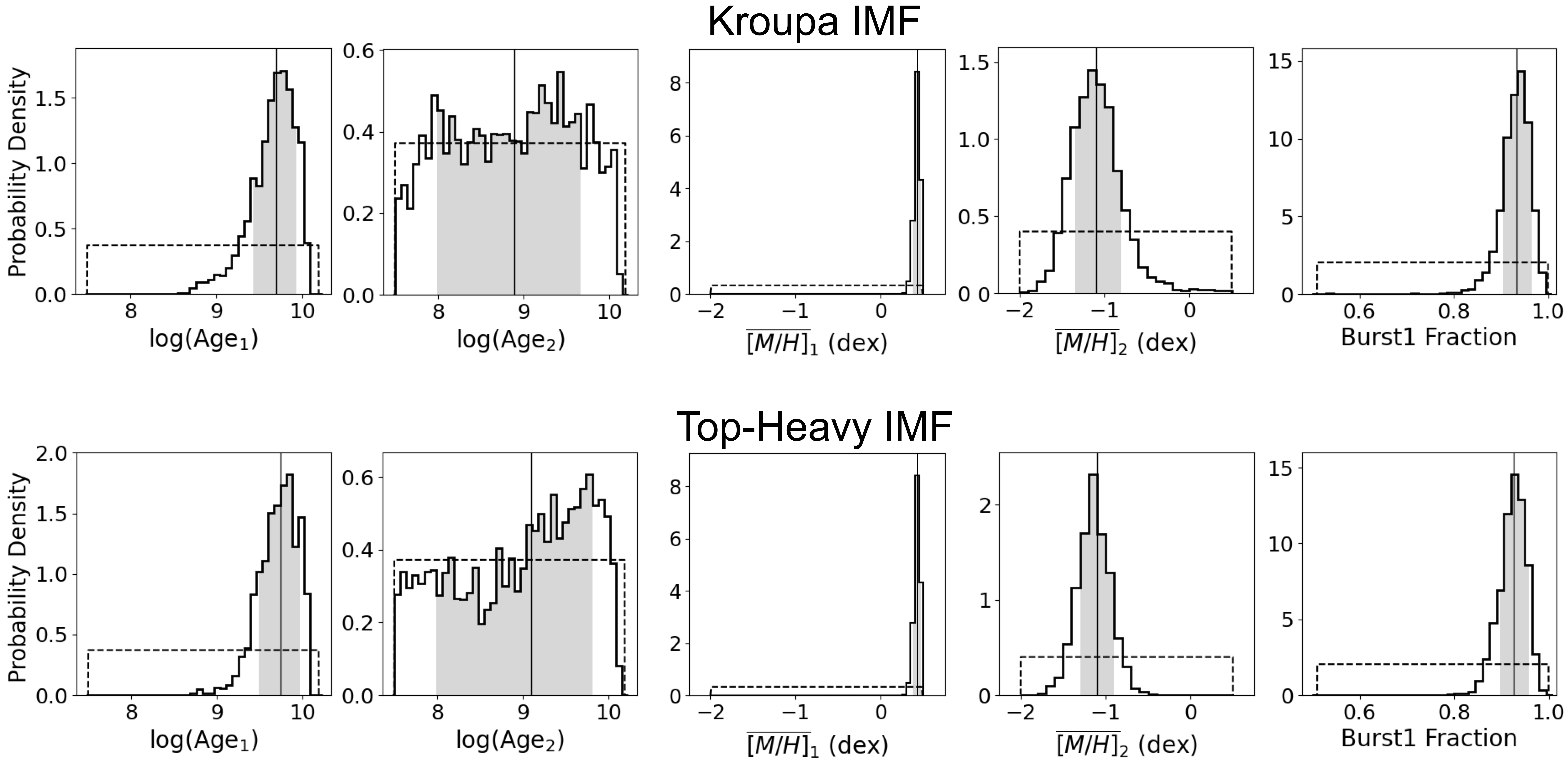}
\centering
\caption{Observed marginalized 1D posterior probability density functions of age and metallicity for each burst and the mass fraction of burst 1, based on our two-bursts star-formation history model fitted to the the \textbf{AO dataset}. The histograms show the results from the MultiNest Bayesian analysis assuming a Kroupa IMF (\textbf{top panels}), and a top-heavy IMF (\textbf{bottom panels}) respectively. The vertical solid line shows the weighted median. The shaded region shows the 68\% (1$\sigma$ equivalent) Bayesian confidence interval. The dashed line shows the adopted prior probability distribution. The resulting constraints on the age of burst 1, metallicity of both bursts, and the mass fraction are significant compared with the prior probability distributions. The constraint on the age of burst 2 is poor. \label{fig:do_1d}}
\end{figure*}

\begin{figure*}[ht!]
\includegraphics[width=180mm]{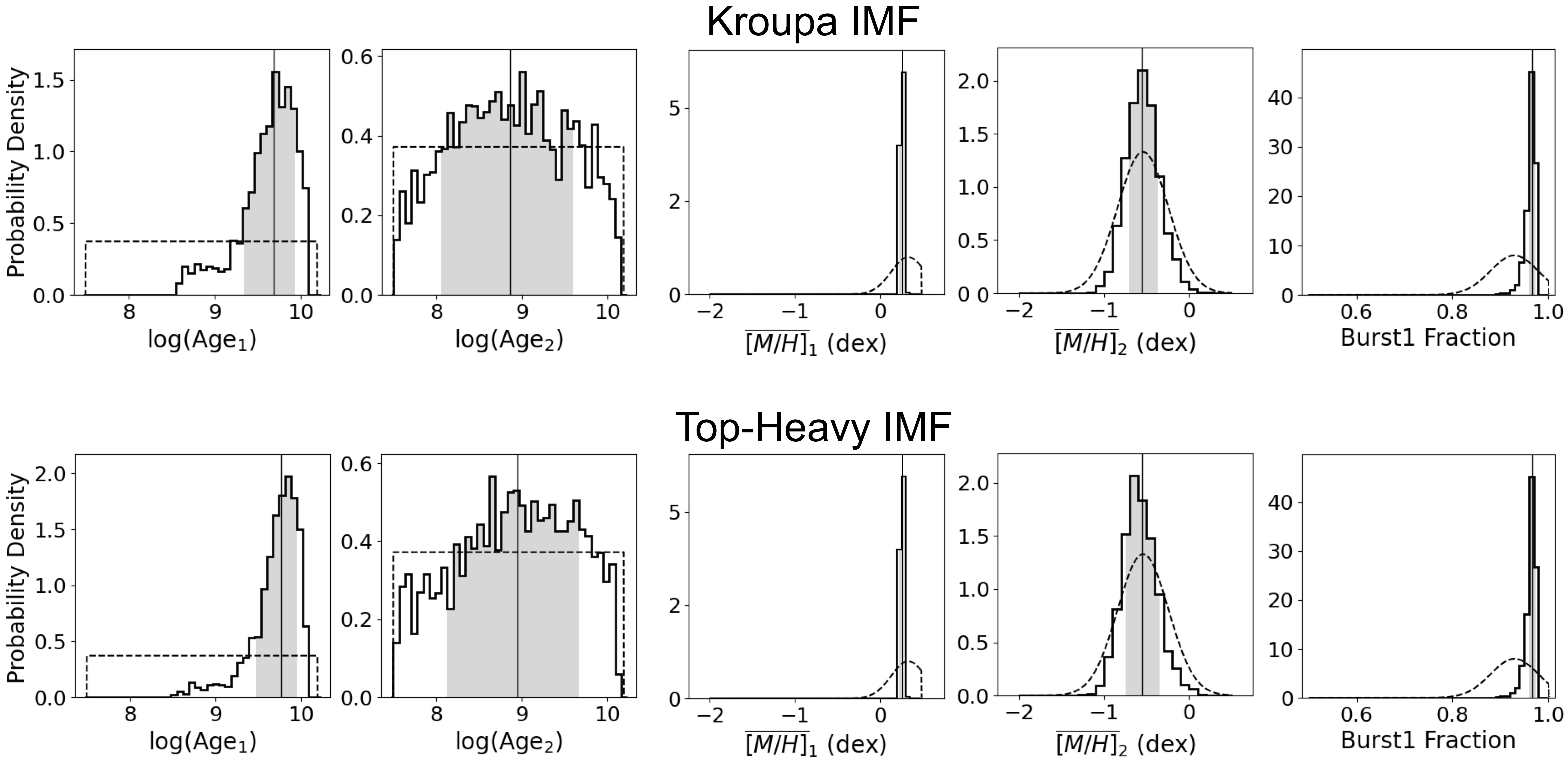}
\centering
\caption{Observed marginalized 1D posterior probability density functions from the two-bursts star-formation history model fitted to the \textbf{seeing-limited dataset}. The resulting constraints on the age, metallicity, and the mass fraction of burst 1 are significant compared with the prior probability distributions. The constraint on the age of burst 2 is poor. The constraint on the metallicity of burst 2 is largely a reflection of the prior.\label{fig:fk_1d}}
\end{figure*}

\begin{deluxetable*}{lcclllccll}
\tablenum{5}
\tablecaption{Fitting results for the AO dataset \label{tab:do_fit}}
\tablehead{
\colhead{} & \multicolumn{4}{c}{Kroupa IMF} & \colhead{}& \multicolumn{4}{c}{Top-heavy IMF} \\
\cline{2-5}
\cline{7-10}
\colhead{Cluster properties} & \colhead{MAP$^{a}$}& \colhead{Median} & \colhead{68\% interval} & \colhead{Prior$^{b}$} & & \colhead{MAP} & \colhead{Median} & \colhead{68\% interval} &  \colhead{Prior}}
\startdata
Mass fraction (burst 1) & 0.93 & 0.93 & [0.90, 0.96] & U(0,1) & & 0.93 & 0.93 & [0.90, 0.96] & U(0,1)\\
Age of burst 1 (Gyr) & 4.7 & 5.0 & [2.7, 8.4] & $U_{log}$(7.5, 10.12) & & 5.6 & 5.5 & [3.0, 8.9] & $U_{log}$(7.5, 10.12)\\
Age of burst 2 (Gyr) & 1.4 & 0.8 & [0.1, 4.6] & $U_{log}$(7.5, 10.12) & & 3.2 & 1.3 & [0.1, 6.0] & $U_{log}$(7.5, 10.12)\\
Metallicity of burst 1 & 0.45 & 0.45 & [0.40, 0.50] & U(-2.0, 0.5) & & 0.45 & 0.45 & [0.40, 0.50] & U(-2.0, 0.5) \\
Metallicity of burst 2 & -1.10 & -1.10 & [-1.35, -0.80] & U(-2.0, 0.5) & & -1.10 & -1.10 & [-1.30, -0.90]  & U(-2.0, 0.5)\\
Cluster mass$^{c}$ ($10^{5}M_{\odot}$) & 0.8 & 1.2 & [0.7, 1.7] & U(0.1, 3.0) & & 4.5  & 4.2 & [2.5, 6.1] & U(0.1, 8.0)\\
Distance (pc) & 8058 & 8031  & [7904, 8158] & G(8030, 200) & & 7966 & 8033  & [7909, 8156] & G(8030, 200)\\
IMF slope ($\alpha$) & -2.37 & -2.26 & [-2.47, -2.07] & G(-2.30, 0.36) & & -1.65 & -1.66 & [-1.77, -1.55] & G(-1.70, 0.20)\\
Average extinction & 2.61 & 2.64 & [2.56, 2.71] & G(2.64, 0.15) & & 2.64 & 2.64 & [2.56, 2.72] & G(2.64, 0.15)\\
Differential extinction & 0.12 & 0.19 & [0.07, 0.37] & U(0, 0.5) & & 0.10 & 0.21 & [0.07, 0.37] & U(0, 0.5)\\
\enddata
\tablecomments{\\
$^{a}$ Property values to get the Maximum A posterior (MAP). \\ 
$^{b}$ U(min, max): Uniform distribution between min and max. G($\mu,\sigma$): Gaussian distribution with mean $\mu$ and standard deviation $\sigma$.\\
$^{c}$ Total initial cluster mass in the observed region.}
\end{deluxetable*}

\begin{deluxetable*}{lcclllccll}
\tablenum{6}
\tablecaption{Fitting results for the seeing-limited dataset \label{tab:fk_fit}}
\tablehead{
\colhead{} & \multicolumn{4}{c}{Kroupa IMF} & \colhead{}& \multicolumn{4}{c}{Top-heavy IMF} \\
\cline{2-5}
\cline{7-10}
\colhead{Cluster properties} & \colhead{MAP}& \colhead{Median} & \colhead{68\% interval} & \colhead{Prior} & & \colhead{MAP} & \colhead{Median} & \colhead{68\% interval} &  \colhead{Prior}}
\startdata
Mass fraction (burst 1) & 0.98 & 0.97 & [0.96, 0.98] & G(0.93, 0.05) & & 0.98 & 0.97 & [0.96, 0.98] & G(0.93, 0.05)\\
Age of burst 1 (Gyr) & 5.0 & 4.9 & [2.7, 8.7] & $U_{log}$(7.5, 10.12) &  & 5.6 & 5.6 & [3.0, 8.9] & $U_{log}$(7.5, 10.12)\\
Age of burst 2 (Gyr) & 0.8 & 0.7 & [0.1, 4.3] & $U_{log}$(7.5, 10.12) & & 0.4 & 0.9 & [0.1, 4.8] & $U_{log}$(7.5, 10.12)\\
Metallicity of burst 1 & 0.30 & 0.30 & [0.25, 0.35] & G(0.33, 0.20) & & 0.30 & 0.30 & [0.25, 0.35] & G(0.33, 0.20)\\
Metallicity of burst 2 & -0.70 & -0.55 & [-0.70, -0.35]  & G(-0.54, 0.30) & & -0.55 & -0.55 & [-0.75, -0.35] & G(-0.54, 0.30)\\
Cluster mass ($10^{6}M_{\odot}$) & 2.1 & 1.9 & [1.3, 2.6]  & U(0.2, 3.5) & & 8.4 & 8.0 & [4.9, 12.2] & U(1.0, 17.0)\\
Distance (pc)  & 8033 & 8041  & [7915, 8162] & G(8030, 200) & & 8041 & 8034 & [7906, 8156] & G(8030, 200)\\
IMF slope ($\alpha$) & -2.28 & -2.31 & [-2.50, -2.12] & G(-2.30, 0.36) & & -1.69 & -1.66  & [-1.78, -1.55] & G(-1.70, 0.20)\\
Average extinction & 2.79 & 2.77 & [2.69, 2.86]  & G(2.76, 0.15) & & 2.81 & 2.77 & [2.69, 2.85] & G(2.76, 0.15)\\
Differential extinction & 0.21 & 0.20  & [0.07, 0.35] & U(0, 0.5) & & 0.09 & 0.20 & [0.08, 0.36] & U(0, 0.5)\\
\enddata
\end{deluxetable*}

\section{Results}\label{sec:results_all}
In this section, we present the fitting results on the AO and seeing-limited datasets independently. Section \ref{sec:res_model} shows that the two-bursts star formation history model is favored. Section \ref{sec:result_ao} and \ref{sec:res_see} present the resulting age, metallicity, and other cluster properties from the two-bursts modeling on each dataset respectively. Section \ref{:sec:feh_constraints} reports the impact of metallicity constraints on the age estimates of the NSC. We report that the most likely age of the main population of the NSC is $\sim$3 Gyr younger than that obtained if one assumes solar metallicity as has been done in earlier studies. Section \ref{sec:res_system} further assesses the systematic uncertainties on the cluster age, and presents arguments for why our reported star formation history is robust.

\subsection{Model selection}\label{sec:res_model}

We present the results based on the CO-$T_{eff}$ and further discuss the Starkit-$T_{eff}$ in section \ref{sec:teffcomp} and Appendix \ref{sec:app-teff}. We modeled the cluster's physical properties using two datasets independently. For each dataset, we fit the parameters listed in Table \ref{tab:SFH_model} for five star formation history models. 

Table \ref{tab:bayes_do} summarizes the $\Delta$BIC between each model for 5 fits on the AO and seeing-limited dataset respectively. The model with the lowest BIC is preferred. For both datasets, the observations show a strong evidence for the two bursts star formation history model (shown in bold in Table \ref{tab:bayes_do}). We will present the results from the two bursts model in the rest of the paper.

\begin{figure*}[ht!]
\includegraphics[width=182mm]{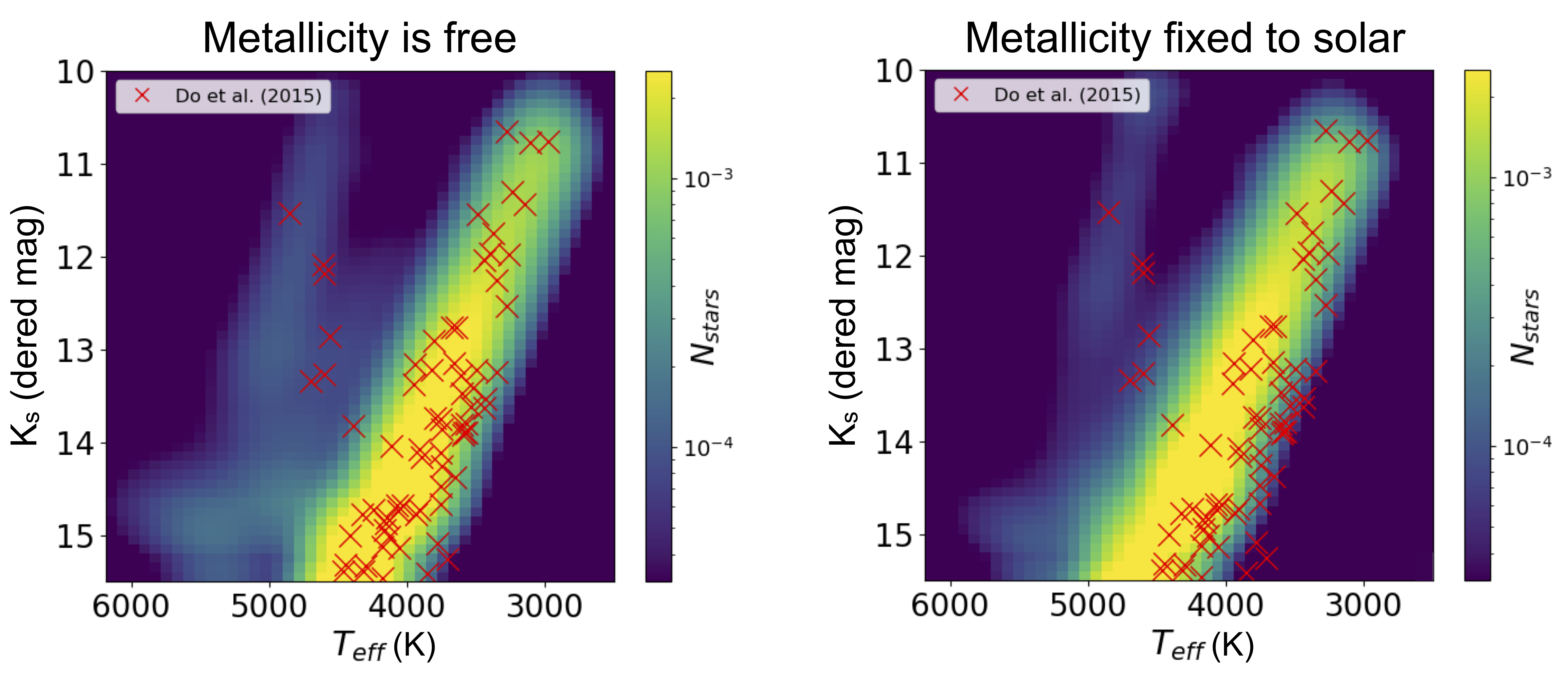}
\centering
\caption{\textbf{Left:} Comparison between the observed AO dataset (\citealt{Do et al. 2015}, red crosses) and the predicted Hess diagram with fitting weights from our best-fit star-formation history using the first metallicity constraints. The observed dataset is differentially dereddened at $K_s$ band. The cluster model has been convolved with observational uncertainties and modified by the completeness correction. The AO dataset is well characterized by the two-bursts model of star formation. The bulk stellar mass ($\sim$ 90\%) is older and metal-rich (bright strip). The minor group is metal-poor (upper left). \textbf{Right:} Comparison between the observed dataset and the predicted Hess diagram with the assumption of solar metallicity ($\overline{[M/H]}$ = 0) for all stars in the NSC. With fixed solar metallicity, the age of the bulk stellar mass was modeled to be $\sim$3 Gyr older. Furthermore, we note that, by including metallicity as a free parameter \textbf{(left panel)}, our models are able to account for low-temperature red giants that were previously difficult to fit. \label{fig:hess_do}}
\end{figure*}

\subsection{AO dataset}\label{sec:result_ao}

Ten free parameters are fitted to the AO dataset with the two-bursts star-formation history model: mass fraction of burst 1, age of burst 1, age of burst 2, metallicity of burst 1, metallicity of burst 2, total initial cluster mass (in the observed region), distance to the cluster, IMF slope ($\alpha$), average extinction, and differential extinction. We report the results based on two IMF scenarios (see section \ref{sec:prior}) with either a Kroupa IMF ($\alpha = -2.3 \pm 0.36$, \citealt{Kroupa 2001}) or a top-heavy IMF ($\alpha = -1.7 \pm 0.20$, \citealt{Lu et al. 2013}).

Figure \ref{fig:do_1d} shows the 1D posterior probability distributions from the Multinest Bayesian analysis for five of the parameters in the two-bursts modeling assuming a Kroupa and a top-heavy IMF respectively. See Table \ref{tab:do_fit} for the fitting results of all parameters with the median and 68\% (1$\sigma$ equivalent) Bayesian confidence intervals, along with the adopted priors. The confidence intervals are calculated by first finding the 50$^{th}$ percentile of the marginalized 1D posterior probability distribution and then stepping away from the center until the integrated probability reaches 68\%. We also report the Maximum A Posterior (MAP) value for each parameter. 

\begin{figure*}[ht!]
\includegraphics[width=182mm]{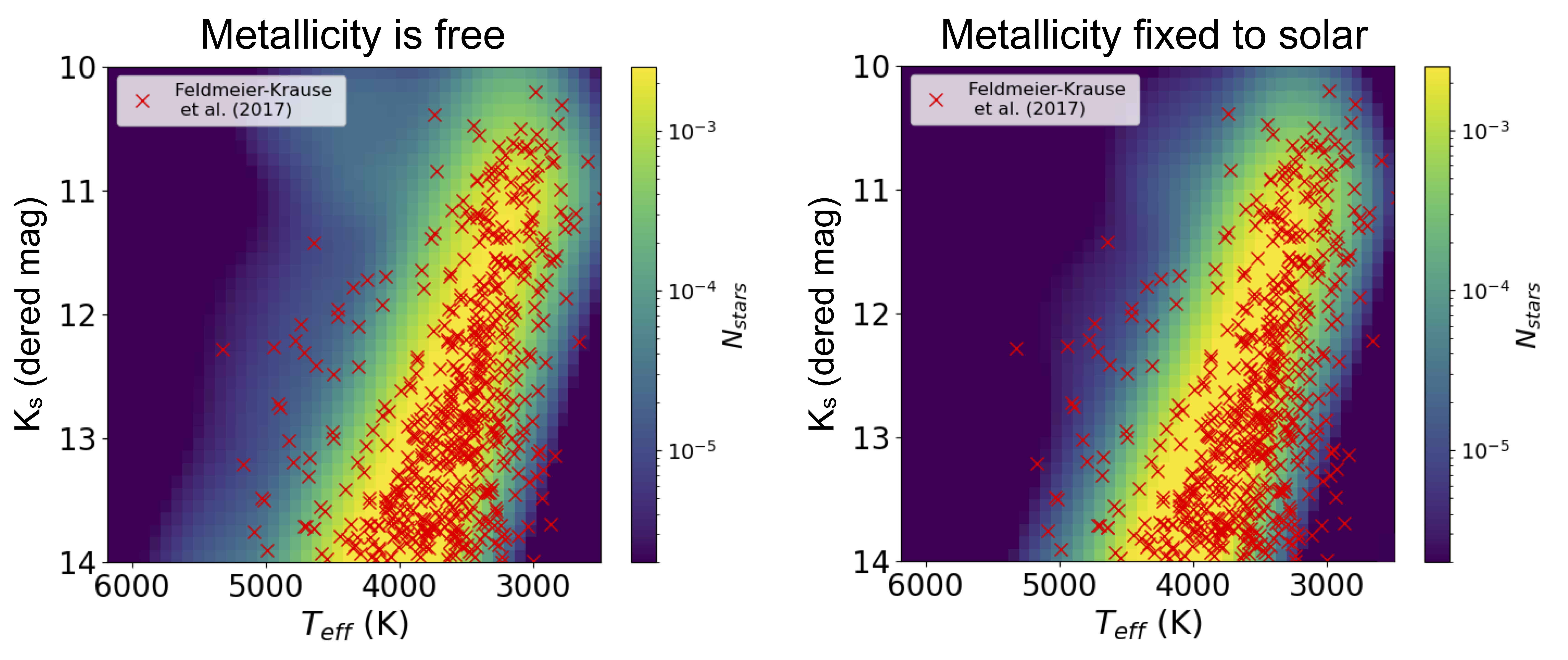}
\centering
\caption{\textbf{Left:} Comparison between the observed seeing-limited dataset (\citealt{Feldmeier-Krause et al. 2017}, red crosses) and the predicted Hess diagram with fitting weights from our best-fit star-formation history with metallicity constraints. The observed dataset is well characterized by the two-bursts star-formation model. \textbf{Right:} Comparison between the observed dataset and the predicted Hess diagram with the assumption of solar metallicity ($\overline{[M/H]}$ = 0) for all stars in the NSC. \label{fig:hess_fk}}
\end{figure*}

\begin{figure*}[ht!]
\includegraphics[width=184mm]{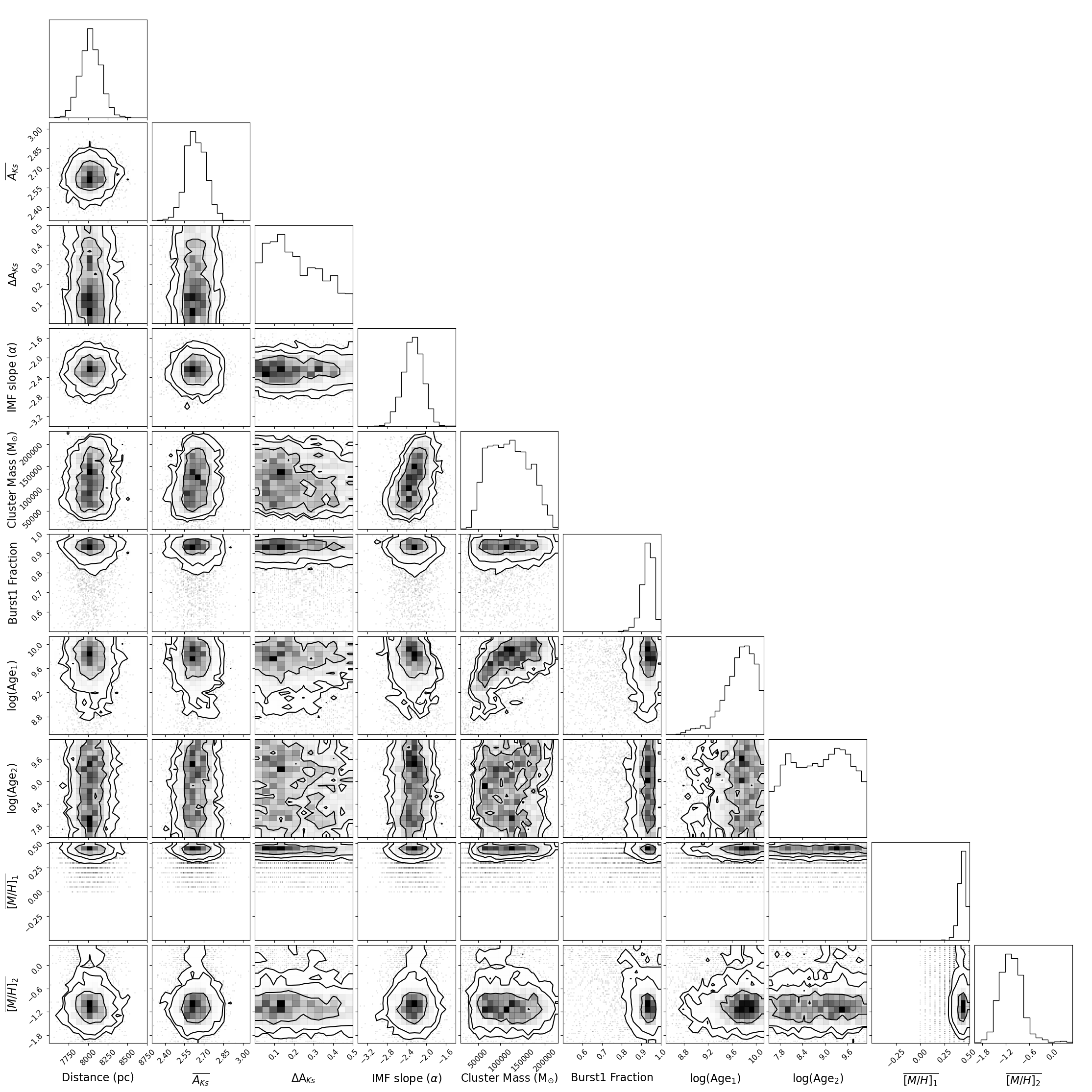}
\centering
\caption{Two-dimentional posterior probability density functions for the observed NSC's properties from modeling the AO dataset \citep{Do et al. 2015} assuming a Kroupa IMF. Here we show the results from the MultiNest Bayesian analysis on the two-bursts star formation model after the model selection. The overplotted contours give 68\%, 95\%, and 99\% confidence intervals.\label{fig:dos_2d}}
\end{figure*}

\begin{figure*}[ht!]
\includegraphics[width=184mm]{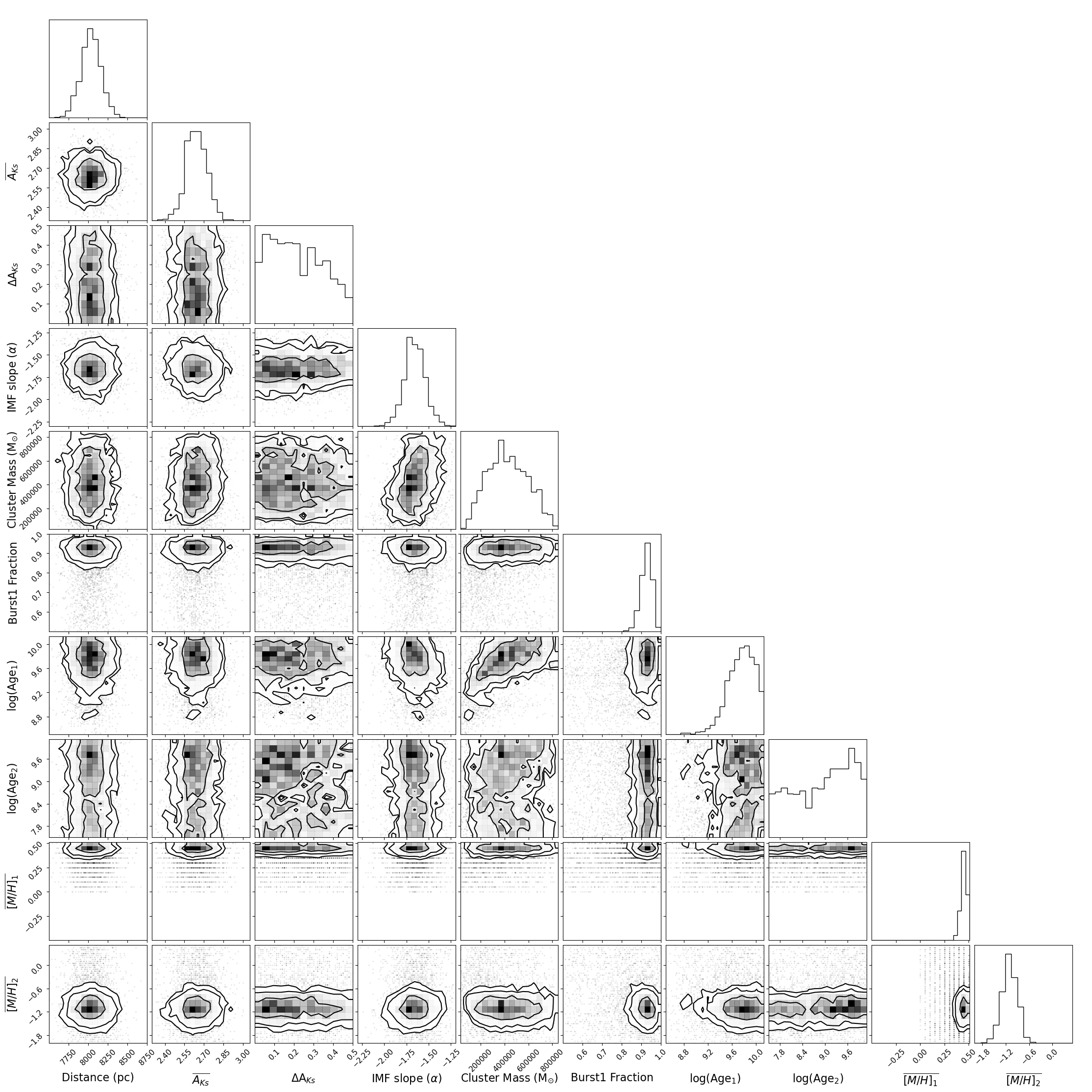}
\centering
\caption{Two-dimentional posterior probability density functions for the observed NSC's properties from modeling the AO dataset \citep{Do et al. 2015} assuming a top-heavy IMF. \label{fig:dol_2d}}
\end{figure*}

Here is the summary of the fitting results. With the assumption of \textbf{(1) the Kroupa IMF:} the bulk of the stellar mass (93\% $\pm$ $3\%$) is modeled to have formed 5.0 $^{+3.4}_{-2.3}$ Gyr ago (Age$_{MAP}$ = 4.7 Gyr), and is metal-rich ($\overline{[M/H]}$ = 0.45 $\pm$ 0.05). The burst 2 with 7\% $\pm$ $3\%$ of the stellar mass is modeled to form 0.8 $^{+3.8}_{-0.7}$ Gyr ago (Age$_{MAP}$ = 1.4 Gyr), and is metal-poor ($\overline{[M/H]}$ = -1.10 $^{+0.30}_{-0.25}$). \textbf{(2) top-heavy IMF:} the bulk stellar mass (93\% $\pm$ $3\%$) is modeled to form 5.5 $^{+3.4}_{-2.5}$ Gyr ago (Age$_{MAP}$ = 5.6 Gyr), and is metal-rich ($\overline{[M/H]}$ = 0.45 $\pm$ 0.05). Burst 2, with 7\% $\pm$ $3\%$ of the stellar mass, is modeled to form 1.3 $^{+4.7}_{-1.2}$ Gyr ago (Age$_{MAP}$ = 3.2 Gyr), and is metal-poor ($\overline{[M/H]}$ = -1.10 $\pm$ 0.20). Of particular note, the age of burst 2 is poorly constrained.  The 1D posterior probability distribution of the age of this burst is nearly flat with respect to the adopted prior probability distribution. This is owing to the small fraction of the total sample size represented by this burst, and consequently the small observed sample size. See Figure \ref{fig:dos_2d} and \ref{fig:dol_2d} for the two-dimensional posterior probability density functions.

As discussed in section \ref{sec:prior}, some properties including the total cluster mass, IMF slope and the age show moderate correlations (also see Figures \ref{fig:dos_2d} and \ref{fig:dol_2d}). At older ages, the most massive stars have disappeared and the total initial cluster mass needs to be increased to match the observed numbers of stars. Assuming a top-heavy IMF results in a higher total cluster mass than when a Kroupa IMF is assumed (4.2 $^{+1.9}_{-1.7}$ $\times$ $10^5$ $M_{\odot}$ and 1.2 $^{+0.5}_{-0.5}$ $\times$ $10^5$ $M_{\odot}$, respectively). In summary, the fitting results from the two IMF profiles show very consistent modeling within the uncertainties on all cluster properties, except for the total cluster mass.

\subsection{Seeing-limited dataset}\label{sec:res_see}

Similarly, ten free parameters are fitted in the two-bursts star-formation history model to the seeing-limited dataset under two assumptions of IMF. Specifically, we include the prior knowledge (see Table \ref{tab:fk_fit}) on the mass fraction and metallicity of each burst from the dynamical modeling on this dataset \citep{Do et al. 2020}. Figure \ref{fig:fk_1d} shows the resulting 1D posterior probability distributions for five of the parameters assuming a Kroupa IMF and a top-heavy IMF respectively. Table \ref{tab:fk_fit} displays the fitting results for all parameters with their median and 68\% (1$\sigma$ equivalent) Bayesian confidence intervals, the calculated MAP value, and the adopted priors.

\begin{figure}[ht!]
\includegraphics[width=85mm]{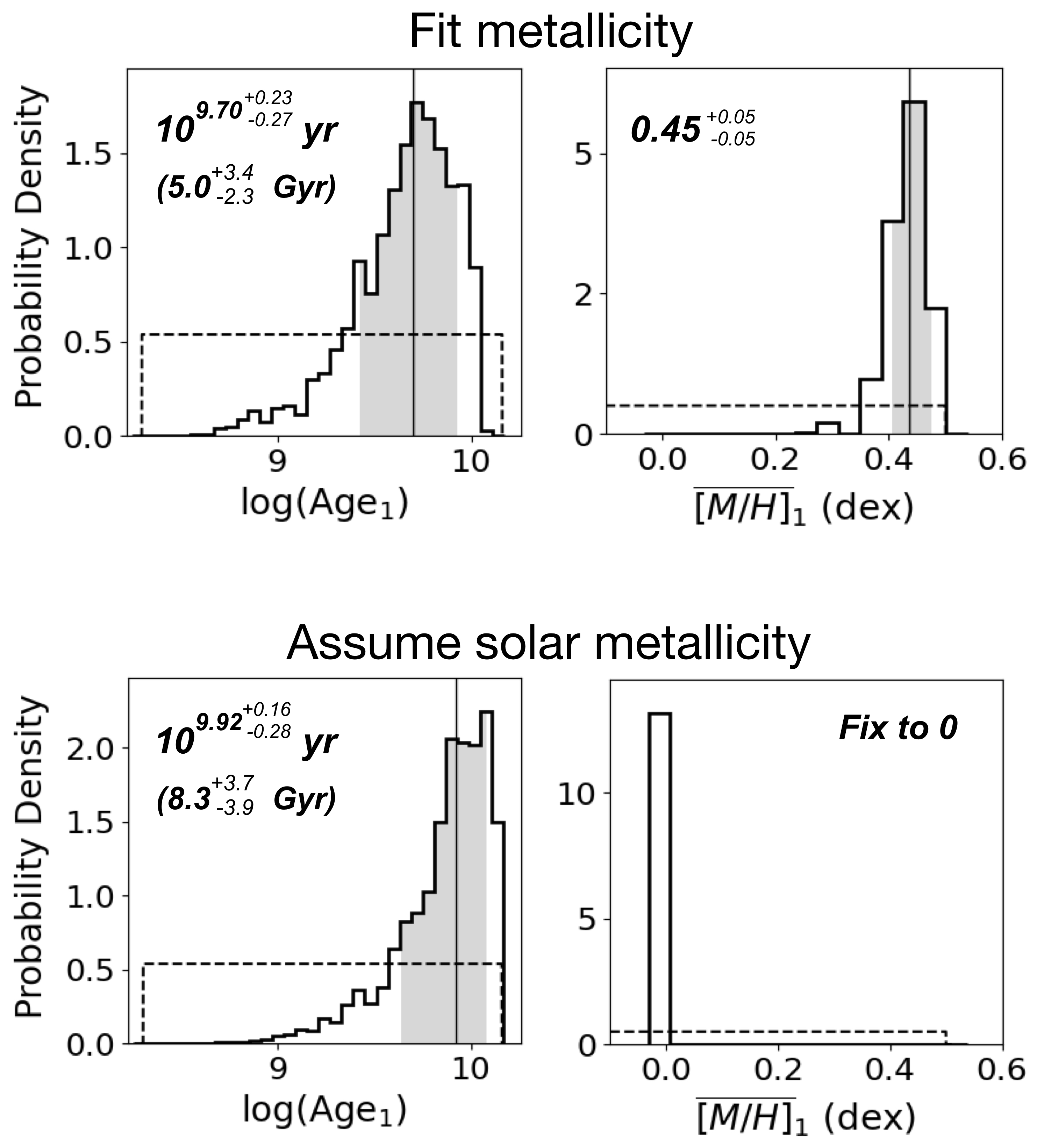}
\centering
\caption{Observed marginalized 1D posterior probability density functions for age and metallicity of NSC stars formed in burst 1, assuming a Kroupa IMF in our modeling of the AO dataset. The vertical solid line shows the weighted median. \textbf{Top panels:} burst metallicity constrained by stellar metallicity measurements, and the bulk of the stellar mass was modeled to be 5.0 $^{+3.4}_{-2.3}$ Gyr old and metal-rich ($\overline{[M/H]}$ = 0.45 $\pm$ 0.05). \textbf{Bottom panels:} assuming that stars have solar metallicity, as done by previous works, yields an age of 8.3 $^{+3.7}_{-3.9}$ Gyr. The most likely age for the main population of the NSC is $\sim$3 Gyr older than our determination if we assume a solar metallicity for all stars. \label{fig:obs_fehcomp}}
\end{figure}

\begin{deluxetable*}{cccccccccc}
\tablenum{7}
\tablecaption{Fitted metallicity vs. fixed solar metallicity for burst 1$^{a}$\label{tab:feh_comp}}
\tablehead{
\colhead{} & \colhead{} & \multicolumn{3}{c}{Fit metallicity} & \colhead{} & \colhead{} & \multicolumn{3}{c}{Fix to solar metallicity} \\
\cline{3-5}
\cline{8-10}
\colhead{Dataset} & \colhead{IMF} & \colhead{Age (Gyr)} & \colhead{$\overline{[M/H]}$} & \colhead{$\Delta$BIC$^{b}$} & & &   \colhead{Age (Gyr)} & \colhead{$\overline{[M/H]}$} & \colhead{$\Delta$BIC$^{b}$}}
\startdata
AO & Kroupa & 5.0 $^{+3.4}_{-2.3}$ & 0.45 $\pm$ 0.05 & 0 & & & 8.3 $^{+3.7}_{-3.9}$ & 0 & 35.8\\
& Top-heavy & 5.5 $^{+3.4}_{-2.5}$ & 0.45 $\pm$ 0.05 & -12.0 & & & 8.4 $^{+3.8}_{-3.5}$ & 0 & 37.3\\
\hline
Seeing-limited & Kroupa & 4.9 $^{+3.8}_{-2.2}$ & 0.30 $\pm$ 0.05 & 0 & & & 7.9 $^{+3.5}_{-3.4}$ & 0 & 146.5\\
& Top-heavy & 5.6 $^{+3.3}_{-2.6}$ & 0.30 $\pm$ 0.05 & 2.2 & & & 8.7 $^{+3.0}_{-3.9}$ & 0 & 155.2\\
\enddata
\tablecomments{\\
$^{a}$ The bulk of the stellar mass of the NSC.\\
$^{b}$ We compare the BIC within each dataset. The model with the lowest BIC is preferred.}
\end{deluxetable*}

\begin{figure*}[ht!]
\includegraphics[width=184mm]{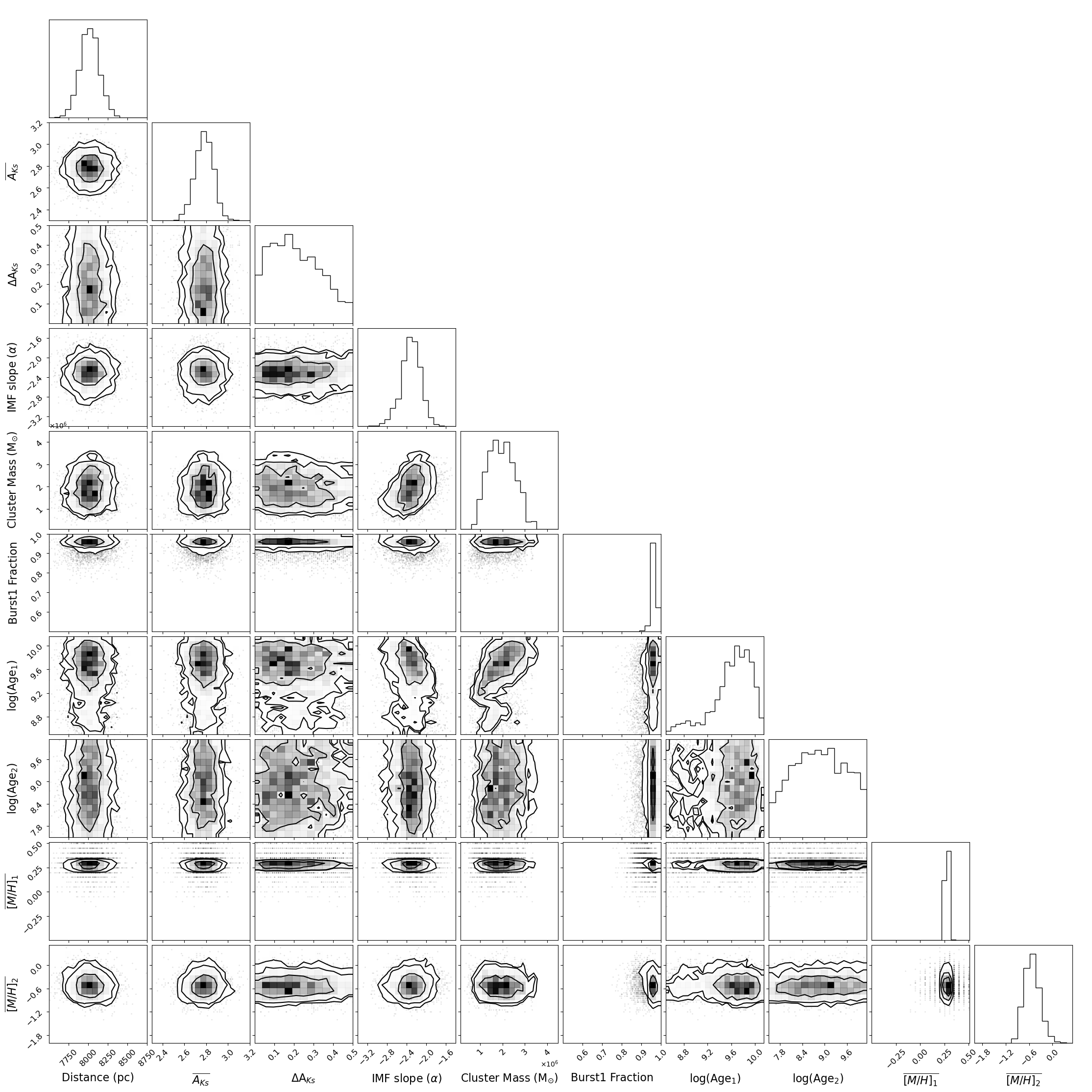}
\centering
\caption{Two-dimentional posterior probability density functions for the observed NSC's properties from modeling the seeing-limited dataset \citep{Feldmeier-Krause et al. 2017} assuming a Kroupa IMF. \label{fig:fks_2d}}
\end{figure*}

\begin{figure*}[ht!]
\includegraphics[width=184mm]{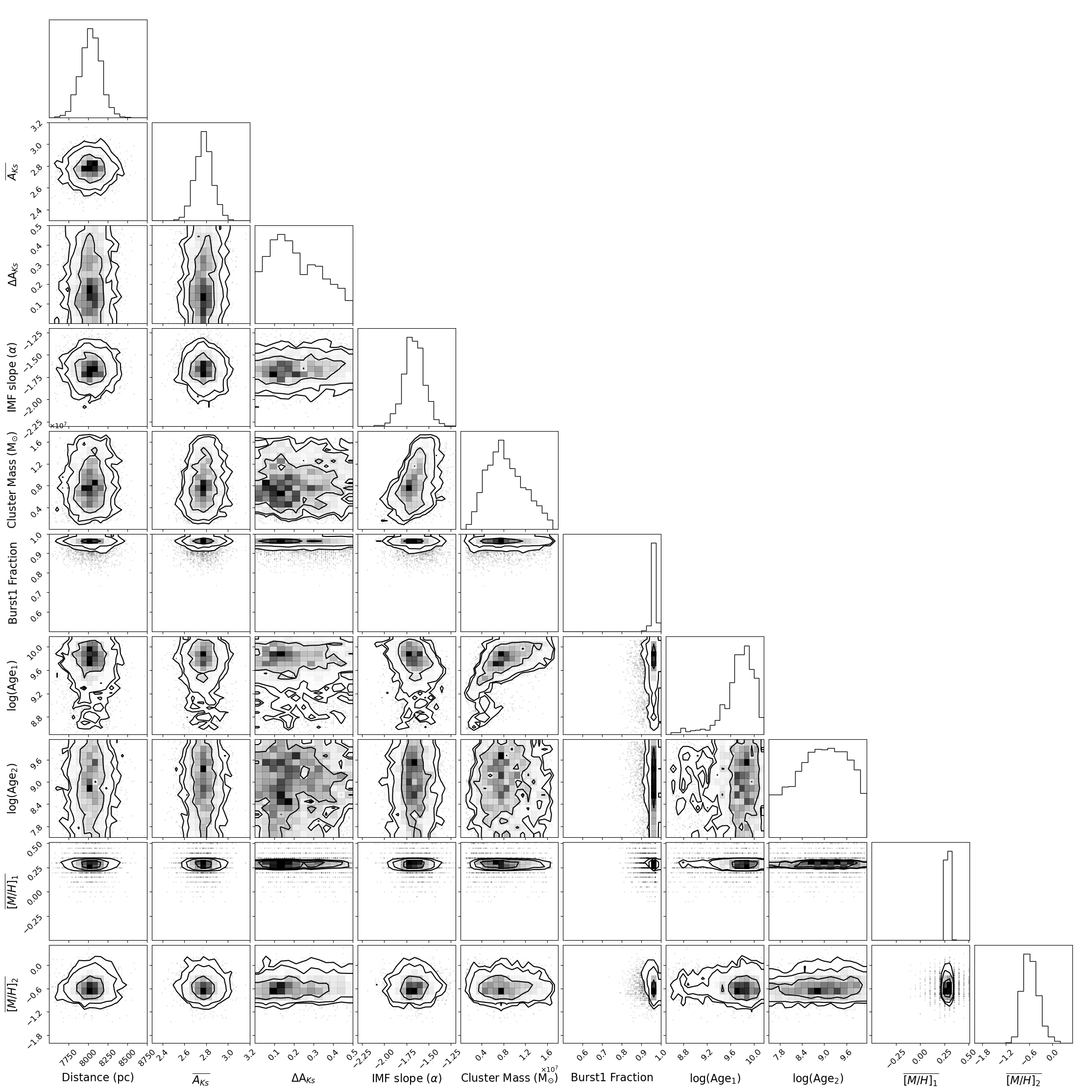}
\centering
\caption{Two-dimentional posterior probability density functions for the observed NSC's properties from modeling the seeing-limited dataset \citep{Feldmeier-Krause et al. 2017} assuming a top-heavy IMF. \label{fig:fkl_2d}}
\end{figure*}

A summary of the fitting results for the seeing-limited dataset is as follows: with the assumption of \textbf{(1) the Kroupa IMF:} the bulk of the stellar mass (97\% $\pm$ $1\%$) is modeled to have formed 4.9 $^{+3.8}_{-2.2}$ Gyr ago (Age$_{MAP}$ = 5.0 Gyr), and is metal-rich ($\overline{[M/H]}$ = 0.30 $\pm$ 0.05). Burst 2, with 3\% $\pm$ $1\%$ of the stellar mass, is modeled to have formed 0.7 $^{+3.6}_{-0.6}$ Gyr ago (Age$_{MAP}$ = 0.8 Gyr), and is metal-poor ($\overline{[M/H]}$ = -0.55 $^{+0.20}_{-0.15}$). \textbf{(2) top-heavy IMF:} the bulk of the stellar mass (97\% $\pm$ $1\%$) is modeled to have formed 5.6 $^{+3.3}_{-2.6}$ Gyr ago (Age$_{MAP}$ = 5.6 Gyr), and is metal-rich ($\overline{[M/H]}$ = 0.30 $\pm$ 0.05). Burst 2 with 3\% $\pm$ $1\%$ of the stellar mass is modeled to have formed 0.9 $^{+3.9}_{-0.8}$ Gyr ago (Age$_{MAP}$ = 0.4 Gyr), and is metal-poor ($\overline{[M/H]}$ = -0.55 $\pm$ 0.20). Similar to the AO dataset, the age of burst 2 is poorly constrained. Figures \ref{fig:fks_2d} and \ref{fig:fkl_2d} provide the two-dimensional posterior probability density functions.

Assuming a top-heavy IMF results in a higher total cluster mass than when a Kroupa IMF is assumed (8.0 $^{+4.2}_{-3.1}$ $\times$ $10^6$ $M_{\odot}$ and 1.9 $^{+0.7}_{-0.6}$ $\times$ $10^6$ $M_{\odot}$, respectively). In summary, the fitting results from the two alternative IMF profiles show very consistent modeling within the uncertainties on all cluster properties, except for the total cluster mass.

A comparison of the fitting results for the seeing-limited and deeper AO datasets shows consistency between them. In particular, we report consistent age estimates from both datasets for each of the star formation bursts. See the following sections \ref{:sec:feh_constraints} and \ref{sec:res_system} for further investigations of the impact of metallicity and systematic uncertainties on the cluster age. Further comparison between the results for each dataset are presented in Appendix \ref{sec:app_2datasets}. The slightly higher mass fraction of burst 1 in the seeing-limited dataset compared to that of the AO dataset is due to the shallower seeing-limited observations and a conservative K magnitude cut (K = 14 mag). A comparison of the observed dataset and the modeled Hess diagram from the inferred parameters of the star-formation history fits is shown in Figure \ref{fig:hess_do} (AO dataset) and Figure \ref{fig:hess_fk} (seeing-limited dataset).

\subsection{Impact of metallicity constraints}\label{:sec:feh_constraints}
In this work, we report the star formation history of the NSC with the first metallicity constraints as obtained from individual stellar metallicity measurements. In order to understand the impact of metallicity measurements on the age estimates of the NSC, we compare the fitting results with and without metallicity constraints. To assess the effect of modeling metallicity as a free parameter, we repeated the fit of the star formation history with the assumption of fixed solar metallicity ($\overline{[M/H]}$ = 0) for all stars in the NSC as has been done by earlier studies (e.g., \citealt{Pfuhl et al. 2011}). All fitting configurations and priors on the rest of the parameters are consistent for fair comparison. Since the age constraints on the minor group (burst 2) are relatively poor, here we only examine the impact of metallicity constraints on the age of the bulk of the stellar mass (burst 1). See Table \ref{tab:feh_comp} for the fitting results and the $\Delta$BIC when we model the metallicity as a free parameter, compared to those with a fixed solar metallicity. The fitted-metallicity models are overwhelmingly preferred over the fixed-solar-metallicity models. When metallicity is fixed to be solar, the median age of the NSC main population increases by $\sim$3 Gyr. For $\sim$90\% of the stellar mass, this assumption results an age of 8.3 $^{+3.7}_{-3.9}$ Gyr (Kroupa IMF) and 8.4 $^{+3.8}_{-3.5}$ Gyr (top-heavy IMF) from the AO dataset, and 7.9 $^{+3.5}_{-3.4}$ Gyr (Kroupa IMF) and 8.7 $^{+3.0}_{-3.9}$ Gyr (top-heavy IMF) from the seeing-limited dataset. The systematic bias to higher ages is due to the fact that high-metallicity stars tend to be cooler and less luminous. 

In summary, the most likely age for the NSC main population reported in this work with metallicity constraints is $\sim$3 Gyr younger than that obtained if one assumes solar metallicity. See Figure \ref{fig:obs_fehcomp} for an example of the comparison with and without metallicity constraints. The comparison of the observed and best-fit modeled Hess diagram from the inferred parameters with and without metallicity constraints are shown in Figures \ref{fig:hess_do} and \ref{fig:hess_fk}. Furthermore, we note that, by including metallicity as a free parameter, our models are able to account for low-temperature red giants that were previously difficult to fit.

\subsection{Systematic uncertainties on the cluster age} \label{sec:res_system}
We further assess the accuracy of our age estimates of the NSC by considering the impact of systematic errors from the following effects: (i) IMF assumptions, (ii) different methods of measuring $T_{eff}$, (iii) priors on the model parameters, (iv) uncertainties in metallicity measurements, (v) limitation of theoretical stellar evolutionary models, (vi) different spectral resolutions and grids, and (vii) contamination from foreground or background sources. Our analyses lead us to conclude that these possible systematic uncertainties do not lead to any substantial bias in the age estimates. The reported star formation history and the impact of metallicity constraints are robust and reliable.

\subsubsection{IMF assumptions}
In order to understand the impacts from the IMF assumptions (see section \ref{sec:prior}), we modeled the star formation history and cluster's physical properties with two IMF scenarios independently: a Kroupa IMF ($\alpha = -2.3$ $\pm$ 0.36, \citealt{Kroupa 2001}) or a top-heavy IMF ($\alpha = -1.7$ $\pm$ 0.20, \citealt{Lu et al. 2013}). Tables \ref{tab:do_fit} and \ref{tab:fk_fit} summarize the comparison of the fitting properties. For both datasets, assuming a top-heavy IMF results in a slightly older age for each burst. However, the age difference (either on the median or MAP value) due to the IMF assumptions is always smaller than 1 Gyr, which is much smaller than the 1$\sigma$ equivalent uncertainty on the age from the 68\% Bayesian confidence intervals. No additional systematic uncertainty (or rescaling of the two IMF assumptions) is suggested by the fits. We also note that the low-metallicity component (burst 2) shows similar values under different assumptions of IMF. No indication of different IMFs for different components is suggested in this work. Furthermore, we investigated the impact of metallicity constraints under the two IMF assumptions independently (also see section \ref{:sec:feh_constraints} and Table \ref{tab:feh_comp}). Any potential age bias attributable to uncertainties associated with the two IMF assumptions is negligible compared to the age difference resulting from imposing the metallicity measurements, compared to assuming solar metallicity. The impact of metallicity constraints that we report is robust. 

\subsubsection{Methods of measuring stellar effective temperature}\label{sec:teffcomp}
We assess the possibility of a systematic offset of cluster age resulting from two different methods of measuring stellar effective temperature: CO-$T_{eff}$, derived from the calibrated $T_{eff}$-$EW_{CO}$ (CO equivalent width) relation \citep{Feldmeier-Krause et al. 2017}; and STARKIT $T_{eff}$, derived from full-spectrum fitting using the STARKIT code \citep{Kerzendorf Do 2015} with synthetic grids. See Appendix \ref{sec:app-teff} for details. We followed the same methodology and repeated the modeling of the star-formation history and other cluster properties with Starkit $T_{eff}$. For both datasets, the differential impact on the fitting results of using the Starkit $T_{eff}$ compared to using CO-$T_{eff}$ is very small (see Table \ref{tab:teff_comp}). No additional systematic uncertainty is suggested by the $T_{eff}$ assumptions adopted for the fits. We further investigated the impact of metallicity constraints using the different $T_{eff}$ independently (see Table \ref{tab:teff_comp}). The age bias due to the different $T_{eff}$ assumptions is negligible, and the impact of metallicity constraints that we present in this work is still robust.

\subsubsection{Priors on model parameters}
Since the extinction and cluster age show a moderate correlation, we further investigate the possible systematic uncertainty from the fitting priors adopted for the average extinction ($\overline{A_{Ks}}$, see section \ref{sec:prior}). We tested the fitting bias by repeating the modelings using a more conservative uniform prior on the $\overline{A_{Ks}}$ covering a 5-$\sigma$ range around the mean of stellar extinction values. The results are consistent and show that no additional systematic bias needs to be considered. 

\subsubsection{Uncertainties in metallicity measurements}

Both AO and seeing-limited datasets have a fraction of very metal-rich stars with metallicity measurements inferred to be higher than $[M/H]$ = +0.5, though they are subject to greater systematic uncertainties (\citealt{Do et al. 2015,Feldmeier-Krause et al. 2017}). We assess the accuracy of our age estimates using simulations that introduce a bias in the observations to see the effect of an artificial tail in the metallicity distribution at high metallicities. See Appendix \ref{sec:app-mh} for detailed simulations. The fitting results on synthetic clusters show that the fitter is still able to recover the input cluster age with no bias. We find that the peak of the metallicity distribution is of the most important factor for our conclusions about the star formation history. The high metallicity tail that we observed does not change the results significantly. The peak at approximately twice solar metallicity is a robust result that is confirmed by multiple studies by different groups using different data and methods (e.g., \citealt{Do et al. 2015, Feldmeier-Krause et al. 2017, Rich et al. 2017, Schodel et al. 2020, Nogueras-Lara et al. 2022}). Super solar metallicity stars in the NSC ($[M/H]$ $>$ +0.3) were also observed in using high-resolution spectroscopic studies (e.g., \citealt{Rich et al. 2017, Do et al. 2018, Thorsbro et al. 2020}). Furthermore, the metallicity calibration used by \citet{Feldmeier-Krause et al. 2017} sample has been confirmed with a larger sample to $[M/H]$ = +0.5 dex \citep{Feldmeier-Krause 2022}, which is beyond the peak of the metallicity distribution. We conclude that no additional systematic bias needs to be considered.

\subsubsection{Limitation of theoretical stellar evolutionary models}

One limitation is that the current upper limit of the metallicity available in all theoretical evolutionary models is $\overline{[M/H]}$ = +0.5 ($\sim$3 times solar). The range is limited by incomplete knowledge of the opacities and the equation of state \citep{Choi et al. 2016}. The posterior distribution of metallicity of burst 1 modeled from the AO dataset has a distribution that peaks near the edge of the grids at $\overline{[M/H]}$ = +0.5. This brings some systematic uncertainties on the resulting cluster age estimates. With improved grids covering a larger metallicity range in the future, we could expect an even younger age of the NSC's main population. Our results present a conservative estimate on the impact of metallicity constraints on the cluster age (see section \ref{:sec:feh_constraints}).

\subsubsection{Different spectral resolutions and grids for the two datasets}

We investigate the possible systematic offsets between two datasets that were observed with different spectral resolution, and analyzed using different spectral grids. As discussed in \citet{Feldmeier-Krause et al. 2017} and \citet{Feldmeier-Krause 2022}, the absolute metallicity measurements above $[M/H]$ = +0.5 and below $[M/H]$ = -0.5 from the seeing-limited observations are difficult to measure and calibrate to higher accuracy due to the lower spatial and spectral resolution than that of AO. The systematic uncertainties for those measurements are potentially underestimated. \citet{Feldmeier-Krause et al. 2017} investigated and claimed that a lower spectral resolution would result in a lower $[M/H]$ measurement by a systematic shift of 0.1 dex. We further investigated such effects by re-fitting the AO spectra using the PHOENIX grid for the 27 common stars of the two datasets. The re-fitted metallicity measurements show that both the spectra resolution and the grids have about the same effect on the overall difference between the two datasets. See Appendix \ref{sec:app-mh-comp} and Figure \ref{fig:mh_grids_comp} for details. In summary, the 27 common stars between the two surveys have consistent metallicity measurements within the uncertainties, indicating that the two datasets with different spectral resolution and grids are in reasonable agreement. 

In this work, the mean metallicity of the bulk of the stellar mass (burst 1) is modeled to be 0.45 $\pm$ 0.05 (AO dataset), and 0.30 $\pm$ 0.05 (seeing-limited dataset). They agree to within 2$\sigma$. In the following, we will use a metallicity of 0.35 $\pm$ 0.05 (2$\sigma$ overlap) for subsequent predictions of compact objects and merger rates. The small offset is a reflection of the systematic effects from the two datasets listed above. The mean metallicity of the metal-poor burst from the two datasets is also consistent within 2$\sigma$. Our reported fitting results represent a robust estimate of the systematic uncertainties inherent in the two datasets. In this work we assume a single metallicity for each burst. We do not model the metallicity dispersion due to the fact that the metallicity spread in the distribution ($\sigma\sim$ 0.35) is roughly comparable to the uncertainties on the individual stellar metallicity measurements ($\Delta\sim$ 0.32). More data with a higher accuracy is needed in the future to measure the intrinsic dispersion of the metallicity of each burst. 

\subsubsection{Contamination from foreground or background sources}\label{sec:sys_contam}

The membership of each star to the Milky Way NSC has been identified using extinction corrected colors and contamination analyses for the AO dataset (see section 6.1 in \citealt{Do et al. 2015}) and seeing-limited dataset (see section 5.4 in \citealt{Feldmeier-Krause et al. 2017}) respectively. In addition, this work includes only the observations within the central 1.5 pc, where the NSC dominates. The number density of stars in the NSC, in this region, is roughly 100 times higher than that of the NSD \citep{Sormani et al. 2022}. We would therefore expect negligible contamination from the NSD in our sample.

In summary, our analyses conclude that these systematic uncertainties do not lead to any substantial bias in the estimation of the cluster age presented in this work. The reported star formation history is robust. 

\begin{figure}[ht!]
\includegraphics[width=85mm]{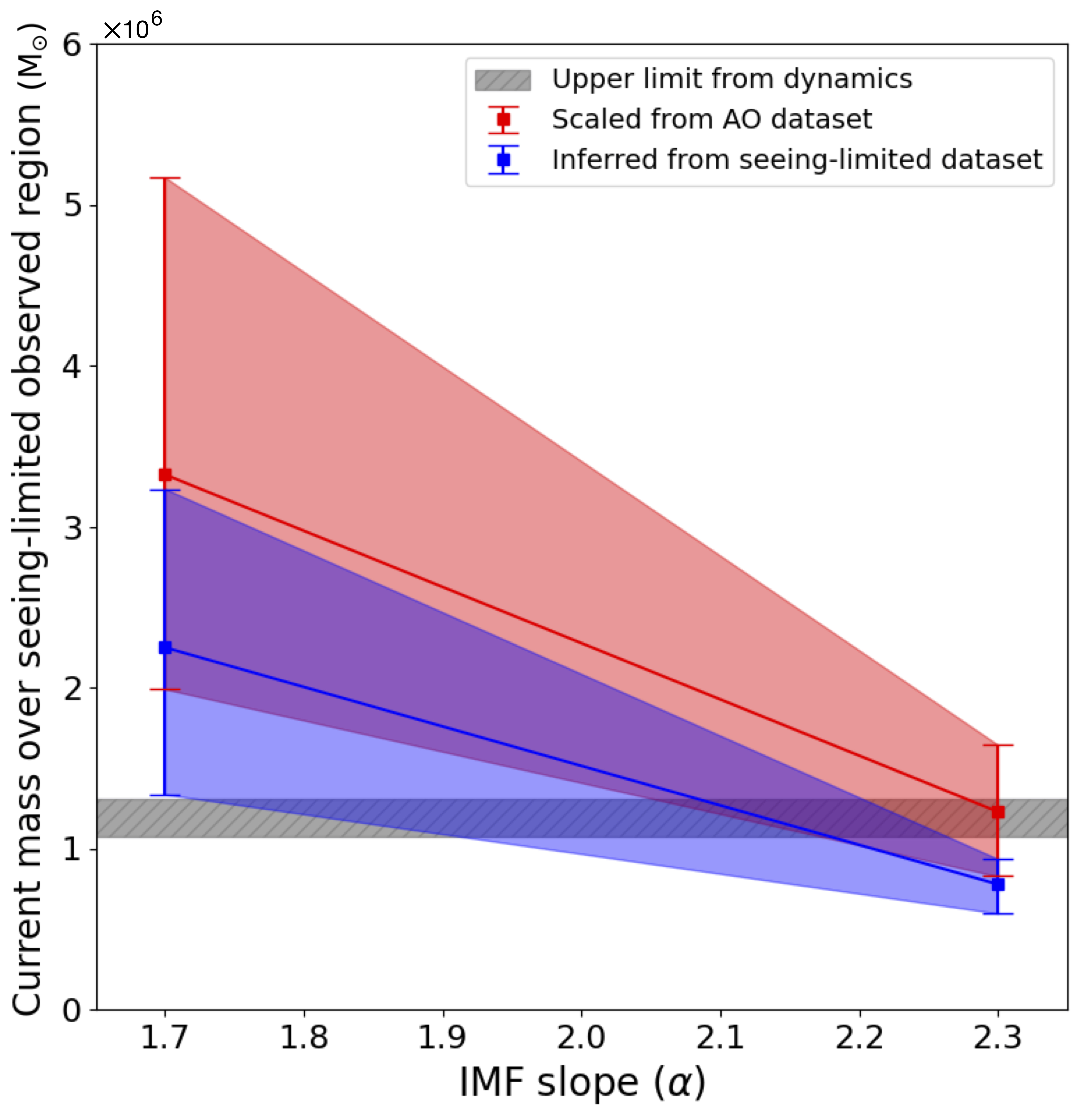}
\centering
\caption{Current cluster mass enclosed in the seeing-limited observed region predicted from our best-fit star formation history as a function of IMF slope. The blue errorbars with band show the enclosed mass inferred from the seeing-limited dataset with weighted median and 68\% (1$\sigma$ equivalent) confidence interval, while the red ones are scaled from the AO dataset. The grey band shows the upper limit of current mass enclosed in this region as estimated from dynamical measurements. The Kroupa IMF is slightly favored though the uncertainties are large to further constrain the cluster IMF. Regardless of which IMF is assumed, our model predictions are in agreement with the dynamical constraints within 2$\sigma$. \label{fig:currmass}}
\end{figure}

\subsection{Mass comparison with dynamical constraints}\label{sec:mass_imf}

For each dataset, the present-day cluster mass enclosed in the observed region is predicted from our best-fit star formation history under two IMF assumptions. The current mass within AO observed region under Kroupa or top-heavy IMF is estimated to be 4.8$\pm$1.6 $\times$ 10$^4$ $M_{\odot}$ or 1.3$\pm$0.6 $\times$ 10$^5$ $M_{\odot}$, respectively. The current mass within seeing-limited observed region is 7.8$\pm$1.7 $\times$ 10$^5$ $M_{\odot}$ or 2.2$\pm$1.0 $\times$ 10$^6$ $M_{\odot}$, respectively. For consistent comparison, the enclosed mass inferred from the AO dataset is scaled to the seeing-limited observed region by multiplying a scale factor of 25.4 (estimated from overlapped stars). The current enclosed mass as predicted from two datasets agree to within 1$\sigma$ regardless of which IMF is assumed. See Figure \ref{fig:currmass}. 

The predictions are then compared with the enclosed stellar mass profile M(r) as modeled from dynamical measurements \citep{Chatzopoulos et al. 2015}. The enclosed mass of the seeing-limited observed region is approximately equivalent to (or slightly lower than owing to asymmetric observed region) that within a spherical radius of r = 1.14 pc ($\sim$29.3 arcsec). We therefore estimate an upper limit from dynamical constraints of the current enclosed mass in this region of 1.2$\pm$0.1 $\times$ 10$^6$ $M_{\odot}$. See Figure \ref{fig:currmass}. 

The enclosed mass predicted inferred from our model is consistent with the dynamical mass estimates, which presents an independent check of the results. By comparing to the dynamical constraints, the Kroupa IMF is slightly favored. However, the uncertainties are too large to further constrain the IMF of the NSC since multiple assumptions have been made such as equivalent radial range, scale factor among observed regions and IMF low-mass cut-off adopted in the work (0.8 $M_{\odot}$, see Table \ref{tab:model_para}). Regardless of which IMF is assumed, the model predictions for the current cluster mass are in agreement with the dynamical measurements within 2$\sigma$.

\subsection{Predicted number of compact objects and their merger rates}\label{sec:compacto_re}

\subsubsection{Compact objects}
One important outcome of the star formation history of the NSC is that it allows us to predict the type and number of compact objects including stellar mass black holes (SBHs), neutron stars (NSs) and white dwarfs (WDs). We calculate the predicted number of compact objects via SPISEA with our derived star formation history, the first metallicity constraints on the NSC, realistic multiplicity properties \citep{Lu et al. 2013} and the metallicity-dependent initial-final mass relation (IFMR) implemented by \citet[the Spera15 IFMR object within SPISEA]{Rose et al. 2022}.

\begin{figure*}
\includegraphics[width=180mm]{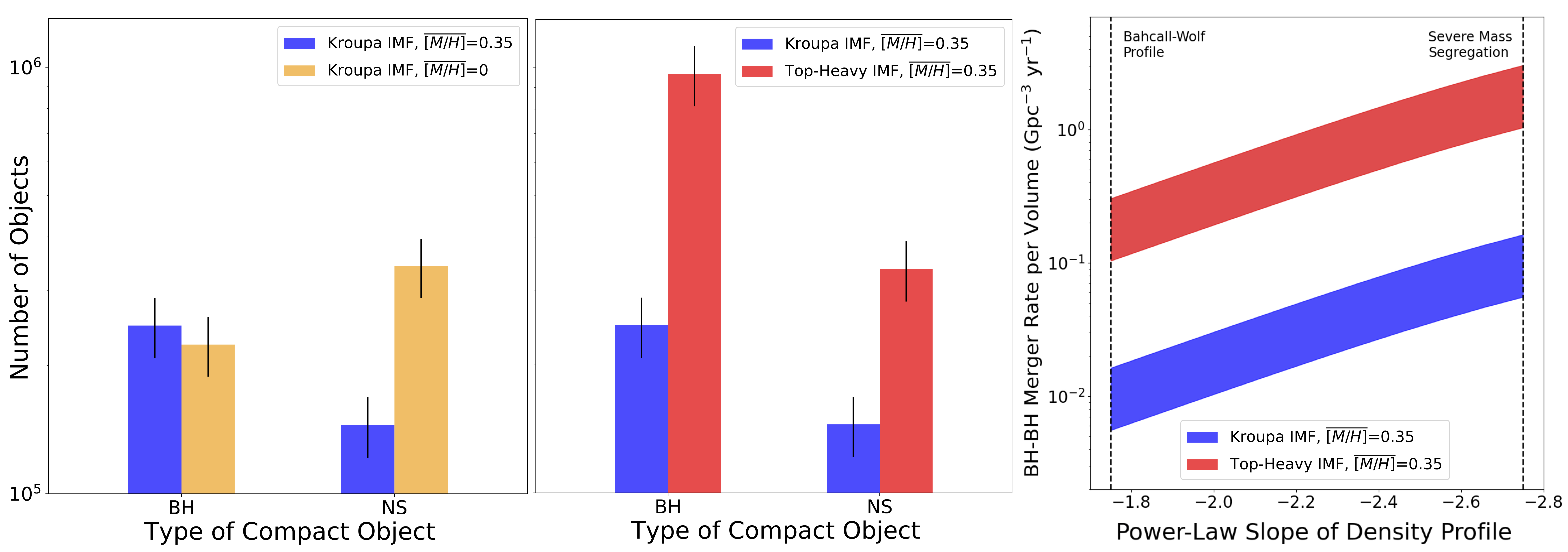}
\centering
\caption{IMF and metallicity are crucial properties for predicting the number of compact objects and their merger rates at the Galactic center. \textbf{Left panel}: The high metallicity of the main population of the NSC ($\overline{[M/H]}$ = 0.35, blue bar) predicts 2 - 4 times fewer neutron stars than those assuming a solar metallicity ($\overline{[M/H]}$ = 0, yellow bar). \textbf{Middle panel}: A cluster with a top-heavy IMF ($\alpha$ = -1.7 $\pm$ 0.2, red) produces 3 - 6 times more black holes and 2 - 3 times more neutron stars than a cluster with a Kroupa IMF. \textbf{Right panel}: Comparison of the predicted BH-BH merger rate per volume assuming different IMFs, as calculated from a range of possible 3-d radial density profile of BHs with a power-law indice range of 7/4 $<$ $\beta$ $<$ 11/4 \citep{Alexander et al. 2009} covering both severe and weak (Buhcall-Wolf profile) mass segregation scenarios. The width in the band corresponds to the assumption of ellipticity of the NSC from 0.1 to 1. The top-heavy IMF predicts the BH-BH mergers with a rate of up to 19 times higher than that with a Kroupa IMF. \label{fig:comp_merger}}
\end{figure*}

We predict 1.0 $\times$ 10$^4$ BHs, 6.0 $\times$ 10$^3$ NSs and 3.8 $\times$ 10$^5$ WDs for every 10$^6$ $M_{\odot}$ ($\sim$18\% uncertainty for each type), with a super-solar metallicity ($\overline{[M/H]}$ = 0.35) and a Kroupa IMF ($\alpha$ = -2.3 $\pm$ 0.36, m $>$ 0.8$M_{\odot}$). See Table \ref{tab:compact}. The fractional uncertainties were estimated by calculating the number of compact objects 500 times and drawing from uncertainties on the IMF and total cluster mass. Assuming the NSC with a total current cluster mass of 2.5 $\times$ 10$^7$ $M_{\odot}$ \citep{Schodel et al. 2014}, we then predict 2.5 $\times$ 10$^5$ BHs, 1.5 $\times$ 10$^5$ NSs and 8.7 $\times$ 10$^6$ WDs in the NSC. Of particular note, the predicted number of neutron stars in this work, when metallicity measurements are included, decreases by a factor of 2 - 4 (see Figure \ref{fig:comp_merger}) compared to earlier predictions, based on the assumption of solar metallicity. 

We also predict the number of compact objects under two IMF assumptions: Kroupa and a top-heavy IMF. The IMF profile of the NSC has a significant impact on the resulting compact remnants populations. For a given metallicity, a top-heavy IMF predicts a factor of 3 - 6 times more BHs and 2 - 3 times more NSs than a Kroupa IMF. See Table \ref{tab:compact} for the summary and Figure \ref{fig:comp_merger} for the comparison of predicted number of compact objects with different IMF profiles. 

If assuming a range of possible radial density profiles (see more details in section \ref{sec:mergerrate}), we estimate 0.5\% - 1.1\% of the total number of compact objects within the AO observed region, and 16\% - 62\% within the seeing-limited observed region. Here the mass segregation has not been included in the predictions, which may change the compact objects radial profiles significantly. We also bring up that the predictions are not including binary stellar evolution, binary dynamical evolution, nor the mergers.

\begin{deluxetable}{ccccc}
\tablenum{8}
\tablecaption{Predicted number of compact objects \label{tab:compact}}
\tablewidth{0pt}
\tablehead{
\colhead{Model} & \colhead{IMF} & \colhead{$\overline{[M/H]}$} & \colhead{$N_{BH}$} & \colhead{$N_{NS}$}}
\startdata
(1) & Kroupa & 0.35 & 1.0$\pm$0.2 $\times$10$^4$ & 0.6$\pm$0.1 $\times$10$^4$\\
(2) & Kroupa & 0 & 0.9$\pm$0.2 $\times$10$^4$ & 1.5$\pm$0.3 $\times$10$^4$ \\
(3) & Top-heavy & 0.35 & 4.1$\pm$0.7 $\times$10$^4$ & 1.4$\pm$0.3 $\times$10$^4$\\
(4) & Top-heavy & 0 & 3.4$\pm$0.6 $\times$10$^4$ & 2.8$\pm$0.5$\times$10$^4$\\
\enddata
\tablecomments{\\
Predictions for every 10$^6$ $M_{\odot}$. For each type of compact objects, we report a $\sim$18\% uncertainty on the predicted number.}
\end{deluxetable}

\subsubsection{BH-BH merger rate}\label{sec:mergerrate}

We calculate the predicted number of BH-BH mergers N$_{merge}$ that has occurred at the Galactic center,
\begin{equation}\label{equ:n_merger}
  \begin{aligned}
    N_{merge} = N_{binary} \cdot f_{stable} \cdot f_{merge}
  \end{aligned}
\end{equation}

\begin{itemize}
 \item {N$_{binary}$}: Number of massive binary star systems that will form BH-BH binaries at the end of stellar evolution. Calculated for a total cluster mass of 2.5 $\times$ 10$^7$ $M_{\odot}$, with the updated star formation history, IMF assumption, realistic multiplicity properties and the adopted IMFRs. 
 \item {f$_{stable}$}: Fraction of BH-BH binary systems that will produce stable binaries (2.5\% - 4.5\%; \citealt{Petrovich et al. 2017}) and not be torn apart by supernovae. 
 \item {f$_{merge}$}: Fraction of stable BH-BH binaries that will eventually merge within 1 Gyr (5.8\% - 17\%; \citealt{Petrovich et al. 2017, Hoang et al. 2018})
\end{itemize}

We predict 2.2 $\times$ 10$^4$ (Kroupa IMF) or 2.3 $\times$ 10$^5$ (top-heavy IMF) BH-BH binaries in the NSC, assuming a total cluster mass of 2.5 $\times$ 10$^7$ $M_{\odot}$.

Currently the fraction of BH-BH mergers modeled from dynamical simulations has only considered the central 0.4 pc (\citealt{Petrovich et al. 2017, Hoang et al. 2018}), where the majority of massive early-type stars are found. We thus scale the total predicted number of BH-BH mergers across the NSC down to only those within r = 0.4 pc, based on the 3-dimensional BH radial density profile. The BH number density distribution generally follows a power-law density cusp with n(r) $\propto$ r$^{-\beta}$ near the central SMBH, with the indice range of 7/4 $<$ $\beta$ $<$ 11/4 covering both severe and weak (Buhcall-Wolf profile) mass segregation scenarios \citep{Alexander et al. 2009}. With the radial density profile, we predict 0.1-1.0 $\times$ 10$^4$ (Kroupa IMF) or 1.0-9.9 $\times$ 10$^4$ (top-heavy IMF) BH-BH binaries in the central 0.4 pc. By applying to the factors of f$_{stable}$ and f$_{merge}$ in equation \ref{equ:n_merger}, we predict a BH-BH merger rate in the range of 0.01-0.16 Gpc$^{-3}$yr$^{-1}$ (Kroupa IMF) or 0.10-3.03 Gpc$^{-3}$yr$^{-1}$ (top-heavy IMF). The rate per volume assumes a number density of galaxies of $\sim$0.02 Mpc$^{-3}$ (e.g., \citealt{Conselice et al. 2005, Kopparapu et al. 2008}). See Figure \ref{fig:comp_merger} (right) for the comparison of the BH-BH merger rates calculated from two IMF assumptions with a range of possible radial density profiles and different ellipticities (from 0.1 to 1) for the NSC. A top-heavy IMF predicts the BH-BH mergers with a rate up to $\sim$19 times higher than that with a Kroupa IMF.

\section{Discussion}\label{sec:discussion}

\subsection{Comparison with previous work}
Previously the star formation history measurements have assumed a solar or a slightly super-solar metallicity and found the NSC to be 5 - 12 Gyr old. \citet{Blum et al. 2003} measured the star formation history from spectroscopy of the most luminous AGB stars in the inner 5 pc, and reported that $\sim$75\% of stars formed more than 5 Gyr ago. \citet{Maness et al. 2007} reported AO spectroscopy of late-type stars in the central 1 pc and favored continuous star formation over the last 12 Gyr with a top-heavy IMF. \citet{Pfuhl et al. 2011} presented AO spectroscopy for late-type stars in the central 1 pc, and reported that $\sim$80\% of the stellar mass formed more than 5 Gyr ago. Due to the limited metallicity measurements, these spectroscopic studies all assumed a solar metallicity for all stars in the NSC, which would bring large bias on the age estimates as a result of age-metallicity degeneracy. \citet{Schodel et al. 2020} presented the star formation history study based on only photometry. They established the K luminosity function for a large sample of stars (down to K$\sim$19 mag) and reported that the age of the bulk of the stellar mass could range from 4 to 12 Gyr depending on metallicity assumptions.

In this work, we include metallicity measurements for the first time in modeling the star formation history of the Milky Way NSC. When metallicity is included as a free parameter, we find that the main population of the NSC is metal-rich and likely younger (5.0 $^{+3.4}_{-2.3}$ Gyr). Including metallicity systematically results in a younger age than previous studies (5 - 12 Gyr), there is some overlap in the certainties with previously reported ages. When metallicity is fixed to be solar, the median age increases by $\sim$3 Gyr. This assumption results an age of 8.3 $^{+3.7}_{-3.9}$ Gyr for $\sim$90\% of the stellar mass, which is in agreement with previous studies with solar metallicity assumption. This bias to higher ages is due to the fact that high metallicity stars tend to be cooler and less luminous. It is therefore important to include metallicity constraints in the star formation history of the NSC.

\citet{Schodel et al. 2020} fit for the fraction of stars formed in 17 age bins ranging from 0.03 to 13 Gyr. While this work did not include metallicity measurements, they did explore different metallicity assumptions. For the highest metallicity assumption of 2 times solar, they find that 30\% of stars formed at 4 Gyr and 50\% of stars formed at 13 Gyr. In this work, we find a young population with an age of $\sim$5 Gyr and a super-solar metallicity, however, find no apparent evidence showing that there is a population older than 10 Gyr. This younger population is consistent with the 4 Gyr population in \citet{Schodel et al. 2020}, but we do not find the older 13 Gyr population as they reported. The differences between the two analyses may be from a number of different factors. \citet{Schodel et al. 2020} used only Ks photometry while our work uses spectroscopically measured temperatures and H and Ks photometry. \citet{Schodel et al. 2020} Ks photometry is deeper (down to K$\sim$19 mag) but properties such as temperature and metallicity require spectroscopy to accurately measure. The two analyses also use different fit parameters. Our work considers additional model variables including the total cluster mass, distance to the cluster, IMF slope and differential extinction. Future deeper spectroscopic observations will help to clarify whether the ancient 13 Gyr burst exists in the NSC.

In this work, we test different star formation models and find that a single burst of star formation can explain the origin of 90\% of the stars. Previous studies (e.g., \citealt{Blum et al. 2003, Maness et al. 2007, Pfuhl et al. 2011, Schodel et al. 2020}) modeled star formation in this region with a fixed number of age bins and fitting the star formation rate in each bin. The inferred star formation rates range from 0.5$\times$10$^{-4}$ to 8$\times$10$^{-4}$ $M_\odot$/yr, but those studies did not do a model comparison to assess whether a single burst can fit most of the data. This is important as star formation in a single burst would imply a star formation rate that could be much higher for a short time. Formation of $\sim10^7$ $M_\odot$ in stars would suggest an extraordinary starburst at the Galactic center during the formation of the NSC. Future work with more stars will be able to test our conclusions with more complex star formation history models.

\subsection{Implication for the co-evolution of the NSC, the SMBH, the NSD and the bulge}

Surveys of galaxies similar to the Milky Way have shown that their galactic nuclei are often occupied by a massive object of either a NSC, a SMBH, or both \citep{Neumayer et al. 2020}. Their NSC mass generally scales with the mass of the bulge ($M_{bulge}$) and the total stellar mass of the host-galaxy ($M_{galaxy}$). The scaling relations between the $M_{galaxy}$ and the mass of the central massive object indicate that the SMBH, the NSC and the bulge are undergoing mutual evolution and linked by similar physical mechanisms (e.g., \citealt{Ferrarese et al. 2006, Georgiev et al. 2016}). Moreover, a NSC and a NSD can co-exist in one galactic nuclei, while they might also exist independently (e.g., \citealt{Bittner et al. 2020}). It is not clear yet how they influence each other. Specifically, the Milky Way galaxy is the best-studied example for the coexistence of NSC and SMBH (e.g. \citealt{Ghez et al. 2008, Gillessen et al. 2009, Schodel et al. 2014, Feldmeier-Krause et al. 2014}) in galaxies with the $M_{galaxy}$ $\sim$ $10^{10} M_{\odot}$, which represents the transition region between the high-mass galaxies with SMBH dominated and the low-mass galaxies with NSC dominated (e.g., \citealt{Graham Spitler 2009, Neumayer Walcher 2012}). The Milky Way galaxy can also examine if the formation and evolution of the NSC and the NSD are connected.

While the existence of a scaling relation between NSC mass and galaxy properties suggest co-evolution of galactic nuclei and their inner bulge regions, our measurement of a younger age of the NSC calls this into question for the Milky Way. We find that roughly 90\% of the stellar mass of the NSC formed 5.0 $^{+3.4}_{-2.3}$ Gyr ago. In comparison, the bulk of the metal-rich stellar population of the Galactic bulge has an age of 10 $\pm$ 2.5 Gyr \citep{Zoccali et al. 2003}. SMBHs are believed to build in at early times and have existed in the galaxy more than 12 Gyr ago (e.g., \citealt{Fan et al. 2001b, Volonteri 2010}). The younger age of the NSC suggests that the NSC, the SMBH and the bulge might not be co-eval. If the NSC and SMBH in the Milky Way is not just two types of a single central massive object, it means that there are likely different physical processes that regulate their growth and evolution.

Furthermore, the decreasing stellar metallicity outwards from the NSC to the NSD suggests that these two components are likely connected via gas inflow from the NSD to the NSC \citep{Feldmeier-Krause 2022}. \citet{Nogueras-Lara et al. 2020} reported that the bulk (over 90\%) of the stars in the Milky Way NSD formed $\gtrsim$8 Gyr ago. Our reported age of the NSC (5.0 $^{+3.4}_{-2.3}$ Gyr) is slightly younger than the reported age of the NSD, while they could still be consistent at $\sim$8 Gyr where the ages overlap. The NSD is likely a consequence of gas funneled towards the Galactic center via the bar/bulge from a few tens of pc to a kpc in radius (e.g., \citealt{Comeron et al. 2010, Sormani Barnes 2019, Bittner et al. 2020}). The nuclear gas inflow, instead, dominates the inner region until a few pc where the bar/bulge inflow becomes inefficient (e.g., \citealt{Tress et al. 2020}). The gas could be triggered towards the inner pc by magnetic fields, a nuclear bar, supernova and stellar feedbacks or external pertubers, and then form stars in-situ. The star formation and following supernova feedback in the NSD might not be the only contribution to trigger the star formation in the NSC. Future work can use the method reported in this work to further constrain the star formation of the NSD with metallicity measurements for comparison.

\subsection{Implication for the formation of the NSC}

A metallicity and age for the Milky Way NSC may offer constraints on its formation mechanisms. The formation of the NSC is still poorly understood, but two main scenarios of the formation process have been proposed. One is in-situ scenario \citep{Milosavljevic 2004}, where gas falls onto the center of the galaxy and then triggers star formation within the cluster or the accretion of star clusters formed in the vicinity. The other is migration scenario \citep{Tremaine et al. 1975}, where globular clusters that formed elsewhere migrate towards the central region through dynamical friction mechanism, and then fall in and merge with each other (\citealt{Andersen et al. 2008, Antonini 2013}). Both scenarios could also operate at the same time. The two formation scenarios imprint specific observable signatures on the ages and metallicities of the stellar population of NSC.
If the in-falling globular clusters were the main contributions to the stars in the NSC, we would expect a large fraction of mass with a very old age and a sub-solar metallicity which are comparable to typical globular clusters. Most globular clusters in the Milky Way (more than 95\%) have a low metallicity with $\overline{[M/H]}$ $<$ -0.3 \citep{Harris 2010}, and an age older than 11.2 Gyr \citep{Krauss Chaboyer 2003}. This work showing a younger age ($\sim$5 Gyr old) and a higher metallicity($\overline{[M/H]}$ $\sim$ 0.35) for the bulk stellar mass of the NSC, is inconsistent with the globular clusters in-falling scenario as a dominant mechanism for the main population of the NSC.

The high-metallicity and relatively young age ($\sim 5$ Gyr) suggests that the bulk of the NSC formed in-situ. Chemical evolution models suggest that the chemical enrichment of the Galactic center can occur very rapidly at time scales of 0.1 - 0.7 Gyr \citep{Grieco et al. 2015}. The higher metallicity of the NSC also follows the trend in the the Galactic inner disk or the Galactic bulge, where the stellar metallicities are generally higher towards the Galactic center (\citealt{Trevisan et al. 2011, Bensby et al. 2013, Feltzing Chiba 2013, Garcia-Perez et al. 2018, Nogueras-Lara et al. 2018b, Schultheis et al. 2021}).

While the bulk of the cluster may have formed in-situ, about 10\% of the stars have metallicity at half-solar or less, which is consistent with an infall of a globular cluster or dwarf galaxy. The presence of the RR Lyrae stars in the NSC suggests that the old metal-poor population could contribute up to 18\% of the total mass of the NSC by globular cluster infall \citep{Dong et al. 2017}. The spatial anisotropy of the sub-solar metallicity stars may indicate a recent star cluster infall event \citep{Feldmeier-Krause et al. 2020}. These lower metallicity stars also appear to have different kinematic signatures than the super-solar metallicity stars, which is the further evidence that the two groups of stars may have different origins  \citep{Do et al. 2020}. Furthermore, the alpha elemental abundances of the low-metallicity population are also consistent with an infalling cluster or dwarf galaxy \citep{Bentley et al. 2022}. Age constraints on the low metallicity stars could help to differentiate between the formation scenarios. Simulations from \citet{Arca Sedda et al. 2020} of an infall of a star cluster in a galactic nucleus using N-body simulations suggest that the infall of a massive star cluster should occur in $\sim$0.1 - 3 Gyr ago to remain the current distinguishable kinematic features as observed. However, our model has poor age constraints on the lower metallicity stars due to their small sample size. Additional age constraints will be important to assess whether this population is consistent with the results of these simulations.

\subsection{Implications of the predicted number of compact objects and their merger rates}

\subsubsection{The ``missing-pulsar problem"}

\begin{figure}[ht!]
\includegraphics[width=86mm]{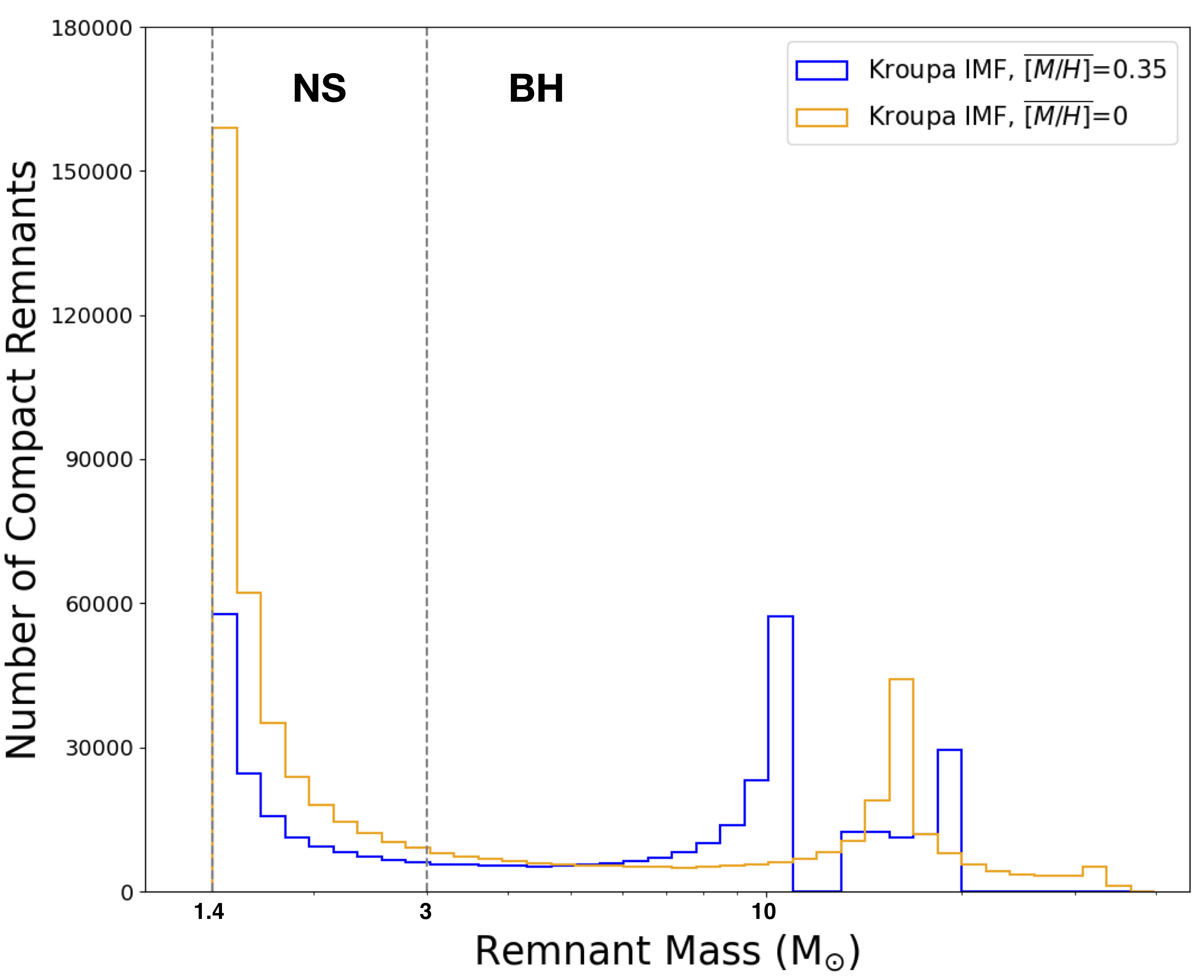}
\centering
\caption{Predicted number of compact remnants as a function of remnant mass with a super-solar metallicity reported in this work ($\overline{[M/H]}$ = 0.35, blue) or a solar metallicity ($\overline{[M/H]}$ = 0, yellow). Dashed lines show the thresholds to differentiate WD and NS (M$_{rem}$ = 1.4 M$_{\odot}$), NS and BH (M$_{rem}$ = 3 M$_{\odot}$). Metallicity impacts both the mass loss by stellar wind and the supernova explosion process, and thus the remnant mass. A high metallicity ($\overline{[M/H]}$ = 0.35) predicts 2 - 4 times fewer neutron stars as a result of smaller remnant masses. Neutron stars occupy in a small range of masses (1.4 - 3 $M_{\odot}$), and thus are more sensitive to fraction changes in the remnant masses than stellar mass black holes. \label{fig:hist_rem}}
\end{figure}

This work predicts 2 - 4 times fewer neutron stars with a super-solar metallicity ($\overline{[M/H]}$ $\sim$ 0.35), compared to earlier predictions assuming a solar metallicity. Two major effects result a smaller number of predicted number of neutron stars at the Galactic center. With a higher metallicity, an increased mass loss by the stellar wind on the main-sequence is expected (e.g., \citealt{Kudritzki et al. 1987, Leitherer et al. 1992, Vink et al. 2001}). Metallicity also impacts the supernova explosion process which determines the remnant mass of the progenitor (e.g., \citealt{Fryer et al. 2012}). With a higher metallicity, higher supernova progenitor masses are necessary to produce neutron stars (e.g., \citealt{Poelarends et al. 2008}). Both of these factors lead to smaller remnant mass, and thus more white dwarfs compared to neutron stars. In addition, neutron stars occupy in a small range of masses (1.4 - 3 $M_{\odot}$), and thus are more sensitive to fraction changes in the remnant masses than stellar mass black holes, which occupy a greater range in masses (Figure \ref{fig:hist_rem}). 

If the Galactic center has fewer neutron stars than expected, then this may help us understand the ``missing-pulsar problem". The astronomical community has surveyed for decades at the Galactic center without detecting a population of pulsars as expected (e.g., \citealt{Johnston et al. 1995, Bates et al. 2011, Torne et al. 2021}). Here we show that the number of pulsars we expect depends on stellar metallicity, which should be considered when evaluating how many pulsars are ``missing" at the Galactic center. 

\subsubsection{Gravitational-wave merger rate}
Since 2016, the Advanced Laser Interferometer Gravitation-Wave Observatory (LIGO) and VIRGO have enabled direct detections of gravitational waves from in-spiraling compact object binaries (\citealt{2016a,2016b,2017,Abbott et al. 2016a,Abbott et al. 2016b,Abbott et al. 2016c, Abbott et al. 2017a,Abbott et al. 2017b}). The location and rate of these gravitational wave sources are important for understanding their nature. Dense star clusters such as NSCs at the center of galaxies are thought to be the major source of these mergers since these regions are expected to be abundant in SBHs and BH-BH binaries with higher merger rates (\citealt{Antonini et al. 2010, Portegies Zwart McMillan 2000, Wen 2003, OLeary et al. 2006, OLeary et al. 2009, OLeary et al. 2016, Kocsis Levin 2012, Antonini Perets 2012, Antonini et al. 2014, Rodriguez et al. 2016b, VanLandingham et al. 2016, Bartos et al. 2017b, Stone et al. 2017, Hoang et al. 2018}). 

The center of the Milky Way offers us the ideal prototype for constraining the compact object population of galactic nuclei and gravitational-wave merger rates. Previous studies reported a BH-BH merger rate in the proximity of galactic nuclei with a range of 0.6-15 Gpc$^{-3}$yr$^{-1}$ \citep{Petrovich et al. 2017} or 1-3 Gpc$^{-3}$yr$^{-1}$ \citep{Hoang et al. 2018}. In this work, with our updated star formation history of the NSC, we predict a BH-BH merger rate in the range of 0.01-0.16 Gpc$^{-3}$yr$^{-1}$ (Kroupa IMF) or 0.10-3.03 Gpc$^{-3}$yr$^{-1}$ (top-heavy IMF). We find that the predicted number of black hole mergers are most sensitive to the IMF, IMFR, and the density profile. The number of black holes are not very sensitive to metallicity, so these values are consistent with previous literature predictions.

\subsubsection{Improving compact object predictions}

The most important factors on the predicted number of compact objects and their merger rates are: the IMF, IFMR, and the compact object density profile. Currently the observations are not deep enough to constrain the cluster IMF simultaneously with the star formation history. Additional observations of stars with metallicity measurements will allow us to fit for the IMF. In addition, the IFMR prescription also affects the number of compact objects predicted for a given star formation history and is a function of stellar properties (e.g., \citealt{Heger et al. 2003, Sukhbold et al. 2018}). Better calibrations of the IFMR will lead directly to more accurate predictions at the Galactic center. Finally, the density profile of compact objects is largely unknown due to the difficulty of observing these objects close to the supermassive black hole. X-ray observations of accreting stellar mass black holes suggest a 3-d radial density profile with a power-law index in a range between 2.1 and 2.7 (\citealt{Hailey et al. 2018,Mori et al. 2021}). Stellar measurements disagree on the density profile, which range from core-like (\citealt{Do et al. 2009}) to cups-like profiles (\citealt{Schodel et al. 2020}). We present here a range of predictions based on different density profiles, but better constraints on this `dark cusp' will help to narrow the range of predictions.

\section{Conclusion}\label{sec:conclusion}

We model the star formation history of the Milky Way NSC, incorporating constraints on the metallicity for the first time from a large sample of stellar metallicity measurements. We use spectroscopy and photometry of 770 late-type giants along with a Bayesian inference methodology to derive the star formation history and global properties of the cluster. We test different star formation models (continuous, single-burst, multiple bursts) and find that a two-bursts star formation model is strongly favored. The bulk of the stars (93\% $\pm$ 3\%) is metal-rich ($\overline{[M/H]}$ = 0.45 $\pm$ 0.05) with an age of 5.0 $^{+3.4}_{-2.3}$ Gyr. The minor group with 7\% $\pm$ 3\% of stellar mass is metal-poor ($\overline{[M/H]}$ = -1.10 $\pm$ $^{+0.30}_{-0.25}$) with age being uncertain (0.1 - 5 Gyr old). By including metallicity as a free parameter, our models are able to account for low-temperature red giants that were previously difficult to fit. The bulk of the stars in the NSC is likely younger than previously reported. We find that the age of the stars is systematically younger by $\sim$3 Gyr when metallicity is included compared to assuming all stars are solar metallicity. This younger age for the NSC could challenge the mutual evolution scenario of the NSC, the central SMBH and the inner bulge. The younger age and the supersolar metallicity for the bulk stellar mass may also challenge the globular clusters in-falling scenario for the main population of the NSC.

This work also updates the predictions of the number of compact objects at the Galactic center and the rate at which they merge using our updated star formation history models. We predict 2.5 $\times$ 10$^5$ BHs, 1.5 $\times$ 10$^5$ NSs and 8.7 $\times$ 10$^6$ WDs in the NSC assuming a total cluster mass of 2.5 $\times$ 10$^7$ $M_{\odot}$. Specifically, when metallicity constraints are included, we predict 2 - 4 times fewer neutron stars compared to earlier predictions, which may introduce to a new path to further understand the so-called ``missing pulsar problem" at the Galactic center. We also predict 2.2 $\times$ 10$^4$ (Kroupa IMF) or 2.3 $\times$ 10$^5$ (top-heavy IMF) BH-BH binaries in the NSC, and a BH-BH merger rate ranging from 0.01-3 Gpc$^{-3}$yr$^{-1}$ depending on the IMF and density profile.

Future deeper spectroscopic observations and larger spatial coverage of the NSC would be crucial to extend our understanding of the star formation history of the NSC. In particular, the detection of a main-sequence turnoff with spectroscopy reaching K $\sim$ 19 mag (predicted for a 5 Gyr population) will greatly improve the age estimate of the NSC. Moreover, higher spatial and spectral resolution observations are required to place tighter constraints on the population of subsolar metallicity stars and their origins. 

\hspace{5mm}

The authors would like to thank Smadar Naoz, Kelly Kosmo O'Neil, Bao-Minh Hoang, and other members of the UCLA Galactic Center Group for providing helpful comments and discussions. The primary support for this work was provided by NSF AAG grant NSF AAG AST-1909554. Additional support was received from the UCLA Galactic Center Star Society and the Brinson Prize Fellowship (held by M.W.H.). This research uses the Galactic Center Orbit Initiative (GCOI) catalogue based on the data obtained from W. M. Keck Observatory. The W. M. Keck Observatory is operated as a scientific partnership among the California Institute of Technology, the University of California, and the National Aeronautics and Space Administration. The authors wish to recognize that the summit of Maunakea has always held a very significant cultural role for the indigenous Hawaiian community. We are most fortunate to have the opportunity to observe from this mountain. The Observatory was made possible by the generous financial support of the W. M. Keck Foundation.  

\bibliographystyle{aasjournal}

\pagebreak

\pagebreak

\pagebreak

\pagebreak

\appendix \label{sec:appendix}
\section{Fitter testings}\label{sec:fitter_testings}

\begin{figure}[ht!]
\includegraphics[width=86mm]{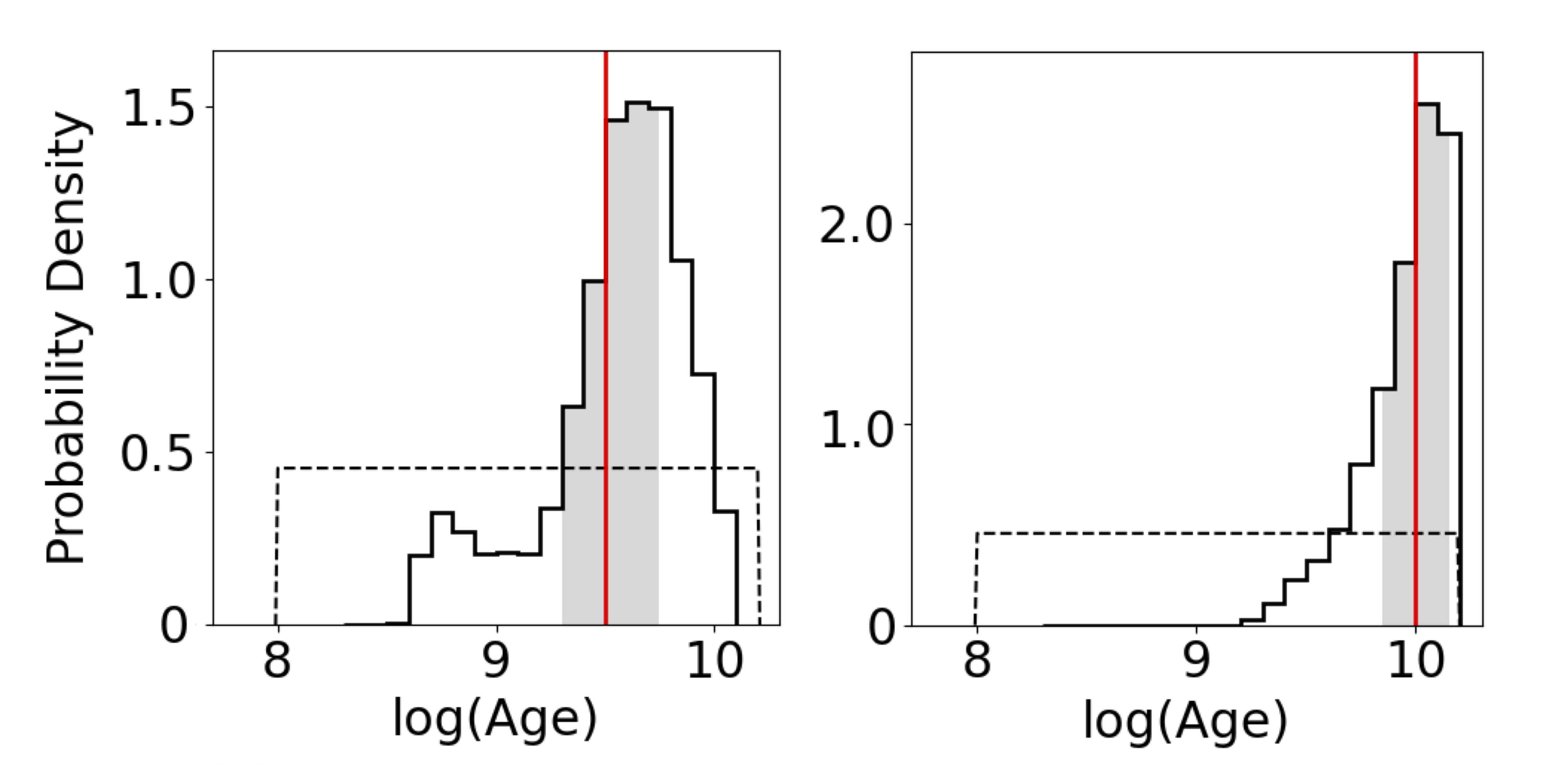}
\centering
\caption{Test cluster's marginalized 1D posterior probability density function for a simulated cluster with an age of 3.2 Gyr (log(Age) = 9.5, left), and an age of 10 Gyr (log(Age) = 10, right). A uniform prior probability distribution is used in the fits (black dashed line). The input age (red line) always falls within the 68\% (1$\sigma$ equivalent) Bayesian confidence interval (grey shaded region). Our Bayesian inference methodology is able to recover both moderate and old cluster ages with no substantial systematic biases. \label{fig:age_comp}}
\end{figure}

\begin{figure*}[ht!]
\includegraphics[width=180mm]{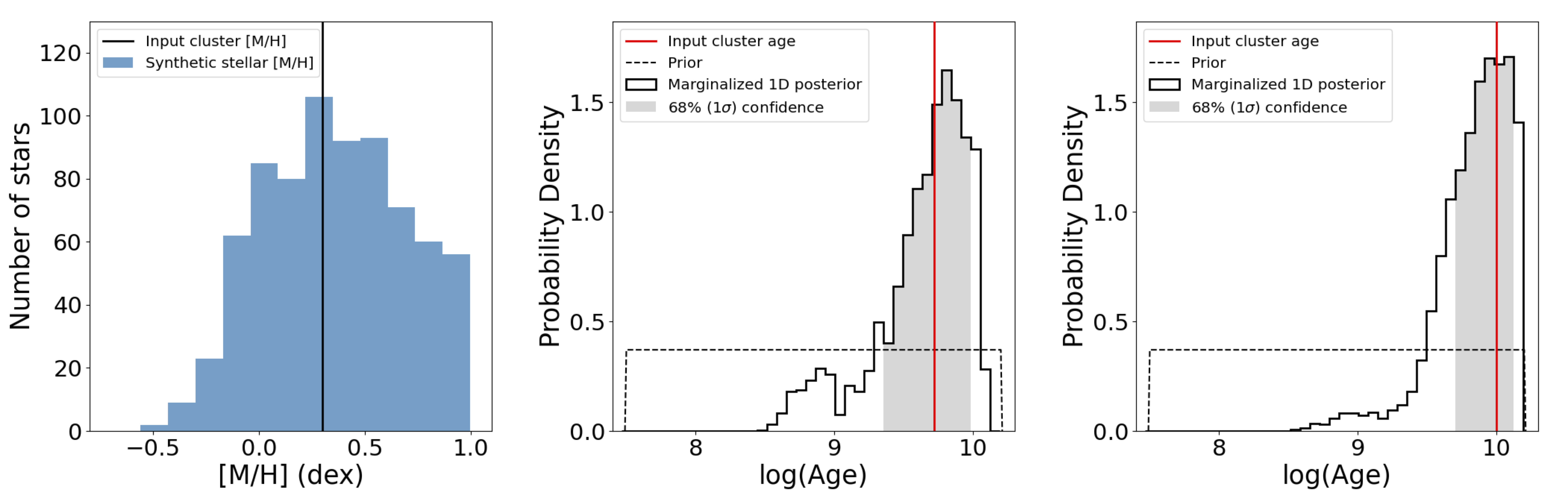}
\centering
\caption{\textbf{Left:} Stellar metallicity distribution from a synthetic cluster. The input cluster metallicity is shown as a vertical line. Uncertainty is added for each star as a Gaussian distribution. A random bias to $[M/H]$ is artificially introduced to simulate an apparent tail in the measured metallicity distribution out to $[M/H]$ $\sim$ +1 dex. \textbf{Middle:} Test-cluster's marginalized 1D posterior probability density function for a simulated cluster with an age of 5 Gyr (log(Age) = 9.7). We find that the best fit cluster age of 5.7 $^{+3.8}_{-3.5}$ Gyr is consistent with the input age (red line). \textbf{Right:} Example for a simulated cluster with an age of 10 Gyr (log(Age) = 10). The best fit cluster age of 8.9 $^{+4.0}_{-3.6}$ Gyr is consistent with the input age (red line). Our methodology is able to recover the cluster age with no substantial systematic biases.
\label{fig:mhfittertest}}
\end{figure*}

\begin{figure*}[ht!]
\includegraphics[width=160mm]{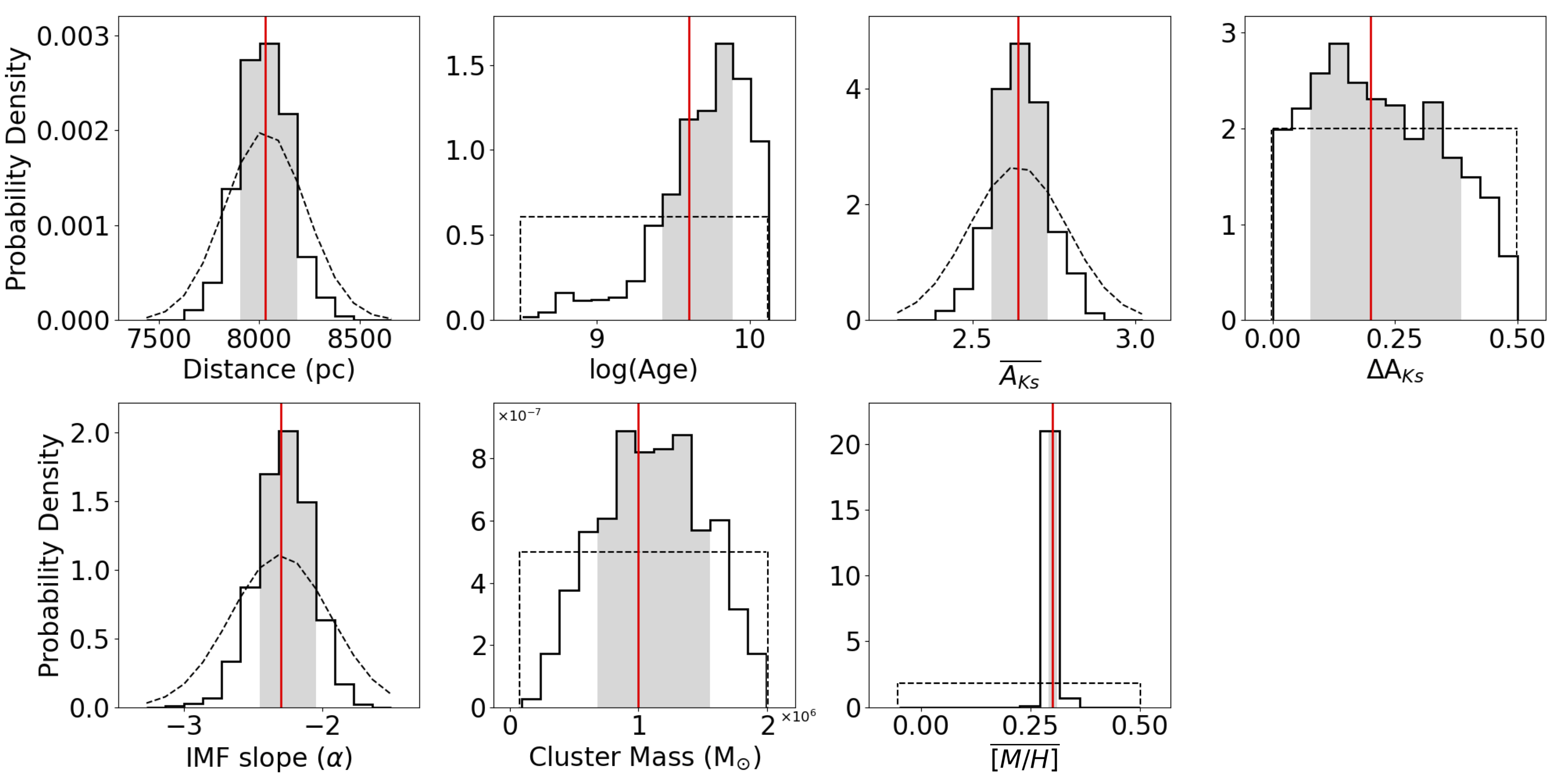}
\centering
\caption{Single age test-cluster's marginalized 1D posterior probability density functions for each fitting parameter. The input values for the single-age cluster's distance, age, average extinction ($A_{Ks}$), differential extinction ($\Delta A_{Ks}$), IMF slope, initial cluster mass, and metallicity are shown as a vertical red line. Each parameter falls well within the 68\% (1$\sigma$ equivalent) confidence interval of the distribution (grey shaded regions). The confidence intervals are calculated by first finding the 50$^{th}$ percentile of the posterior in probability distribution and then stepping away from the center until the integrated probability reached 68\% (1$\sigma$ equivalent). \label{fig:test_single}}
\end{figure*}

\begin{figure*}[ht!]
\includegraphics[width=180mm]{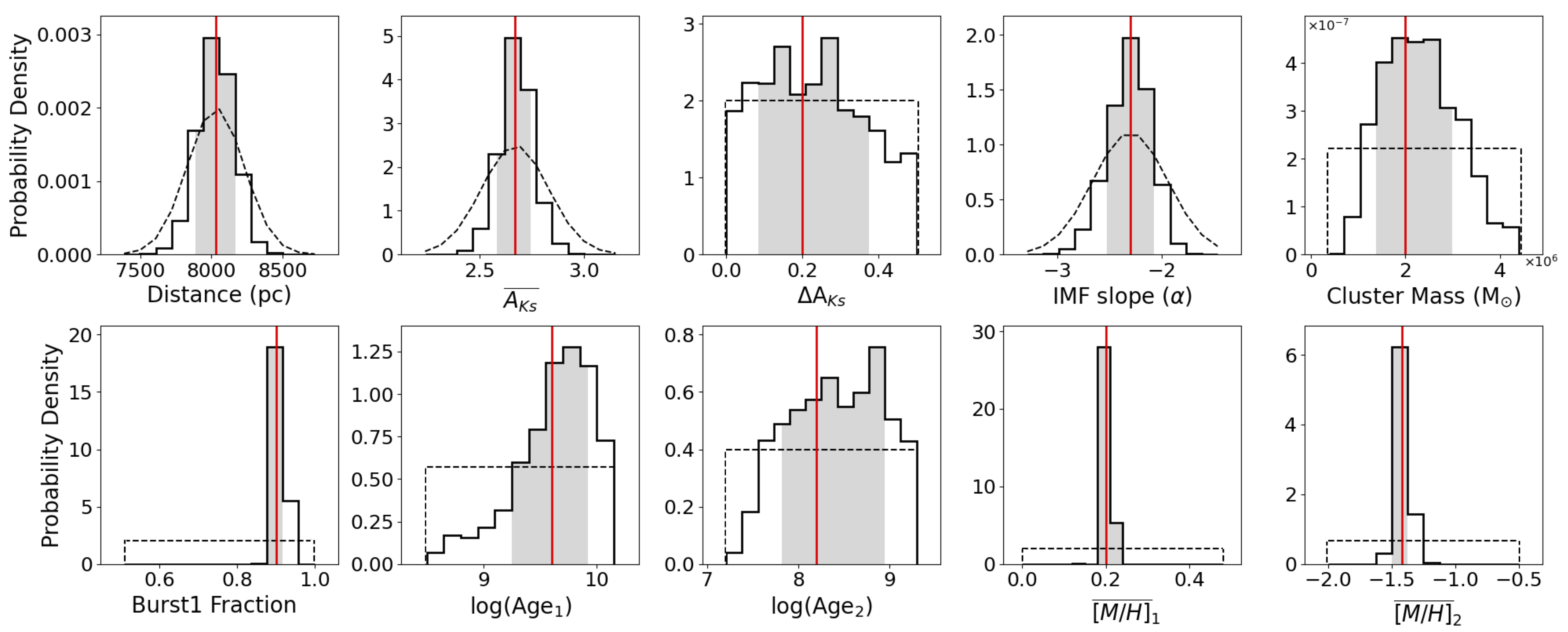}
\centering
\caption{Two-bursts star formation history model testing. The input values for the cluster including two bursts of star formation are: distance to the cluster, average extinction ($A_{Ks}$), differential extinction ($\Delta A_{Ks}$), IMF slope, total cluster mass, mass fraction of burst 1, age of burst 1, age of burst 2, metallicity of burst 1, and metallicity of burst 2 (see vertical red lines). Each parameter falls well within the 68\% (1$\sigma$ equivalent) confidence interval of the distribution (grey shaded regions). \label{fig:2pop}}
\end{figure*}

\begin{figure*}[ht!]
\includegraphics[width=180mm]{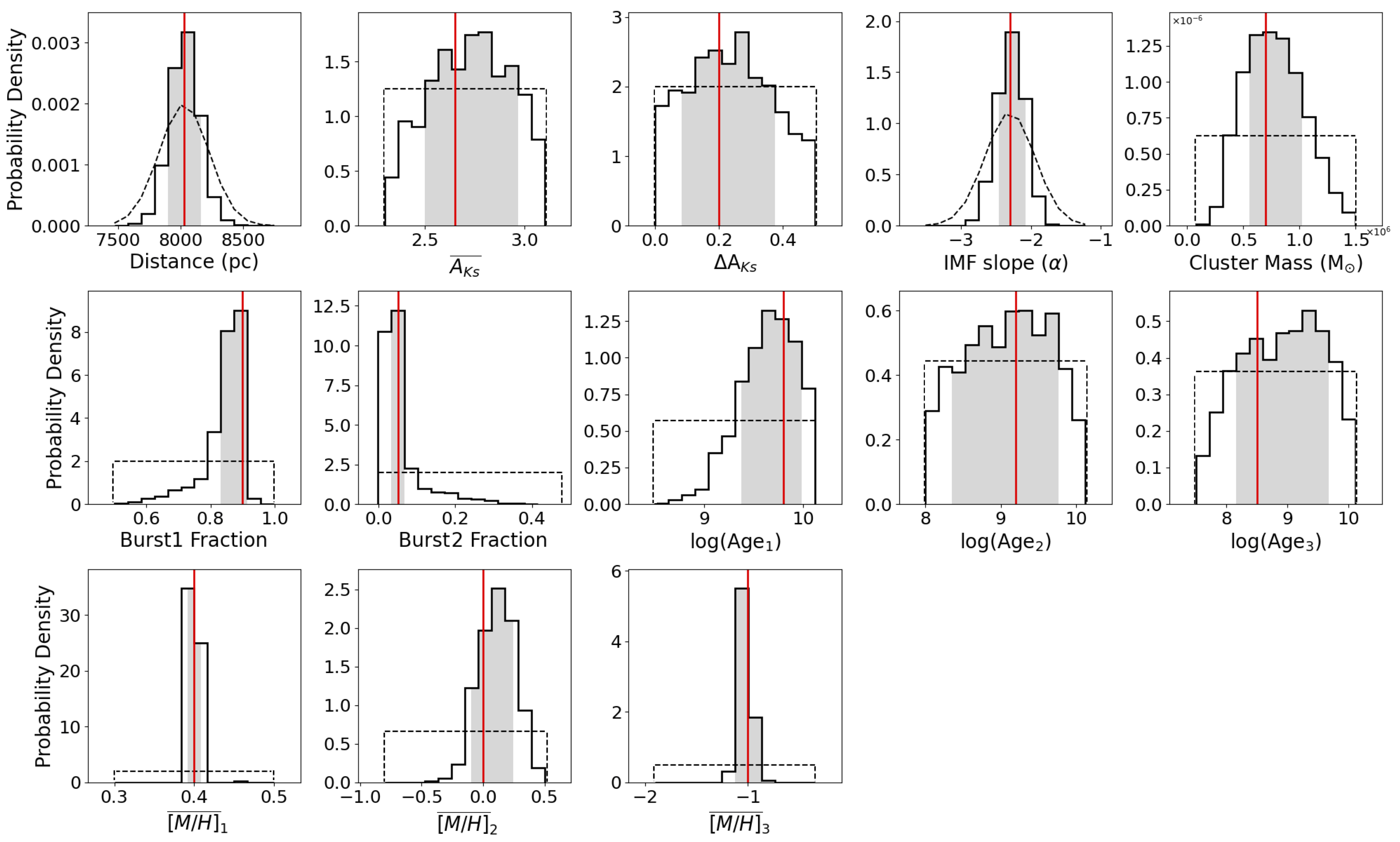}
\centering
\caption{Three-bursts star formation history model testing. The input values for the cluster including three bursts of star formation are: distance to the cluster, average extinction ($A_{Ks}$), differential extinction ($\Delta A_{Ks}$), IMF slope, total cluster mass, mass fraction of burst 1, mass fraction of burst 2, age of burst 1, age of burst 2, age of burst 3, metallicity of burst 1, metallicity of burst 2, and metallicity of burst 3 (see vertical red lines). Each parameter falls well within the 68\% (1$\sigma$ equivalent) confidence interval of the distribution (grey shaded regions). \label{fig:3pop}}
\end{figure*}

\begin{figure*}[ht!]
\includegraphics[width=180mm]{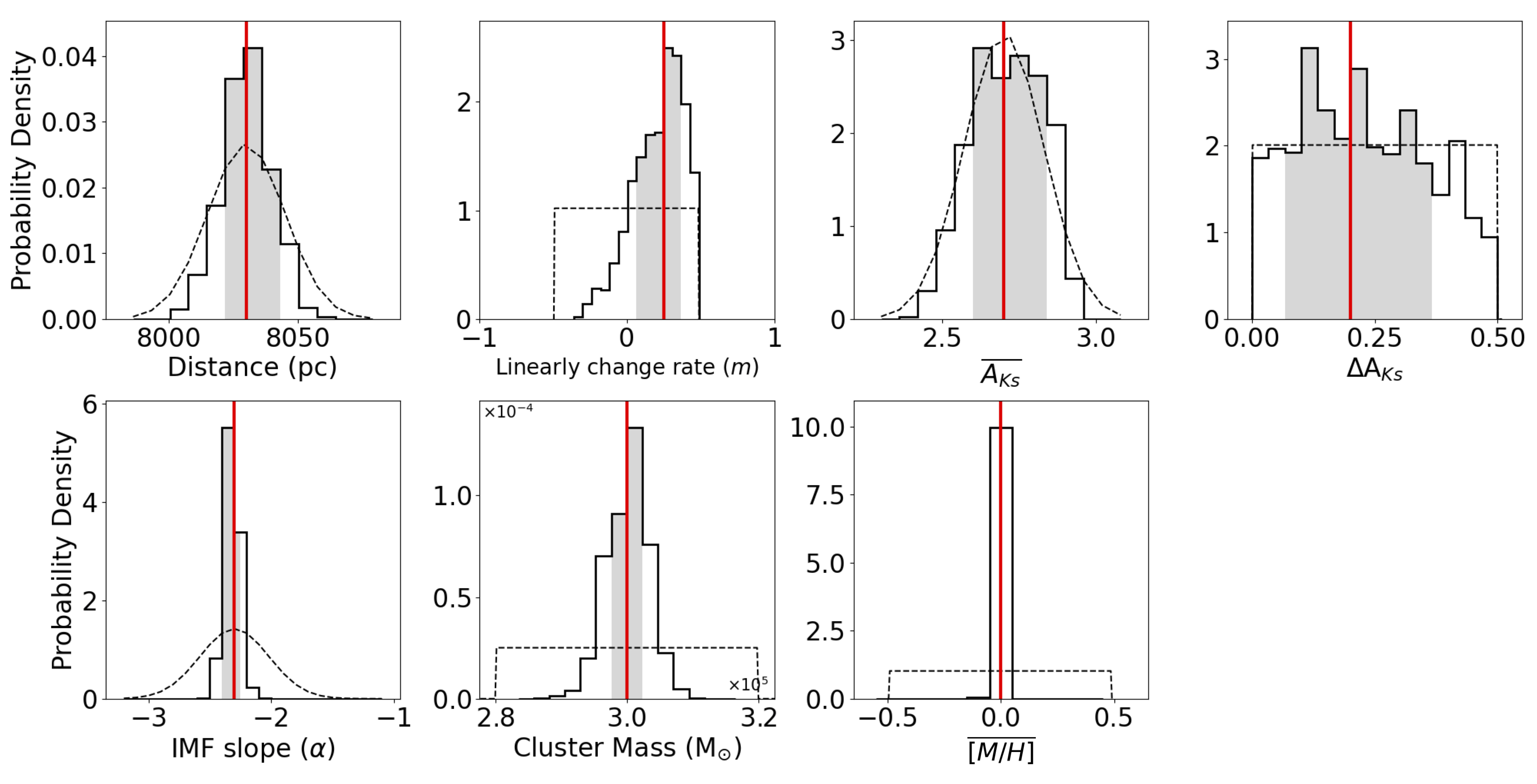}
\centering
\caption{Linear SFR model testing. The input values for the continuous star formation with a linearly increasing/decreasing SFR are: distance to the cluster, linearly change rate ($m$), average extinction ($A_{Ks}$), differential extinction ($\Delta A_{Ks}$), IMF slope, initial cluster mass, and metallicity (see vertical red lines). Each parameter falls well within the 68\% (1$\sigma$ equivalent) confidence interval of the distribution (grey shaded regions). \label{fig:linear}}
\end{figure*}

\begin{figure*}[ht!]
\includegraphics[width=180mm]{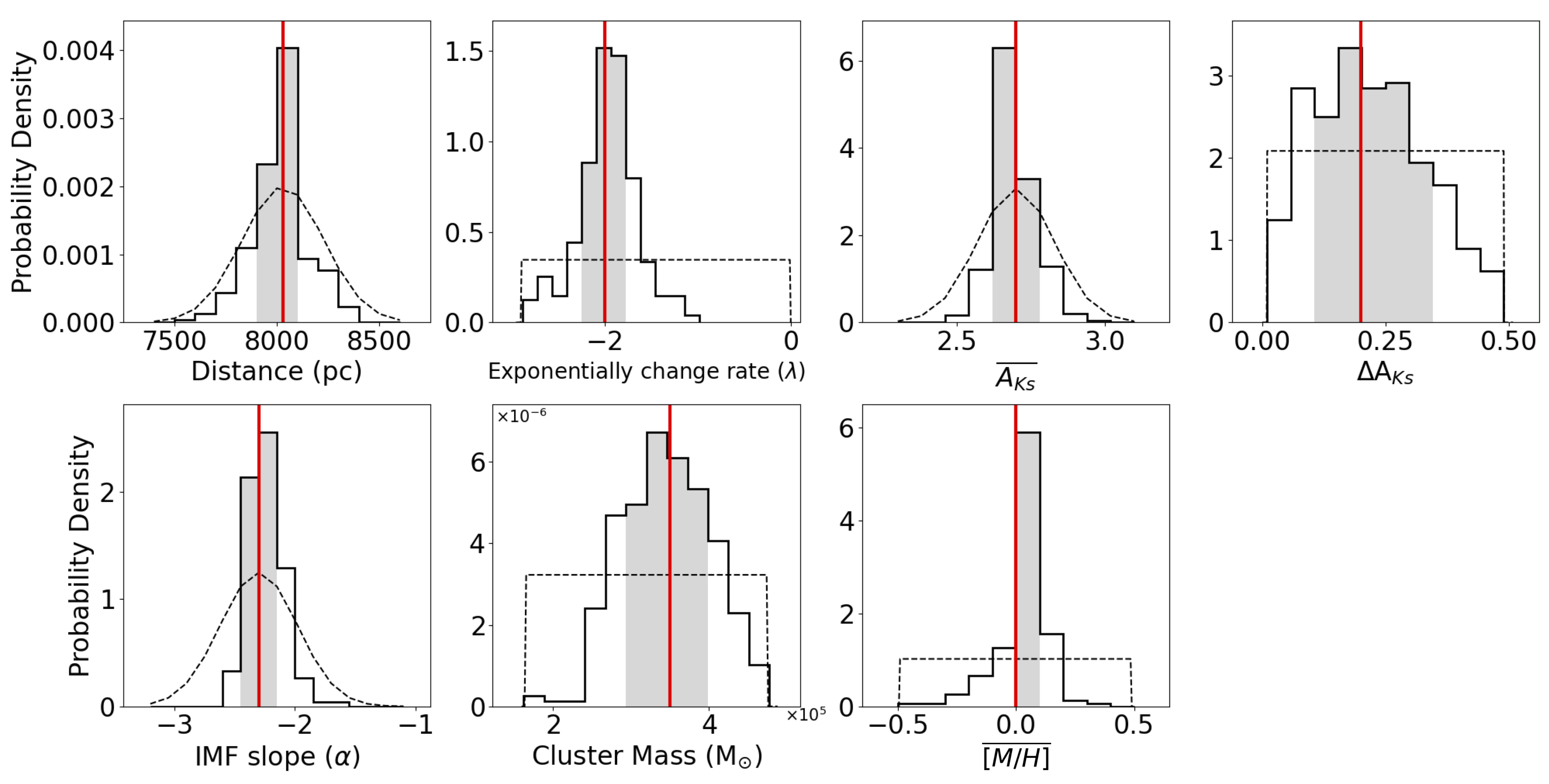}
\centering
\caption{Exponential SFR model testing. The input values for the continuous star formation with an exponentially increasing/decreasing SFR are: distance to the cluster, exponentially change rate ($\lambda$), average extinction ($A_{Ks}$), differential extinction ($\Delta A_{Ks}$), IMF slope, initial cluster mass, and metallicity (see vertical red lines). Each parameter falls well within the 68\% (1$\sigma$ equivalent) confidence interval of the distribution (grey shaded regions). \label{fig:exp}}
\end{figure*}

We test our Bayesian methodology by generating a synthetically ``observed" cluster, and throwing the simulated sample back to the fitter to derive the probability distribution function for each parameter using the Bayesian inference techniques as described in section \ref{sec:bayesian}. All clusters are generated at a distance of 8030 pc, an extinction of $A_{Ks}$ = 2.7, a differential extinction of $\Delta A_{Ks}$ = 0.2, and with a cluster mass aiming to result in a similar number of late-type stars to our observed sample ($\sim$80 stars for AO observations, and $\sim$700 stars for seeing-limited observations). Photometric and spectroscopic uncertainties for simulated cluster stars are added as the Gaussian distribution from the observational uncertainties. We examine the fitter on synthetic clusters with different ages, IMFs, multiplicity, metallicity properties, and star formation history models. Our Bayesian inference methodology is always able to recover the input properties with no substantial systematic biases in the tests on synthetic clusters. We present detailed fitter testings on the cluster age, metallicity, and each star formation history model in the following sections. 

\subsection{Age}\label{sec:age}

Cluster age is correlated with several parameters in the model fitting. We note moderate correlations between the cluster age, IMF slope, cluster mass and average extinction. Here to understand the correlations between parameters and the reliability of the fitter on estimating the age, we simulate clusters at different ages ranging from 0.2 Gyr to 10 Gyr. Each simulated cluster is fitted using our Bayesian inference methods, and then used to examine the fitting results on cluster properties. Figure \ref{fig:age_comp} shows two examples of resulting probability distributions for ages of the simulated clusters with a moderate age of 3.2 Gyr (log(Age) = 9.5), and an old age of 10 Gyr (log(Age) = 10). Through the whole test range of age, the input age and other cluster properties are always recovered within the 68\% (1$\sigma$ equivalent) confidence region of the fitting distribution. Our Bayesian inference methodology is able to recover the age of the cluster with no significant systematic biases. 

\subsection{Metallicity}\label{sec:app-mh}

The measured metallicity distribution of each dataset shows a super-solar peak ($[M/H]$ $\gtrsim$ +0.3) with a long tail towards higher $[M/H]$ close to +1 dex. We further test the fitter using simulations that introduce a bias in the observations to see the effect of an artificial tail in the metallicity distribution at high metallicities. We simulated clusters at different ages with all stars at the peak of the metallicity distribution ($[M/H]$ $\sim$ 0.3 dex). For each cluster, we then artificially introduced a random bias to $[M/H]$ to simulate an apparent tail in the measured metallicity distribution out to $[M/H]$ $\sim$ +1 dex. See Figure \ref{fig:mhfittertest} left for the artificially shifted metallicities. We then proceeded to fit the star formation history as with the real data. Figure \ref{fig:mhfittertest} middle panel shows the resulting probability distribution for a synthetic cluster with an age of 5 Gyr (log(Age) = 9.7). We find that the best fit cluster age of 5.7 $^{+3.8}_{-3.5}$ Gyr is consistent with the input age. Figure \ref{fig:mhfittertest} right panel shows the example of a synthetic cluster with an age of 10 Gyr (log(Age) = 10). The best fit cluster age of 8.9 $^{+4.0}_{-3.6}$ Gyr is consistent with the input age. The fitter is still able to recover the input cluster age with no substantial bias. We find that the peak of the metallicity distribution is much more important to the cluster age estimate than the spread.

\begin{deluxetable*}{lcccccc}
\tablenum{9}
\tablecaption{Results from CO-$T_{eff}$ vs. Starkit $T_{eff}$ \label{tab:teff_comp}}
\tablehead{
\colhead{} & \colhead{} & \multicolumn{2}{c}{Age (Gyr) in this work} & \colhead{}& \multicolumn{2}{c}{Age (Gyr) with fixed $\overline{[M/H]}$} \\
\cline{3-4}
\cline{6-7}
\colhead{Dataset} & \colhead{IMF} & \colhead{with CO-$T_{eff}$} & \colhead{with Starkit $T_{eff}$} & &  \colhead{with CO-$T_{eff}$} & \colhead{with Starkit $T_{eff}$} }
\startdata
AO & Kroupa & 5.0 $^{+3.4}_{-2.3}$ & 4.5 $^{+3.8}_{-2.4}$ & & 8.3 $^{+3.7}_{-3.9}$ & 7.4 $^{+3.4}_{-3.9}$\\
& Top-heavy & 5.5 $^{+3.4}_{-2.5}$ & 4.8 $^{+3.4}_{-2.3}$ & & 8.4 $^{+3.8}_{-3.5}$ & 7.6 $^{+3.1}_{-4.3}$\\
\hline
Seeing-limited & Kroupa & 4.9 $^{+3.8}_{-2.2}$ & 4.8 $^{+3.2}_{-1.8}$  & & 7.9 $^{+3.5}_{-3.4}$ & 7.4 $^{+3.3}_{-4.0}$\\
& Top-heavy & 5.6 $^{+3.3}_{-2.6}$ & 5.5 $^{+2.8}_{-2.3}$ & & 8.7 $^{+3.0}_{-3.9}$ & 8.3 $^{+3.2}_{-3.0}$\\
\enddata
\end{deluxetable*}

\subsection{Star formation history models}
We explore the reliability of the fitter by testing the Bayesian inference methods on all star formation history models summarized in section \ref{sec:sfh}. 

\textbf{(1) Single burst:} The fits on the single burst star formation history model have been well tested through single-age synthetic cluster modelings. See Figure \ref{fig:test_single} for one example. A handful of similar cluster tests were performed with different ages, masses, IMF slopes, parameter priors, and the input and output parameters always agree very well within the 68\% (1$\sigma$ equivalent) confidence intervals.

\textbf{(2) Two bursts:} For multiple bursts, we assume that all stars in the NSC from different bursts have the same observational physical conditions including the same distance (d), average extinction ($A_{Ks}$), differential extinction ($\Delta A_{Ks}$) and a constant IMF slope for all subgroups of the NSC. For $i^{th}$ burst, we model the age log($t_i$), the metallicity $\overline{[M/H]}_i$ and the mass fraction of the single star burst (also see Table \ref{tab:SFH_model}). 

Figure \ref{fig:2pop} shows one example of the output posterior probability distributions on fitting one synthetic cluster including two bursts: \textbf{burst 1} with an age of 4 Gyr (log(Age$_1$) = 9.6), mass fraction of 90\% and metallicity of $\overline{[M/H]}_{1}$ = 0.3; \textbf{burst 2} with an age of 0.13 Gyr (log(Age$_2$) = 8.1), mass fraction of 10\% and metallicity of $\overline{[M/H]}_{2}$ = -1.45. The total cluster mass of the two bursts was set to produce comparable total number of stars as observed in the dataset. Similar cluster tests were performed with different ages, mass fraction and metallicity of the two bursts, and the input and output parameters always agree very well within the 68\% (1$\sigma$ equivalent) confidence interval.

\textbf{(3) Three bursts:} Similar to the two bursts model, Figure \ref{fig:3pop} shows one example of the output posterior probability distributions on fitting one synthetic cluster including three bursts: \textbf{burst 1} with an age of 6.3 Gyr (log(Age$_1$) = 9.8), mass fraction of 70\% and metallicity of $\overline{[M/H]}_{1}$ = 0.45; \textbf{burst 2} with an age of 1.6 Gyr (log(Age$_2$) = 9.2), mass fraction of 21\% and metallicity of $\overline{[M/H]}_{2}$ = 0; \textbf{burst 3} with an age of 0.32 Gyr (log(Age$_3$) = 8.5), mass fraction of 9\% and metallicity of $\overline{[M/H]}_{3}$ = -1.05. Similar cluster tests were performed with different ages, mass fraction and metallicity of the three bursts, and the input and output parameters always agree very well within the 68\% (1$\sigma$ equivalent) confidence interval. The fitter is reliable to characterize the multiple bursts signatures from our observed sample. 

\textbf{(4) Continuous star formation with a linear SFR:} Continuous star formation between 30 Myr and 10 Gyr ago, with a linearly increasing/decreasing SFR(t) $\propto$ $mt$. For continuous star formation, we assume that all stars in the NSC have the same metallicity. Figure \ref{fig:linear} shows one example of the output posterior probability distributions on fitting the synthetic cluster from continuous star formation with a linearly increasing SFR ($m$ = 0.25). Similar tests were performed with different linearly change rates (either increasing or decreasing), and the input and output parameters always agree very well within the 68\% (1$\sigma$ equivalent) confidence interval.

\textbf{(5) Continuous star formation with an exponential SFR}: Continuous star formation between 30 Myr and 10 Gyr ago, with an exponentially increasing/decreasing SFR(t) $\propto$ $e^{\lambda t}$.  Similarly, we assume that all stars in the NSC have the same metallicity. Figure \ref{fig:exp} shows one example of the output posterior probability distributions on fitting the synthetic cluster from continuous star formation with an exponentially increasing SFR ($\lambda$ = -2.0). Similar tests were performed with different exponentially change rates (either increasing or decreasing), and the input and output parameters always agree very well within the 68\% (1$\sigma$ equivalent) confidence interval.

\section{Measurements of stellar effective temperature}\label{sec:app-teff}

Two different methods are used to measure the stellar effective temperature $T_{eff}$ from the spectra: 

1) CO-$T_{eff}$: derived from the well-calibrated relation of the $T_{eff}$ with the CO equivalent width $EW_{CO}$ using the stars of the spectral library \citep{Feldmeier-Krause et al. 2017}, where the $EW_{CO}$ was defined by \citet{Frogel et al. 2001} 
\begin{equation}\label{equ:coteff}
  \begin{aligned}
    T_{eff} = 5677^{\pm 21}K - 106.3^{\pm 3.0}K {\AA}^{-1} \times EW_{CO} 
  \end{aligned}
\end{equation}
where $EW_{CO}$ is in \AA, and $T_{eff}$ in K. The uncertainties are the formal fit uncertainties by fitting the template stars and the residual scatter is 163 K. The uncertainties on the CO indices $\sigma_{EW_{CO}}$ are computed based on 500 Monte Carlo runs of adding the noise. The statistical uncertainty on the effective temperature $\sigma_{T_{eff}, stats}$ = 106.3 $\times$ $\sigma_{EW_{CO}}$, and the systematic uncertainty is $\sigma_{T_{eff}, sys}$ $\sim$163 K. The total uncertainty $\sigma_{T_{eff}, tot}$ is then calculated by adding statistical and systematic uncertainties in quadrature. 

2) Starkit-$T_{eff}$, derived from full spectrum fitting using STARKIT code \citep{Kerzendorf Do 2015} with synthetic grids. The code interpolates on a grid of synthetic spectra and then utilizes the Bayesian sample MultiNest in the fits. The AO observed spectra were fitted to a MARCS grid \citep{Gustafsson et al. 2008} of synthetic models while the seeing-limited spectra were fitted to a PHOENIX grid \citep{Husser et al. 2013}. Several sources of uncertainties are considered including the statistical uncertainty, interpolation uncertainty between spectra grids, and systematic uncertainty by comparing to standard spectral library in the literature. The total uncertainty $\sigma_{T_{eff}, tot}$ is then calculated by adding all these in quadrature. 

See Table \ref{tab:do_obs} and \ref{tab:fk_obs} for the summary of the stellar effective temperature measurements and the uncertainties from the two different methods for each dataset respectively. 

\begin{figure}[ht!]
\includegraphics[width=87mm]{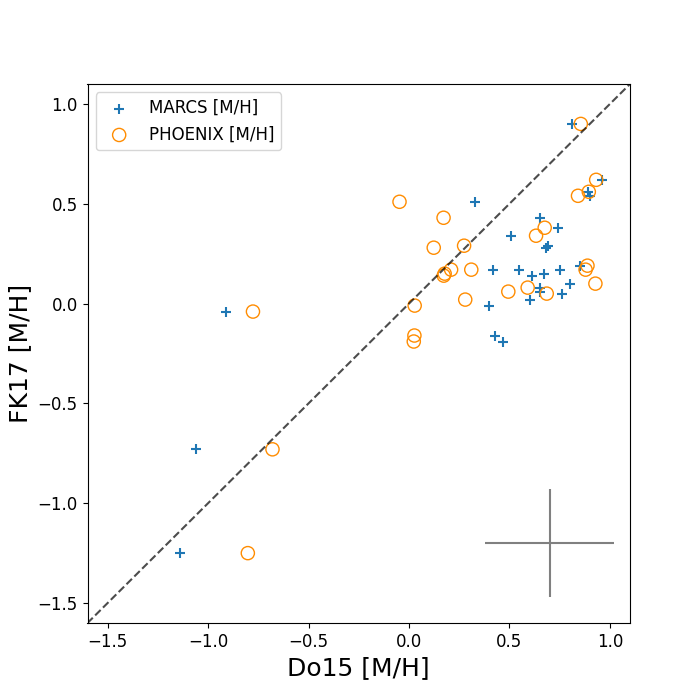}
\centering
\caption{Comparison of the metallicity measurements for the 27 common stars from both datasets. Blue crosses show the metallicities as measured from the AO spectra using MARCS grid \citep{Do et al. 2015}, compared to those as measured from the seeing-limited spectra using PHOENIX grid \citep{Feldmeier-Krause et al. 2017}. We investigated the effects of spectral resolution and grids on the measurements by re-fitting the AO spectra using the PHOENIX grid (see orange open circles with updated x-axis values). The grey error bar on the lower right shows average uncertainties. \label{fig:mh_grids_comp}}
\end{figure}

\section{Measurements of stellar metallicity}\label{sec:app-mh-comp}

The two datasets used in this work were observed with different spectral resolution, and analyzed using different spectral grids. The AO spectra (R $\sim$ 5,000) were fitted with the MARCS spectral grid, while the seeing-limited spectra (R $\sim$ 3310 - 4660) were fitted with the PHOENIX spectra grid. We compare the metallicity measurements for the 27 common stars from both datasets. See Figure \ref{fig:mh_grids_comp}. The median difference is 0.31 dex with a standard deviation of 0.35 dex. We note that very metal-rich stars with $[M/H]$ $>$ +0.5 dex generally show a larger discrepancy between the two measurements owing to the greater systematic uncertainties in the high-metallicity range. See Appendix \ref{sec:app-mh} for further discussion on how these metal-rich stars may affect the fitting results on the cluster age.

We further investigated the effects of spectral resolution and spectral grids on the metallicity measurements by re-fitting the original AO spectra using the PHOENIX grid (low-spectral-resolution) with the same settings as the seeing-limited dataset. See Figure \ref{fig:mh_grids_comp} for the measurements. The median difference between the MARCS-grid and PHOENIX-grid AO measurements is 0.14 dex. The median difference between the PHOENIX-grid AO measurements and the PHOENIX-grid seeing-limited measurements is 0.17 dex. The re-fitted metallicity measurements show that both the resolution of the spectra and the grids have about the same effect on the overall difference between the two datasets. Adding them together results in the total difference between the two datasets. The 27 common stars between the two surveys have consistent metallicity measurements within the uncertainties of each of the method, indicating that the two datasets with different spectral resolution and grids are in reasonable agreement.

\section{Comparison between the AO \& seeing-limited results}\label{sec:app_2datasets}

The star formation history results that are modeled from two datasets with different detection depth and spatial coverage are in great agreement. The star formation history of the NSC shows no substantial discrepancy at different distances to the Galactic center within the central $\sim$1.5 pc. In total we have included $\sim$25\% of the total cluster mass covering the central $\sim$4 pc$^2$ at a distance of 8 kpc. By modeling the two datasets independently, we can investigate if there are systematic differences between the datasets and assess the accuracy of our results. We fitted independently the two datasets that were observed using different telescopes and instruments, and analyzed using different spectral grids. Importantly, we obtain consistent age estimates for both bursts from the star formation history modelings. The possible systematic uncertainties we discussed (see section \ref{sec:res_system}) have been well represented in our reported 68\% confidence interval, and thus our reported results are robust and confident. 

The deeper AO dataset seems to be more useful in modeling the star formation history. The NIFS AO observations with higher spatial and spectral resolution are more sensitive to differentiate supersolar and subsolar metallicity stars (see details in \citealt{Do et al. 2015, Feldmeier-Krause et al. 2017}). Furthermore, the deeper AO observations are able to detect the fainter and low-metallicity stars below K = 14 mag, enabling a more intrinsic estimate of the mass fraction of each burst and a better constraint on the age of the metal-poor population. In the future, AO observations with wider area coverage, as well as a higher spectral resolution, will be helpful to further constrain the star formation history. In addition, the James Webb Space telescope will have the ability to obtain spectra with increased depth and wavelength coverage, and thus will largely increase the number of the observed subsolar metallicity stars and help to place constraints on the origin of their progenitors.

\tablenum{10}
\begin{deluxetable*}{lcccccccccccccc}
\centering
\tablecolumns{14} 
\tablewidth{0pc} 
\tablecaption{Summary of AO Observations}
\tablehead{ 
    \colhead{Name$^{a}$} &
	\colhead{R.A.} &
	\colhead{Dec.} &
	\colhead{$K_S^{b}$} &
	\colhead{$K_{S, err}$} &
	\colhead{$H^{b}$} &
	\colhead{$H_{err}$} &
    \colhead{$A_{K_S}^{b}$} &
	\colhead{$T_{eff, CO}^{c}$} &
    \colhead{$\sigma_{T_{eff, CO}}$} &
    \colhead{$T_{eff, *}^{d}$} &
    \colhead{$\sigma_{T_{eff, *}}$} &
	\colhead{[M/H]$^{d}$} &
	\colhead{$\sigma_{[M/H]}$}\\
	\colhead{} & 
	\colhead{($^{\circ}$)} & 
	\colhead{($^{\circ}$)} & 
	\colhead{(mag)} & 
	\colhead{(mag)} & 
	\colhead{(mag)} & 
	\colhead{(mag)} & 
	\colhead{(mag)} & 
	\colhead{(K)} & 
	\colhead{(K)} & 
	\colhead{(K)} & 
	\colhead{(K)}
}
\startdata
E5-1-001 & 266.421656 & -29.007947 & 12.01 & 0.01 & 14.19 & 0.01 & 2.66 & 3260 & 171 & 3497 & 413 & 0.96 & 0.32 \\
E5-1-002 & 266.421449 & -29.007402 & 12.61 & 0.01 & 14.51 & 0.01 & 2.47 & 3673 & 168 & 3671 & 414 & 0.55 & 0.32 \\
E5-1-003 & 266.421601 & -29.007516 & 13.15 & 0.01 & 15.11 & 0.01 & 2.50 & 3611 & 170 & 3597 & 414 & 0.85 & 0.32 \\
\enddata
\tablenotetext{a}{Name from \citet{Stostad et al. 2015}.}
\tablenotetext{b}{$K_S$- and $H$-band photometry and $A_{K_S}$ extinction taken from \cite{Schodel et al. 2010}; see section \ref{sec:dataset1}. This is the first time the reported matches to those stars were made between the catalogs.}
\tablenotetext{c}{Effective temperature as derived from the calibrated $T_{eff}$-$EW_{CO}$ relation; see Appendix \ref{sec:app-teff} for details.}
\tablenotetext{d}{Effective temperature and metallicity derived from full spectrum fitting using the STARKIT code, \cite{Kerzendorf Do 2015}.}
\tablenotetext{}{(The full table is available online)}
\label{tab:do_obs}
\end{deluxetable*}

\tablenum{11}
\begin{deluxetable*}{ccccccccccccccc}
\centering
\tablecolumns{14} 
\tablewidth{0pc} 
\tablecaption{Summary of seeing-limited Observations}
\tablehead{ 
    \colhead{Name$^{a}$} &
	\colhead{R.A.} &
	\colhead{Dec.} &
	\colhead{$K_S^{b}$} &
	\colhead{$K_{S, err}$} &
	\colhead{$H^{b}$} &
	\colhead{$H_{err}$} &
    \colhead{$A_{K_S}^{b}$} &
	\colhead{$T_{eff, CO}^{c}$} &
    \colhead{$\sigma_{T_{eff, CO}}$} &
    \colhead{$T_{eff, *}^{d}$} &
    \colhead{$\sigma_{T_{eff, *}}$} &
	\colhead{[M/H]$^{d}$} &
	\colhead{$\sigma_{[M/H]}$}\\
	\colhead{} & 
	\colhead{($^{\circ}$)} & 
	\colhead{($^{\circ}$)} & 
	\colhead{(mag)} & 
	\colhead{(mag)} & 
	\colhead{(mag)} & 
	\colhead{(mag)} & 
	\colhead{(mag)} & 
	\colhead{(K)} & 
	\colhead{(K)} & 
	\colhead{(K)} & 
	\colhead{(K)}
}
\startdata
1 & 266.41675 & -29.010296 & 9.95 & 0.01 & 12.07 & 0.01 & 2.51 & 2982 & 251 & 3190 & 209 & 0.87 & 0.31		\\
5 & 266.41571 & -29.012167 & 10.54 & 0.01 & 12.80 & 0.01 & 2.66 & 3119 & 164 & 3301 & 206 & 0.13 & 0.25		\\
6 & 266.42401 & -29.003611 & 10.57 & 0.90 & 12.65 & 0.90 & 2.78 & 3408 & 181 & 3374 & 205 & 0.14 & 0.25		\\
\enddata
\tablenotetext{a}{Name from \citet{Feldmeier-Krause et al. 2017}.}
\tablenotetext{b}{$K_S$- and $H$-band photometry and $A_{K_S}$ extinction taken from \cite{Schodel et al. 2010, Nogueras-Lara et al. 2018a} and \cite{Nishiyama et al. 2006}; see section \ref{sec:dataset2}. This is the first time the reported matches to those stars were made between the catalogs.}
\tablenotetext{c}{Effective temperature as derived from the calibrated $T_{eff}$-$EW_{CO}$ relation; see Appendix \ref{sec:app-teff} for details.}
\tablenotetext{d}{Effective temperature and metallicity derived from full spectrum fitting using the STARKIT code, \cite{Kerzendorf Do 2015}.}
\tablenotetext{}{(The full table is available online)}
\label{tab:fk_obs}
\end{deluxetable*}

\end{document}